\documentclass[10pt,sigconf]{acmart}
\usepackage{subcaption}
\usepackage{graphicx}

\usepackage{enumitem}
\usepackage{algorithm2e}
\usepackage{array}
\usepackage{soul, color}

\usepackage[all]{nowidow}

\usepackage{thmtools,thm-restate}

\definecolor{lightyellow}{rgb}{1,1,.2}
\definecolor{lightgreen}{rgb}{.85,1,.8}
\sethlcolor{lightyellow}
\setulcolor{blue}

\soulregister\cite7
\soulregister\ref7
\soulregister\pageref7

\newtheorem{lemma}{Lemma}
\newtheorem{definition}{Definition}
\newtheorem{property}{Property}
\newtheorem{corollary}{Corollary}

\newcommand{\eat}[1]{} 
\newcommand{\cut}[1]{} 

\newcommand{\tpcds}[0]{TPC-DS\xspace}
\newcommand{\job}[0]{JOB\xspace}

\newcommand{\mssales}[0]{CUSTOMER\xspace}
\newcommand{\sqlserver}[0]{DBMS-X\xspace}
\renewcommand{\sqlserver}[0]{Microsoft SQL Server\xspace}

\newcommand{\associativity}{\textit{associativity property}\xspace}
\newcommand{\commutativity}{\textit{commutativity property}\xspace}
\newcommand{\redundancy}{\textit{redundancy property}\xspace}
\newcommand{\reduction}{\textit{reduction property}\xspace}
\newcommand{\absorption}{\textit{absorption rule}\xspace}

\newcommand{\cameraready}{}

\ifdefined\cameraready
\newcommand{\todo}[1]{}
\newcommand{\changed}[1]{#1}
\newcommand{\done}[1]{}
\newcommand{\outline}[1]{}
\newcommand{\point}[1]{}
\else
	\ifdefined\submit
	\newcommand{\todo}[1]{}
	\newcommand{\changed}[1]{{\color{blue}#1}}
	\newcommand{\done}[1]{}
	\newcommand{\outline}[1]{}
	\newcommand{\point}[1]{}
	\long\def\tocut#1{}
	\else
		\ifdefined\revision
			\newcommand{\todo}[1]{\textcolor{red}{\bf [TODO!: #1]}}
			\newcommand{\changed}[1]{{\color{blue}#1}}
			\newcommand{\done}[1]{}
			\newcommand{\outline}[1]{}
			\newcommand{\point}[1]{}
		\else
			\newcommand{\todo}[1]{\textcolor{red}{\bf [TODO!: #1]}}
			\newcommand{\changed}[1]{{\color{blue}#1}}
			\newcommand{\tocut}[1]{\textcolor{red}{\it\st{#1}}}
			\newcommand{\done}[1]{\textcolor{blue}{\bf [DONE!]}}
			\newcommand{\outline}[1]{\textcolor{blue}{\bf [OUTLINE!: #1]}}
			\newcommand{\point}[1]{\textcolor{teal}{\bf [POINT!: #1]}}
		\fi
	\fi
\fi

\newcommand{\maxworkloadreduction}{64\%\xspace}
\newcommand{\avgworkloadreduction}{37\%\xspace}
\newcommand{\maxjoinreductionworkload}{\job}
\newcommand{\maxjoinreductionworkloadbefore}{0.50\xspace}
\newcommand{\maxjoinreductionworkloadafter}{0.24\xspace}
\newcommand{\maxjoinreduction}{52\%\xspace}

\AtBeginDocument{%
	\providecommand\BibTeX{{%
			\normalfont B\kern-0.5em{\scshape i\kern-0.25em b}\kern-0.8em\TeX}}}

\copyrightyear{2020}
\acmYear{2020}
\setcopyright{acmcopyright}
\acmConference[SIGMOD '20] {2020 ACM SIGMOD International Conference on Management of Data}{June 14--19, 2020}{Portland, OR, USA}
\acmBooktitle{2020 ACM SIGMOD International Conference on Management of Data (SIGMOD'20), June 14--19, 2020, Portland, OR, USA}
\acmPrice{15.00}
\acmDOI{10.1145/3318464.3389769}
\acmISBN{978-1-4503-6735-6/20/06}

\renewcommand\footnotetextcopyrightpermission[1]{} 

\thanks{*~This technical report is an extended version of the ACM SIGMOD 2020 paper \textit{Bitvector-aware Query Optimization for Decision Support Queries}~\cite{bqo20sigmod}.}

\begin{document}
	
\fancyhead{}
	
	%
	\title{Bitvector-aware Query Optimization for Decision Support Queries (extended version)*}
	
	%
	\cut{	\author{Anonymous Authors}
	\affiliation{%
	}}
	
		\author{Bailu Ding}
	\affiliation{%
		\institution{Microsoft Research}
	}	
	\email{badin@microsoft.com}
	
			\author{Surajit Chaudhuri}
	\affiliation{%
		\institution{Microsoft Research}
	}
	\email{surajitc@microsoft.com}
	
			\author{Vivek Narasayya}
	\affiliation{%
		\institution{Microsoft Research}
	}
	\email{viveknar@microsoft.com}
	
	%
	\renewcommand{\shortauthors}{Bailu Ding, et al.}
	
	%
	\begin{abstract}

Bitvector filtering is an important query processing technique that can significantly reduce the cost of execution, especially for complex decision support queries with multiple joins. Despite its wide application, however, its implication to query optimization is not well understood.

In this work, we study how bitvector filters impact query optimization. We show that incorporating bitvector filters into query optimization straightforwardly can increase the plan space complexity by an exponential factor in the number of relations in the query. We analyze the plans with bitvector filters for star and snowflake queries in the plan space of right deep trees without cross products. Surprisingly, with some simplifying assumptions, we prove that, the plan of the minimal cost with bitvector filters can be found from a linear number of plans in the number of relations in the query. This greatly reduces the plan space complexity for such queries from \changed{exponential} to linear.

Motivated by our analysis, we propose an algorithm that \changed{accounts for the impact of bitvector filters in query optimization. Our algorithm optimizes the join order for an arbitrary decision support query} by choosing from a linear number of candidate plans in the number of relations in the query. We implement our algorithm in \sqlserver as a transformation rule. Our evaluation on both industry standard benchmarks and customer workload shows that, compared with the original \sqlserver, our technique reduces the total CPU execution time by 22\%-\maxworkloadreduction for the workloads, with up to two orders of magnitude reduction in CPU execution time for individual queries.

\cut{
Bitvector filters\cut{implement semi-join reductions to} prune out tuples that will not qualify join conditions early in the query execution pipeline. They are an important query processing technique that can significantly reduce the cost of execution, especially for complex decision support queries with multiple joins. Despite their wide application, however, their implication to query optimization is not well understood.

In this work, we study how bitvector filters impact query optimization. We show that a straightforward integration of bitvector filters with the query optimizer can increase the plan space complexity by an exponential factor w.r.t. the number of relations in the query. 

Instead, we propose an algorithm that \changed{accounts for the impact of bitvector filters in query optimization and optimizes the join order for an arbitrary decision support query} by choosing from a linear number of candidate plans w.r.t. the number of relations in the query. We further prove that, surprisingly, \changed{for important classes of decision support queries in the plan space of right deep trees without cross products, our algorithm is optimal under simplifying assumptions.} This greatly reduces the plan space complexity for such queries from \changed{exponential} to linear. We implement our algorithm in a commercial database \sqlserver as a transformation rule. Our evaluation on both standard and customer benchmarks shows that, compared with \sqlserver, our technique reduces the total CPU execution time by 22\%-\maxworkloadreduction for the workloads, with up to two orders of magnitude reduction in CPU execution time for individual queries.
}

\cut{
Bitvector filters\cut{implement semi-join reductions to} effectively prune out tuples that will not qualify join conditions early in the query execution pipeline. They have been used as important query execution techniques to reduce the cost of execution, especially for complex decision support queries. Despite their wide applications and decades of research, however, bitvector filters stay as query execution techniques, and their implication to query optimization is not well understood.

In this work, we study how bitvector filters impact the landscape of query optimization. We show that the existing top-down or bottom-up query optimization framework is not able to incorporate bitvector filters due to violating substructure optimality in dynamic programming, and a naive integration can increase the plan space complexity by an exponential factor w.r.t. the number of relations in the query. 

Despite the hardness of integrating bitvector filters into general query optimization, we focus on \changed{join order optimization for} decision support queries in the plan space of right deep trees without cross products. We prove that, surprisingly, \changed{for important classes of decision support queries}, the number of candidate plans with minimal cost is only linear w.r.t the number of relations with bitvector filters. This greatly reduces the plan space complexity for such queries from \changed{exponential} to linear. Based on our findings, we propose an algorithm to optimize join orders for  \changed{arbitrary} decision support queries.\cut{ \changed{We further optimize our algorithm by adding bitvector filters selectively based on estimated benefit.}} We implement our algorithm in \sqlserver as a transformation rule. Our evaluation on both industry benchmarks and customer workload shows that, compared with \sqlserver, our technique reduces the total CPU execution time by 22\%-\maxworkloadreduction for the workloads, with up to two orders of magnitude reduction in CPU execution time for individual queries.
}

\end{abstract}

	

	\maketitle

	\section{Introduction}
\outline{Describe what is bitvector filter}

\point{What is bitvector filter}
Bitvector filters, including bitmap or hash filter~\cite{bernstein1981using, graefe1993query, bloom1970space}, \changed{Bloom} filter and its variants~\cite{bloom1970space, lang2019performance, fan2014cuckoo, putze2007cache,almeida2007scalable}, perform 'probabilistic' semi-join reductions to effectively prune out rows that will not qualify join conditions early in the query execution pipeline.
\point{bitvector filters are widely used in commercial systems}
Because they are easy to implement and low in overhead, bitvector filters are widely used in commercial databases~\cite{galindo2008optimizing, Das:2015:QOO:2824032.2824074, Hsiao:1994:PEM:191839.191879, lahiri2015oracle}.

\cut{
\point{What is semi-join}
Semi-join\cite{bernstein1981using} ($\ltimes$) is a relational operator that takes the join of two relations, $R$ and $S$, and returns the tuples in $R$ which join with at least one tuple in $S$. In other words, $R \ltimes S$ filters out the tuples in $R$ that will not qualify the join condition with the natural join of $R$ and $S$.
}

\cut{
\point{Bitvector filters are probabilistic semi-join}
A bitvector filter implements semi-join by creating a bitvector to compactly encode the values of join columns on $S$ and filtering the tuples from $R$ with the bitvector. Because there can be collisions of values in the bitvector, \cut{a tuple in $R$ that qualifies the bitvector may not actually join with a tuple in $S$,} i.e., false positives\cut{. Thus}, bitvector filtering is 'probabilistic'. In practice, a bitvector can compactly encode a large domain of value with a small amount of memory at low false positives rate.
}

\point{Describe how bitvector filters are implemented and the algorithm of bitvector push down for hash join in query processing}

\cut{
\begin{figure*}
	\centering
	\begin{minipage}{.39\linewidth}
		\begin{subfigure}[b]{.32\linewidth}
			\centering
			\includegraphics[width=.9\linewidth]{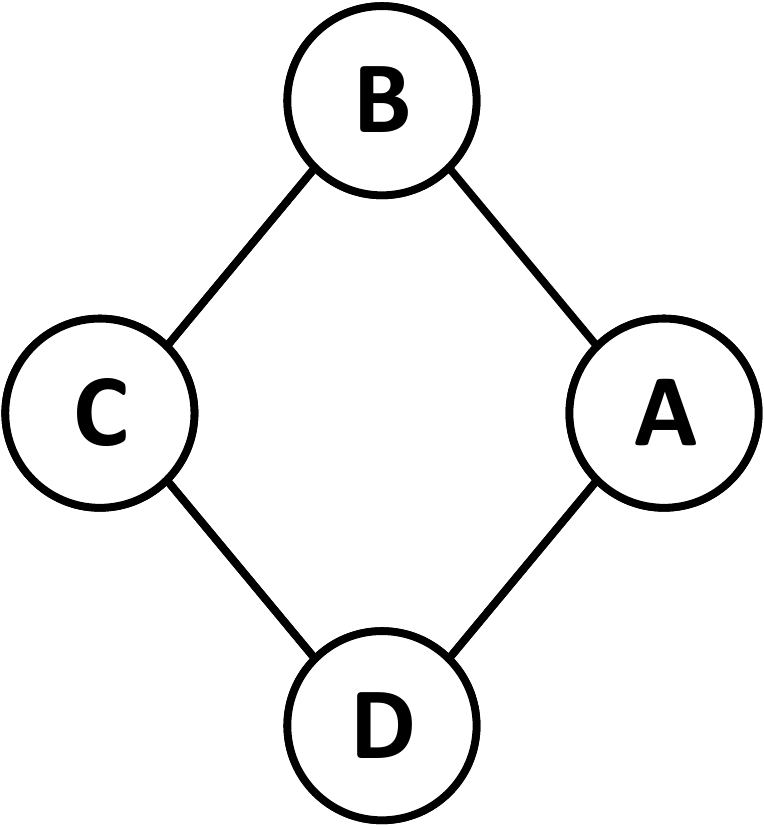}
			\caption{Join graph}
			\label{fig:bitvector_pushdown:join_graph}
		\end{subfigure}
		\begin{subfigure}[b]{.66\linewidth}
			\centering
			\includegraphics[width=.65\linewidth]{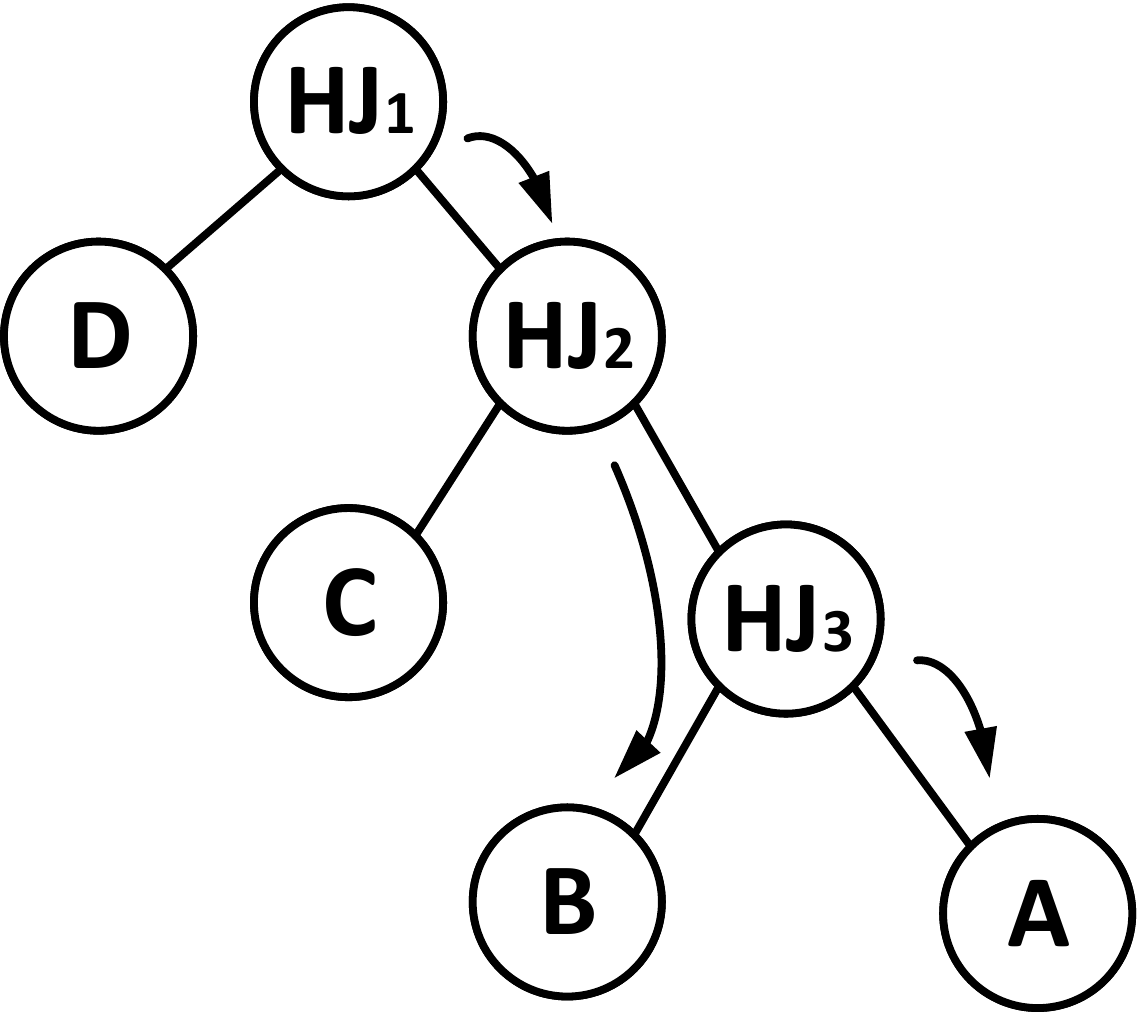}
			\caption{\changed{Plan with bitvector filters}}
			\label{fig:bitvector_pushdown:plan}
		\end{subfigure}
		\caption{Example of pushing down bitvector filters}
	\label{fig:bitvector_pushdown}
	\end{minipage}
\quad
	\begin{minipage}{.58\linewidth}
		\begin{subfigure}[b]{.25\linewidth}
			\centering
			\includegraphics[width=.6\linewidth]{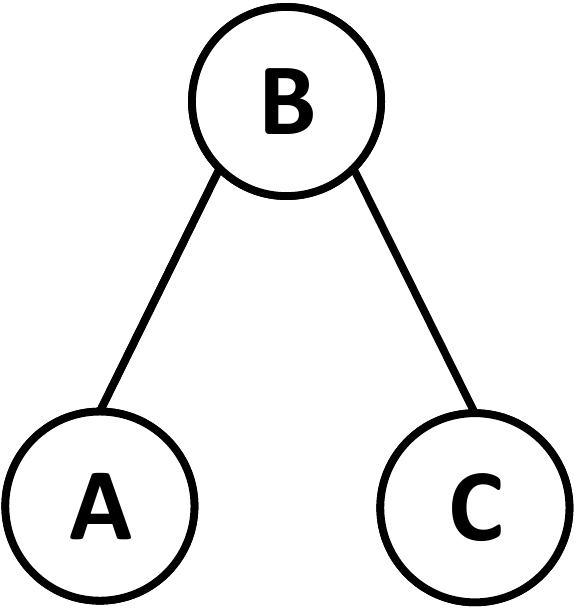}
			\caption{\changed{Join graph}}
			\label{fig:bitvector_plan:join_graph}
		\end{subfigure}
		\begin{subfigure}[b]{.35\linewidth}
			\centering
			\includegraphics[width=.65\linewidth]{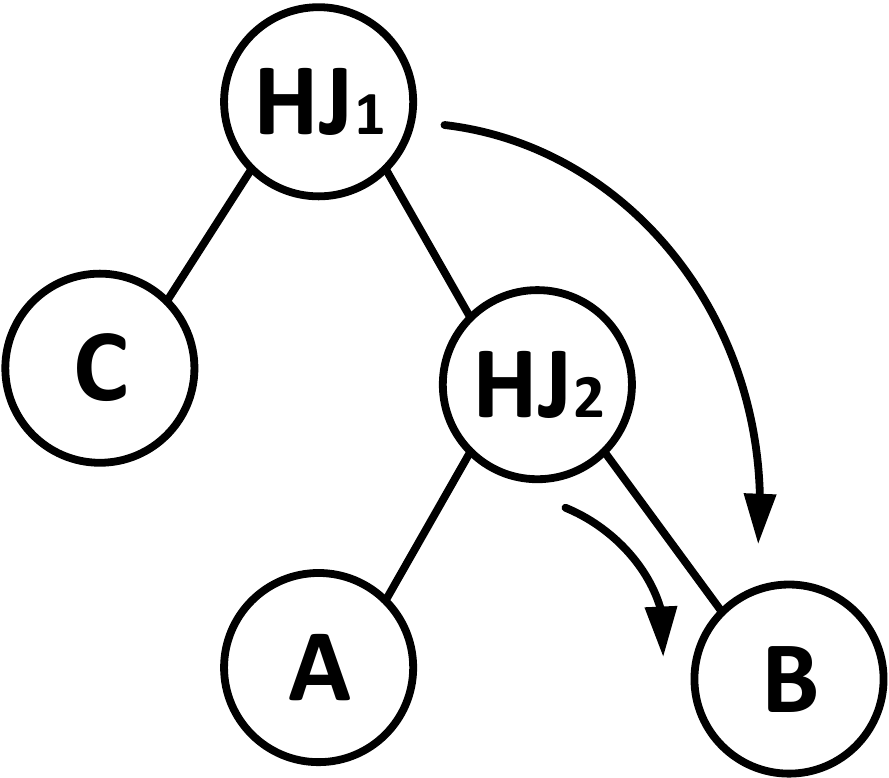}
			\caption{\changed{Best plan without bitvector filters}}
			\label{fig:bitvector_plan:plan_0}
		\end{subfigure}
		\begin{subfigure}[b]{.35\linewidth}
			\centering
			\includegraphics[width=.65\linewidth]{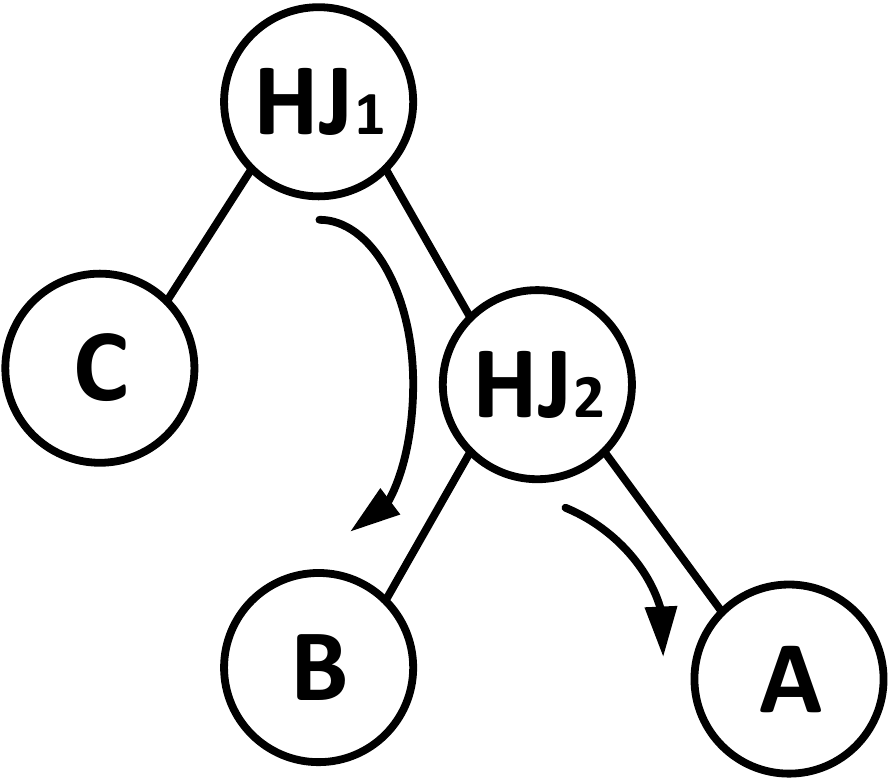}
			\caption{\changed{Best plan with bitvector filters}}
			\label{fig:bitvector_plan:plan_1}
		\end{subfigure}
		\caption{Example of ignoring bitvector filters in query optimization resulting in a suboptimal plan}
		\label{fig:bitvector_plan}
	\end{minipage}

\end{figure*}
}

\begin{figure}
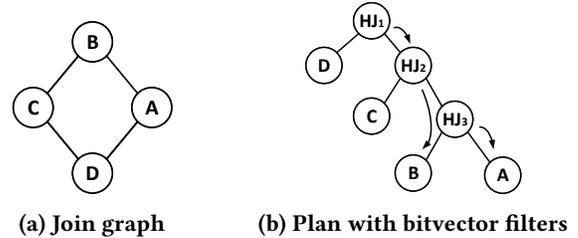

	\begin{subfigure}[b]{.2\textwidth}
		\centering
		\includegraphics[width=.6\linewidth]{./figures/bitvector_pushdown_0}
		\caption{Join graph}
		\label{fig:bitvector_pushdown:join_graph}
	\end{subfigure}
	\begin{subfigure}[b]{.27\textwidth}
		\centering
		\includegraphics[width=.6\linewidth]{./figures/bitvector_pushdown_1}
		\caption{\changed{Plan with bitvector filters}}
		\label{fig:bitvector_pushdown:plan}
	\end{subfigure}
	\caption{Example of pushing down bitvector filters for a query plan joining relations $A, B, C, D$}
	\label{fig:bitvector_pushdown}
\end{figure}

\cut{At a high level, bitvector filters are constructed from the build side relation at a hash join operator and pushed down to the probe side. Since bitvector filters can be constructed as a by-product while building the hash table with the build side relation, the overhead of creating bitvector filters is low.}
\point{Prior work on bitvector filters focus on variants of bitvector filters and how to push down bitvector filters to the query plan}
Prior work on using bitvector filters has heavily focused on optimizing its effectiveness and applicability for query processing.
One line of prior work has explored different schedules of bitvector filters for various types of query plan trees to optimize its effect on query execution~\cite{chen1993applying, chen1997applying, chen1992interleaving}. Many variants of bitvector filters have also been studied that explore the trade-off between the space and accuracy~\changed{\cite{bloom1970space, chan1998bitvector, fan2014cuckoo, lang2019performance, almeida2007scalable,putze2007cache}}.

\changed{In query processing, bitvector filters are mostly used in hash joins~\cite{chen1993applying, chen1997applying, chen1992interleaving}}. \changed{Specifically, the commercial database DBMS-X implements the bitvector filter scheduling algorithm following~\cite{graefe1993query} (Section~\ref{sec:bitvector_algorithm}). At a high level, a \emph{single} bitvector filter is created with the equi-join columns at a hash join operator and is pushed down to the \emph{lowest} possible level of the subplan rooted at the probe side.}
Figure~\ref{fig:bitvector_pushdown} shows an example of applying bitvector filters to a query plan \cut{joining relations $A, B, C$, and $D$}. Figure~\ref{fig:bitvector_pushdown:join_graph} shows the join graph of the query and Figure~\ref{fig:bitvector_pushdown:plan} shows its query plan, \changed{where the arrow in Figure~\ref{fig:bitvector_pushdown:plan} points from the operator that creates the bitvector filter to the operator where the bitvector filter is pushed down to.}  As shown in Figure~\ref{fig:bitvector_pushdown:plan}, a bitvector filter is created from the build side of each hash join operator ($HJ_1$, $HJ_2$, and $HJ_3$). Since $C$ only joins with $B$, the bitvector filter created from $HJ_2$ bypasses $HJ_3$ and is pushed down to $B$. Similarly, because $D$ joins with both $A$ and $C$, the bitvector filter created from $HJ_1$ consists of columns from both $A$ and $C$. Thus, the lowest possible level to push down this bitvector filter is $HJ_2$. \cut{Section~\ref{sec:bitvector_algorithm} describes the bitvector creation and push-down algorithm in detail.} Bitvector filters can also be adapted for merge joins.

\outline{Describe how bitvector filters can reduce query execution cost: join and other operators}
\cut{
\point{Bitvector filters effectively reduces the cost of join operations}
Bitvector filters can effectively reduce the cost of join operators. \cut{Because the cost of a join operator can heavily depend on the cardinalities of the relations involved in the join, eliminating unnecessary tuples that do not contribute to the output of the query from the input relations reduces the cost of the join operation.} In particular, for complex multi-join queries, the bitvector filter can be created from a higher-level join and pushed down to a relation that is multiple levels lower in the query plan. Eliminating unnecessary tuples from input relations as early as possible has a cascading effect in reducing the cost of multiple join operators \changed{as well as other operators (e.g., sort, aggregates, user defined functions)} in the pipeline.
\point{Other impact of bitvector filters}
\changed{Furthermore, if statistical metadata is available for data segments, bitvector filters can also be used to eliminate unqualified data segments to reduce I/O scans as well as data decompression if column stores are used. Finally, bitvector filters can also be adapted for merge joins.
}
}

\cut{
\point{Bitvectors can reduce the I/O scan cost if metadata is available, especially for columnstores}
Bitvector filters can also be used to reduce I/O scan cost, if statistical metadata is available. As DBMS stores data in segments or partitions, together with the statistical metadata of the segment, e.g., min and max values or bit vectors of the values.
When scanning a data segment, the bitvector filters can be first applied to the metadata and filter out a segment if no value in the segment qualifies the bitvector filter. Thus, bitvector filters eliminate reading unnecessary data segments and reduce I/O cost. Such cost reduction is in particular prominent in databases with column stores, where scans dominate data accesses.
}

\point{Little prior work on bitvector filters with QO}
Surprisingly, despite the wide application of and decades of research on bitvector filters for query processing, the impact of bitvector filters on query optimization is not well understood. To the best of our knowledge, most state-of-the-art DBMSs add bitvector filters to the query plans produced by the query optimizer as a post-processing step.

\point{Ignoring bitvector filters in QO can lead to suboptimal query plans}

\cut{
\begin{figure}
	\begin{subfigure}[b]{.12\textwidth}
		\centering
		\includegraphics[width=.6\linewidth]{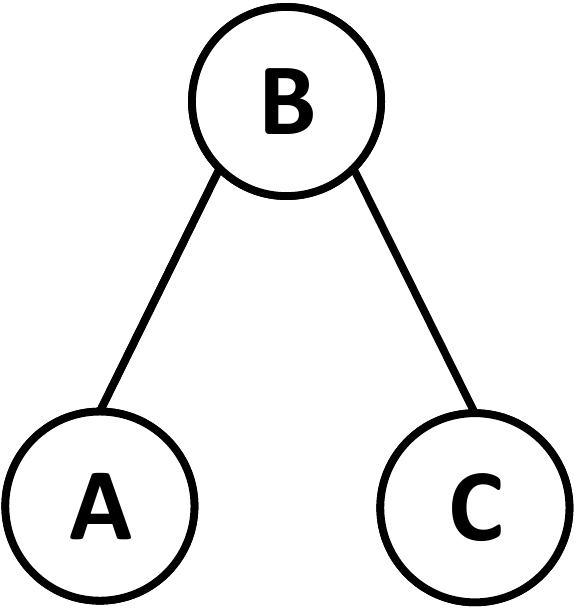}
		\caption{\changed{Join graph}}
		\label{fig:bitvector_plan:join_graph}
	\end{subfigure}
	\begin{subfigure}[b]{.18\textwidth}
		\centering
		\includegraphics[width=.6\linewidth]{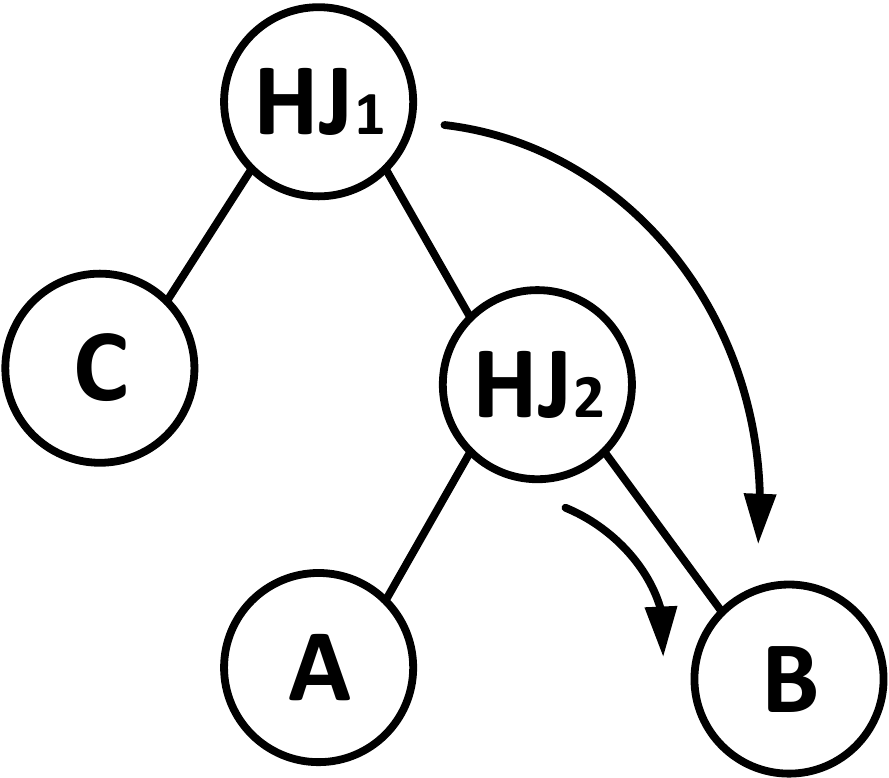}
		\caption{Best plan without bitvector filters}
		\label{fig:bitvector_plan:plan_0}
	\end{subfigure}
	\begin{subfigure}[b]{.19\textwidth}
		\centering
		\includegraphics[width=.6\linewidth]{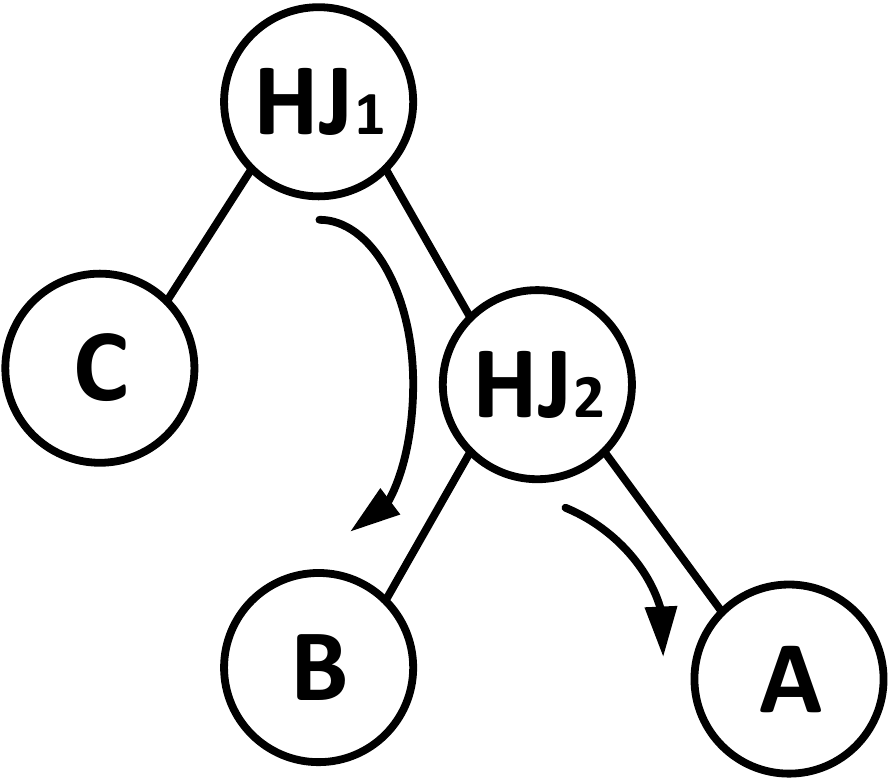}
		\caption{Best plan with bitvector filters}
		\label{fig:bitvector_plan:plan_1}
	\end{subfigure}
	\caption{Example of ignoring bitvector filters in query optimization resulting in a suboptimal plan}
	\label{fig:bitvector_plan}
\end{figure}
}

\begin{figure*}
	\begin{subfigure}[b]{.11\textwidth}
	\centering
	\includegraphics[width=\linewidth]{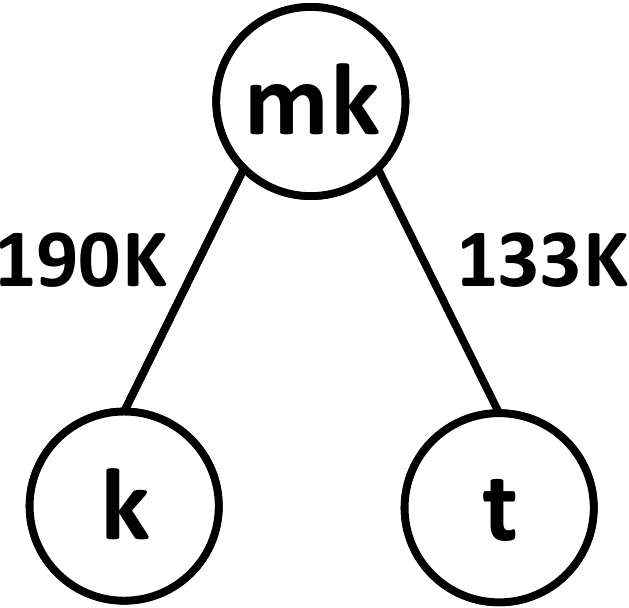}
	\caption{\changed{Join graph w/ cardinality}}
	\label{fig:job_bitvector_plan:join_graph}
	\end{subfigure}
\quad
	\begin{subfigure}[b]{.2\textwidth}
	\centering
	\includegraphics[width=\linewidth]{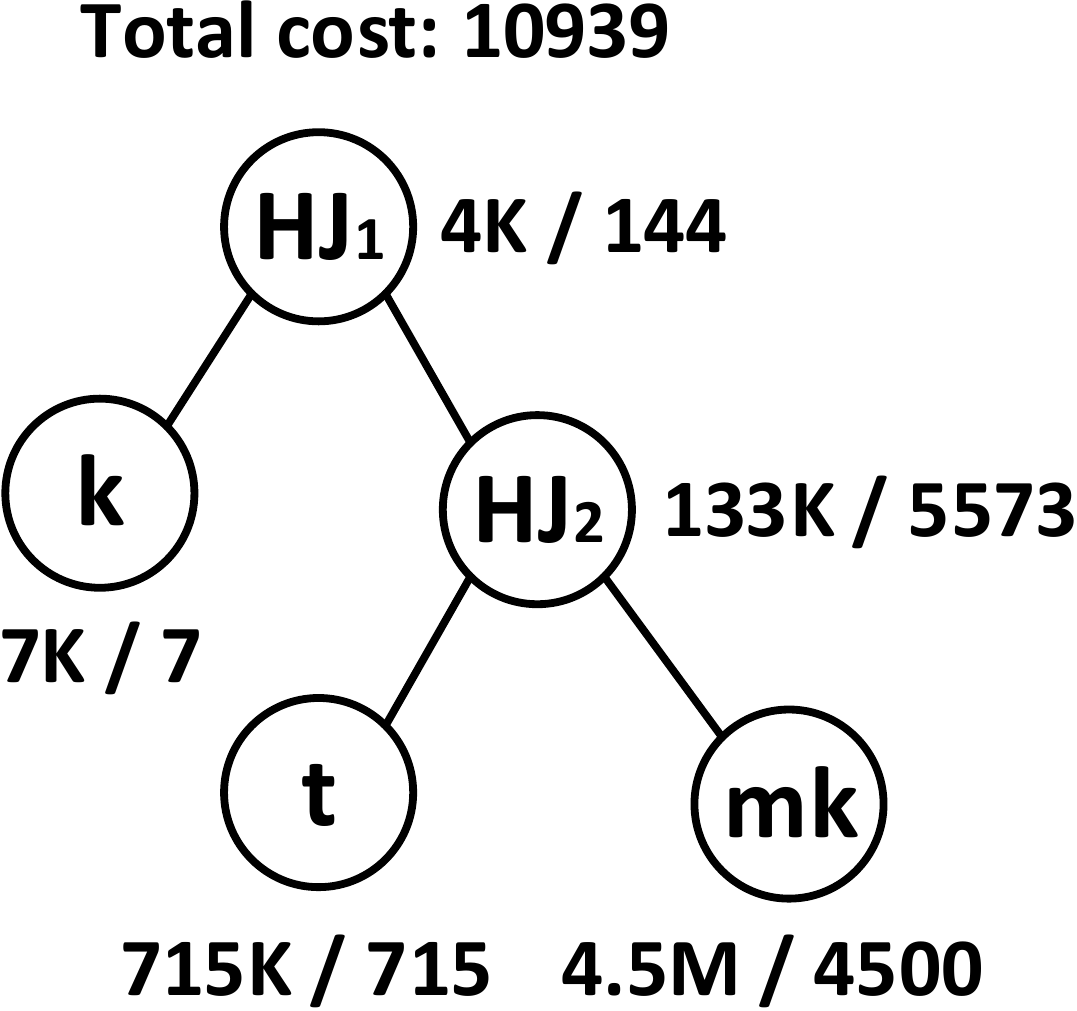}
	\caption{\changed{Best plan $P_1$ without bitvector filters}}
	\label{fig:job_bitvector_plan:plan_0}
	\end{subfigure}
\quad
\begin{subfigure}[b]{.208\textwidth}
	\centering
	\includegraphics[width=\linewidth]{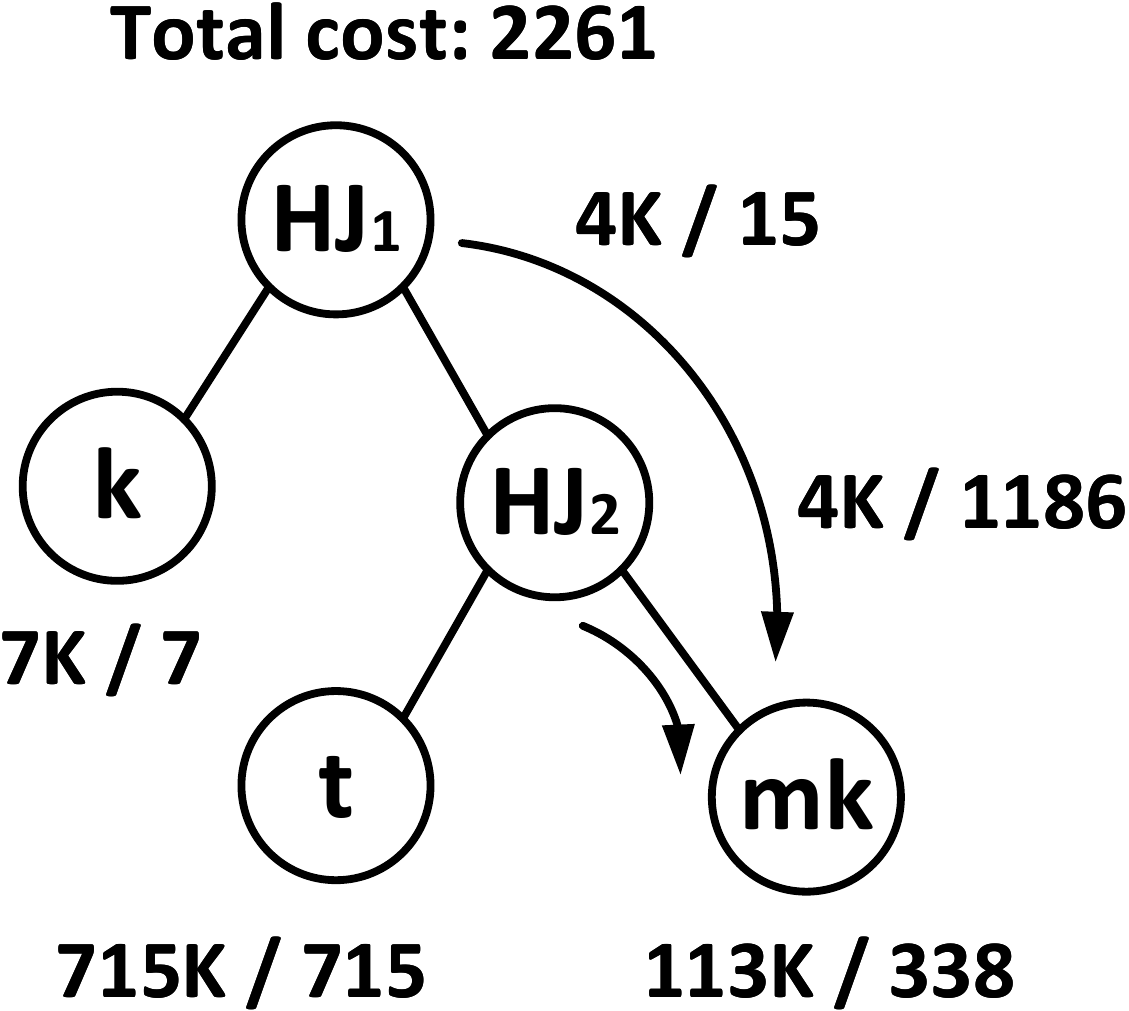}
	\caption{\changed{Post-process $P_1$ by adding bitvector filters}}
	\label{fig:job_bitvector_plan:bitvector_plan_0}
\end{subfigure}
\quad
	\begin{subfigure}[b]{.182\textwidth}
	\centering
	\includegraphics[width=\linewidth]{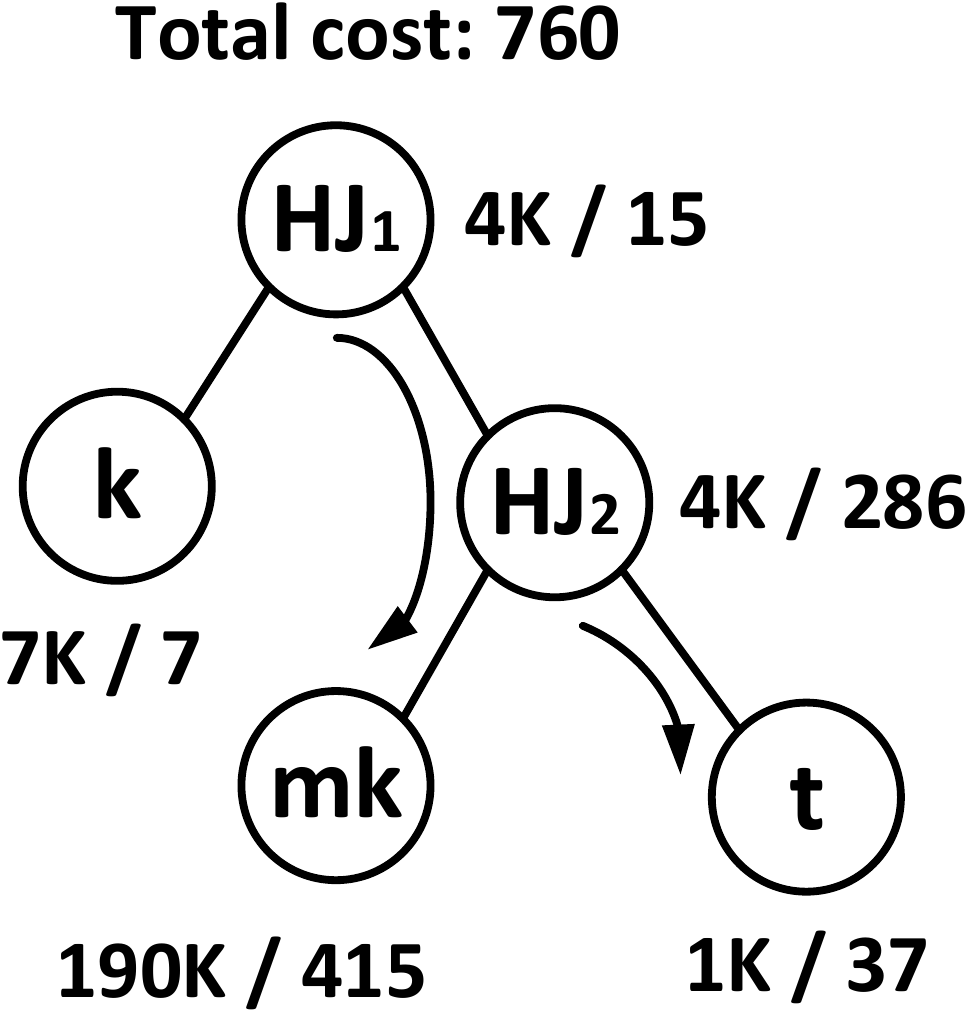}
	\caption{\changed{Best plan $P_2$ with bitvector filters}}
	\label{fig:job_bitvector_plan:bitvector_plan_1}
\end{subfigure}
	\begin{subfigure}[b]{.2\textwidth}
	\centering
	\includegraphics[width=\linewidth]{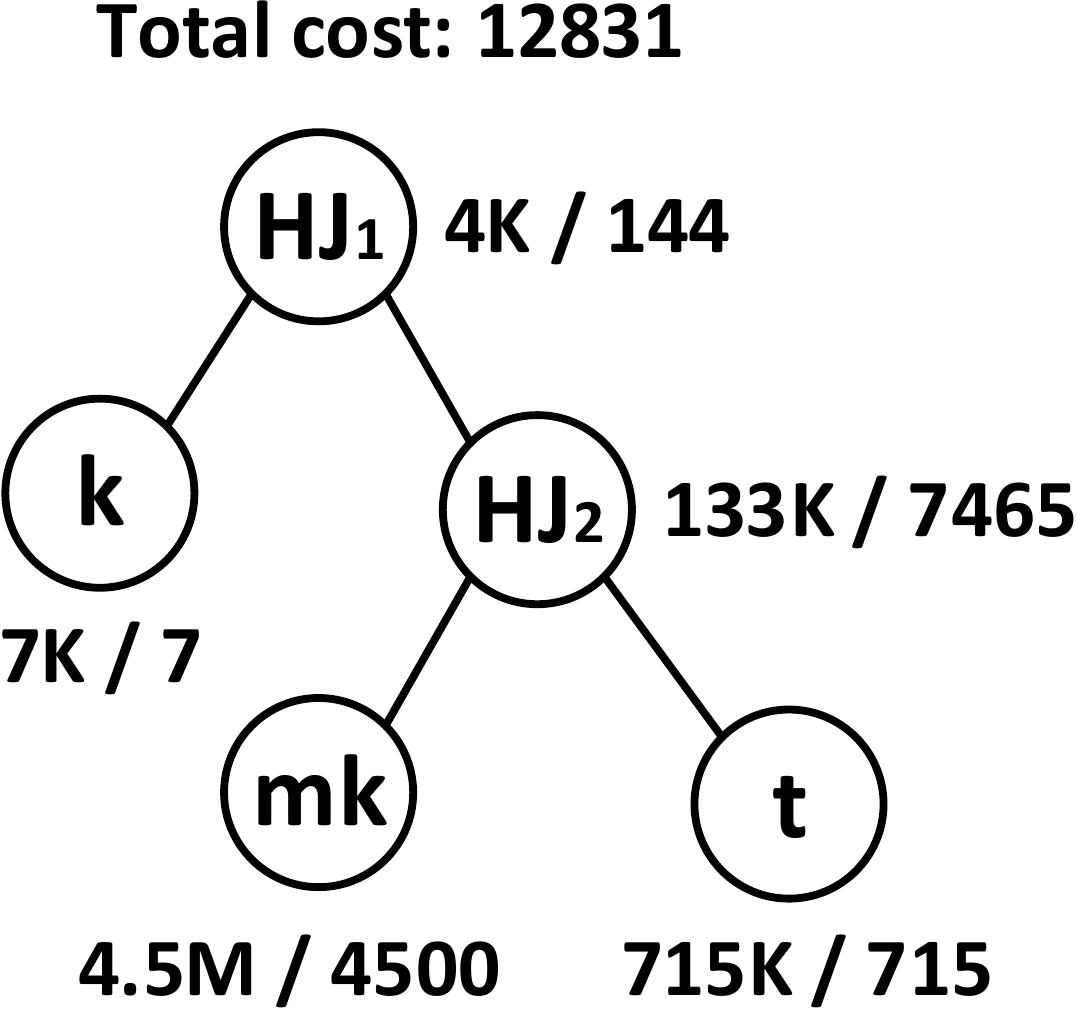}
	\caption{\changed{$P_2$ is suboptimal without bitvector filters}}
	\label{fig:job_bitvector_plan:plan_1}
	\end{subfigure}
\quad
	\caption{\changed{Example of ignoring bitvector filters in query optimization results in a suboptimal plan}}
	\label{fig:bitvector_plan}
\end{figure*}

Neglecting the impact of bitvector filters in query optimization can miss out opportunities of significant plan quality improvement.
\changed{Figure~\ref{fig:bitvector_plan} shows an example of such an opportunity with a query using the \job~\cite{leis2018query} benchmark schema:}

\cut{Figure~\ref{fig:bitvector_plan} shows an example of two plans for a query joining three relations $A$, $B$, and $C$, \changed{where the join graph is shown in Figure~\ref{fig:bitvector_plan:join_graph}. Assume $A$ and $C$ join $B$ with their primary keys,  $|B|>|A|>|B\ltimes C|>|B\ltimes A|>|C|>|(B\ltimes A)\ltimes C|$, and there are columnstore indexes.} Without being aware of bitvector filters, the query optimizer finds an `optimal' query plan $P_1$ that joins $A$ and $B$ with relation $A$ as the build side in hash join $HJ_2$. Then the bitvector filters are added to $P_1$ as optimization for query execution (Figure~\ref{fig:bitvector_plan:plan_0}). After applying the bitvector filter created from $C$ and eliminating the tuples from $B$ that do not qualify the join condition of $B$ joining $C$, however, the input cardinality from $B$ to $HJ_2$ is smaller than $|A|$. Because it is cheaper to build the hash table of $HJ_2$ with $B$, $P_1$ becomes suboptimal. The overall optimal plan when considering the effect of bitvector filters turns out to be $P_2$ (Figure~\ref{fig:bitvector_plan:plan_1}), which puts $B$ as the build side of $HJ_2$, even though the base table cardinality of $B$ is larger than $A$.
}

\begin{verbatim}
SELECT COUNT(*)
FROM movie_keyword mk, title t, keyword k
WHERE mk.movie_id = t.id AND mk.keyword_id = k.id
AND t.title LIKE '%(' AND k.keyword LIKE '%ge%'
\end{verbatim}


\changed{
Figure~\ref{fig:job_bitvector_plan:join_graph} shows the join graph of the query, where each edge is annotated with the join cardinality of the corresponding tables.
Figure~\ref{fig:job_bitvector_plan:plan_0} shows the best query plan $P_1$ without using bitvector filters. Each operator is annotated with the number of tuples after filter predicates being applied and the operator cost ($tuple / cost$). \cut{For simplicity, we assume the cost of a hash join operator is $HJ(R, S)=1.5\times |R|+|S|$, where $R$ is the build side, the cost of filter predicates is $0$, and the cost of a leaf operator is $|\sigma(R)|$.}

Figure~\ref{fig:job_bitvector_plan:bitvector_plan_0} shows the query plan after adding bitvector filters to $P_1$ as a post-processing step. Although the cost of $P_1$ is reduced after adding bitvector filters, it still costs $3\times$ as much as the best plan when the impact of bitvector filters is considered during query optimization (Figure~\ref{fig:job_bitvector_plan:bitvector_plan_1}).

Because $P_2$ is more expensive than $P_1$ without using bitvector filters (Figure~\ref{fig:job_bitvector_plan:plan_1}), the optimizer will choose $P_1$ as the best plan if it neglects the impact of bitvector filters during query optimization. Therefore, the optimizer will choose a much worse plan (Figure~\ref{fig:job_bitvector_plan:bitvector_plan_0}) if the bitvector filters are only considered as a post-processing step after query optimization.}

\point{Show DP based framework cannot easily work with bitvectors}

Incorporating bitvector filters into query optimization is surprisingly challenging. Existing top-down or bottom-up dynamic programming (DP) based query optimization framework cannot directly integrate the bitvector filters into its optimization, because the effect of bitvector filters can violate the substructure optimality property in DP. 
\point{Describe how DP works}
In a DP-based query optimization framework, either top-down or bottom-up, an optimal subplan is stored for each subset $\mathcal{A}$ of relations involved in a query.\cut{ The optimal subplan only depends on the relations in $\mathcal{A}$, and this optimal subplan is then used to construct the optimal subplans for sets of relations that are supersets of $\mathcal{A}$.}
\point{Describe bitvector filters violate the optimal subplans}
With bitvector filters, however, in addition to the relations in $\mathcal{A}$, the optimal subplan also depends on what bitvector filters are pushed down to $\mathcal{A}$ and how these bitvector filters apply to the relations in $\mathcal{A}$ based on the structure of the subplan. \changed{For example, Figure~\ref{fig:job_bitvector_plan:bitvector_plan_0} and Figure~\ref{fig:job_bitvector_plan:bitvector_plan_1} both contain a subplan of joining \{$mk$, $t$\}. The cost of the two subplans, however, is more than $3\times$ different due to the different bitvector filters pushed down to the subplan.}

\point{Incorporate bitvector filters can increase the complexity of substructures exponentially}
Incorporating bitvector filters into query optimization straightforwardly can be expensive. Similar to supporting interesting orders in query optimization~\cite{Simmen:1996:FTO:235968.233320}, the number of optimal substructures can increase by an exponential factor in the number of relations to account for the impact of various combinations of bitvector filters.

\outline{Two contributions of this work}

\changed{
Surprisingly, prior work has shown that, under limited conditions, different join orders results in similar execution cost when bitvector filters are used.
LIP~\cite{zhu2017looking} analyzes the impact of \changed{Bloom} filters for star schema with a specific type of left deep trees, where the fact table is at the bottom. They observe that, if bitvector filters created from dimension tables are pushed down to the fact table upfront, plans with different permutations of dimension tables have similar cost.}

\changed{
Motivated by this observation, we study the impact of bitvector filters on query optimization.}
We focus on an important class of queries, i.e., complex decision support queries, and the plan space of right deep trees without cross products, which is shown to be an important plan space for such queries~\cite{galindo2008optimizing, chen1997applying}. \textbf{Our first contribution} is to systematically analyze the impact of bitvector filters on optimizing \changed{the join order of star and snowflake queries} with \changed{primary-key-foreign-key (PKFK) joins} in the plan space of right deep trees without cross products \changed{(Section~\ref{sec:bitvector_aware_qo}-\ref{sec:snowflake})}. Prior work has shown that, without bitvector filters, the number of plans for \changed{star and snowflake} queries in this plan space is \changed{exponential} in the number of relations in the query~\cite{ono1990measuring}. Intuitively, the plan space complexity should further increase with bitvector filters integrated into query optimization due to violation of substructure optimality. \textbf{Our key observation} is that, when the bitvector filters have no false positives, certain join orders can be equivalent or inferior to others with respect to the cost function $C_{out}$~\cite{neumann2009query,neumann2013taking}, regardless of the query parameters or the data distribution. By exploiting this observation, we prove that, with some simplifying assumption, for star and snowflake queries with PKFK joins, the plan of the minimal $C_{out}$ with bitvector filters can be found by choosing from a \textbf{linear} number of  plans in the number of relations in the query in this plan space. 
To the best of our knowledge, this is the first work that analyzes the interaction between bitvector filters and query optimization for a broad range of decision support queries and a wide plan search space.

\cut{
Prior work has shown that, without bitvector filters, the number of plans for \changed{star and snowflake} queries in this plan space is \changed{exponential} in the number of relations in the query~\cite{ono1990measuring}. Intuitively, the plan space complexity should further increase by incorporating the bitvector filters into query optimization due to violation of substructure optimality. }
\cut{
\changed{
\textbf{Our key observation} is that, when the bitvector filters have no false positives, certain join orders can be equivalent or inferior to others with respect to the cost function $C_{out}$~\cite{neumann2009query,neumann2013taking}, regardless of the query parameters or the data distribution. By exploiting this observation, we prove that, with some simplified assumption, only a linear number of candidate plans can be potentially with some simplified assumption, for star and snowflake queries with PKFK joins, the plan with the minimal $C_{out}$ in this plan space can be found by only looking at a \textbf{linear} number of candidate plans in the number of relations in the query. }
\cut{The key insight is to leverage the properties of bitvector filters, the query graph, and the plan space.}
}

\cut{
\point{Importance of this work}
In this work, we study the impact of bitvector filters on query optimization. 
We focus on an important class of queries: complex decision support queries. We focus on analyzing the plan space of right deep trees without cross products, which is shown to be an important plan space for decision support queries~\cite{galindo2008optimizing, chen1997applying}. Prior work has shown that the number of plans for \changed{star and snowflake} queries in this plan space is \changed{exponential} w.r.t. the number of relations in the query~\cite{ono1990measuring}. Intuitively, the plan space complexity should further increase by incorporating the bitvector filters into query optimization due to violation of substructure optimality. 
}

\cut{
\changed{
The key insight is to leverage the properties of bitvector filters, the query graph, and the plan space. LIP~\cite{zhu2017looking} is the closest prior work to this work. They analyze the impact of \changed{Bloom} filters for star schema with a specific type of left deep trees, where the fact table is at the bottom. They observe that different permutations of dimension tables result in similar execution cost, if filters created from dimension tables are pushed down to the fact table upfront.
}
}

\cut{
\textbf{Our first contribution} is to systematically analyze the impact of bitvector filters on optimizing \changed{the join order of star and snowflake queries} with \changed{(PKFK) joins} in the plan space of right deep trees without cross products \changed{(Section~\ref{sec:bitvector_aware_qo}, Section~\ref{sec:star}, and Section~\ref{sec:snowflake})}. 
To the best of our knowledge, this is the first work that analyzes the interaction between bitvector filters and query optimization for a broad range of decision support queries and a general plan search space.
}

\cut{
By analyzing the plan space complexity with bitvector filters, we prove that, counter-intuitively, the plan with the minimal cost (subject to cost function $C_{out}$~\cite{neumann2009query,neumann2013taking}) in this plan space can be found by only looking at a linear number of candidate plans w.r.t. the number of relations in the query. Our key insight is to leverage the properties of bitvector filters and query patterns to derive a small set of candidate plans that have lower or equivalent cost compared with other plans in this plan space.
}

\cut{
LIP~\cite{zhu2017looking} is the closest to this work, where they analyze the impact of \changed{Bloom} filters for star schema with a specific type of plans.\changed{They observe that different permutations of join orders in their specified plan space does not change the query execution cost much,} and their conclusion of robustness of LIP plans can be directly derived from our analysis.
}

\point{First contribution: analyze the plan space complexity for complex decision support queries for right deep tree without cross products}

\cut{
\begin{figure}
\cut{	\centering
	\begin{subfigure}[b]{.2\textwidth}
		\centering
		\raisebox{2em}{
			\includegraphics[width=.8\linewidth]{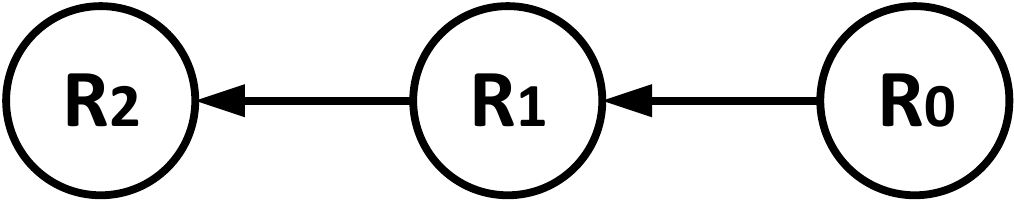}
		}
		\caption{Chain query graph of relations $R_0, R_1, R_2$}
		\label{fig:chain_query}
	\end{subfigure}	
	\quad}
	\begin{subfigure}[b]{.4\textwidth}
		\centering
		\includegraphics[width=.5\linewidth]{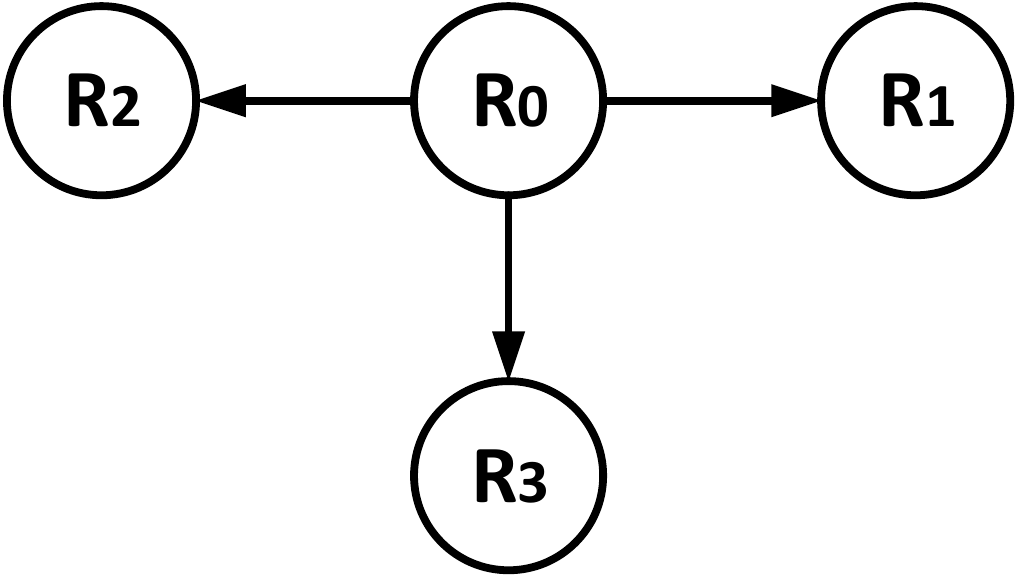}
		\caption{Star query graph with fact table $R_0$ and dimension tables $R_1, R_2, R_3$}
		\label{fig:star_query}
	\end{subfigure}
	\begin{subfigure}[b]{.4\textwidth}
		\centering
		\includegraphics[width=.7\linewidth]{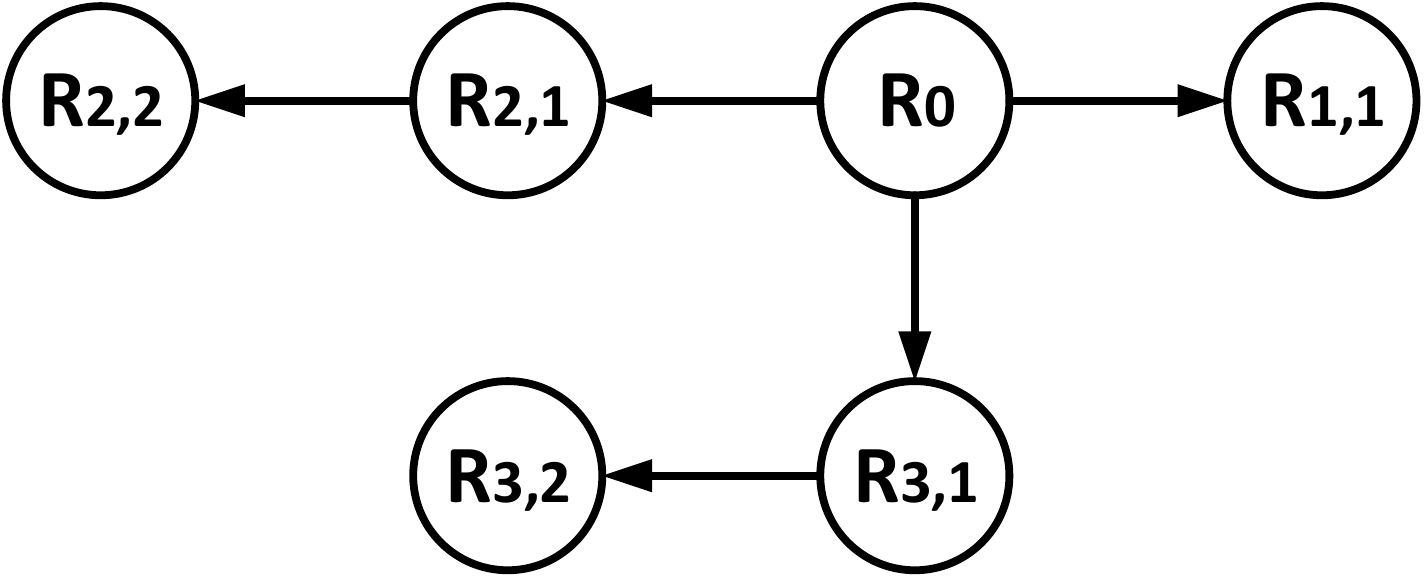}
		\caption{Snowflake query graph with fact table $R_0$ and branches $\{R_{1,1}, R_{1,2}\}, \{R_{2,1}, R_{2,2}\}, \{R_3\}$}
		\label{fig:snowflake_query}
	\end{subfigure}
	
	\caption{Join graph for star and snowflake queries}
	\label{fig:query_graph}
\end{figure}
}

\cut{
Despite the general hardness of incorporating bitvector filters into query optimization, we focus on an important class of queries: complex decision support queries. \textbf{Our first contribution} is to systematically analyze the impact of bitvector filters on optimizing \changed{the join order of star and snowflake queries} with \changed{primary-key-foreign-key (PKFK) joins (Section~\ref{sec:bitvector_aware_qo})}. \cut{Star and snowflake queries describe common join graph patterns as shown in Figure~\ref{fig:query_graph}. Each node in the join graph is a relation in the query; each edge in the join graph is a primary-key foreign-key join, where the join condition includes an equal-join between the primary-key in a dimension table and a foreign-key in a fact table.}
}

\cut{
We focus on analyzing the plan space of right deep trees without cross products, which is shown to be an important plan space for decision support queries~\cite{galindo2008optimizing, chen1997applying}. Prior work has shown that the number of plans for \changed{star and snowflake} queries in this plan space is \changed{exponential} w.r.t. the number of relations in the query~\cite{ono1990measuring}. The plan space complexity further increases by incorporating the bitvector filters into query optimization. \cut{We follow prior work on join order enumeration and adapt the cost function measuring intermediate result sizes ($C_{out}$) to quantify the plan quality~\cite{neumann2009query,neumann2013taking}.} By analyzing the plan space complexity with bitvector filters, we prove that, counter-intuitively, the plan with the minimal cost (subject to cost function $C_{out}$~\cite{neumann2009query,neumann2013taking}) in this plan space can be found by only looking at a linear number of candidate plans w.r.t. the number of relations in the query. Our key insight is to leverage the properties of bitvector filters and query patterns to derive a small set of candidate plans that have lower or equivalent cost compared with other plans in this plan space.
}

\point{Second contribution: propose an algorithm to integrate general query optimization with snowflake queries}
While \changed{star and snowflake} queries are common patterns for decision support queries, in practice, the join graphs can include multiple fact tables and non-PKFK joins. \textbf{Our second contribution} is to propose an algorithm \changed{that optimizes the join order} for \changed{arbitrary} decision support queries motivated by our analysis \changed{(Section~\ref{sec:general_qo})}. \changed{Our technique applies to queries with arbitrary join graphs.} \cut{We propose heuristics to handle additional complexities of their join graphs, such as joining conditions between branches in snowflake queries, joining of multiple fact tables, and joining dimension tables that are larger than the fact table.}\changed{\cut{We describe how to optimize the join order of a snowflake query with non-PKFK joins and multiple fact tables.} Since creating and applying bitvector filters adds overhead, we further optimize our algorithm by selectively adding bitvector filters based on their estimated benefit (Section~\ref{subsec:cost_based_bitvector}).} 
\point{Integration into DBMS}
Our algorithm can be integrated into a query optimization framework as a transformation rule~\cite{graefe1993volcano, graefe1995cascades}.\cut{similar to other dedicated transformation rules that optimize snowflake queries~\cite{Weininger:2002:EEJ:564691.564754, galindo2008optimizing, Antova:2014:OQO:2588555.2595640}.} \cut{Since bitvector filters logically implement semi-joins, which are already supported in DBMSs, our algorithm can leverage the existing mechanism to estimate the selectivity of bitvector filters. }\changed{Depending how a DBMS handles bitvector filters in query optimization, we propose three options to integrate our technique into the DBMS (Section~\ref{sec:integration}).} \cut{Leveraging the existing mechanism to detect such patterns, we can replace the existing transformation rule for snowflake queries with our algorithm. Since bitvector filters logically implement semi-joins, which are already supported in DBMSs, our algorithm can leverage the existing cardinality estimation mechanism to estimate the selectivity of bitvector filters.}

\cut{
\todo{}
Our algorithm can serve as a standalone heuristic to produce a join order for complex decision support queries. This join order can be further optimized with additional rules in the query optimizer to produce a complete query plan. This query plan can either be output as the final query plan, or it can serve as a competing query plan in a full query optimization process to help with cost-based pruning.
}

\point{Describe implementation and evaluation result}
We implement our algorithm in \sqlserver (Section~\ref{sec:implementation}). \cut{We leverage the existing logic in \sqlserver to detect snowflake query patterns and estimate selectivity of bitvector filters. }We evaluate our technique on industry benchmarks TPC-DS~\cite{tpcds} and JOB~\cite{leis2018query} as well as a customer workload (Section~\ref{sec:experiment}). We show that, comparing to the query plans produced by the original \sqlserver, our technique reduces the total CPU execution time of a workload by 22\% to \maxworkloadreduction, with up to two orders of magnitude reduction in CPU execution time for individual queries. \changed{We show that our technique is especially effective in reducing execution cost for expensive queries with low selectivity, where right deep trees is a preferable plan space~\cite{galindo2008optimizing, chen1997applying}.}

\cut{
\point{Summarize contributions}
In summary, we make the following contributions:
\begin{itemize}[leftmargin=*]
	\item We show that the effect of bitvector filters violates the optimal substructure property in the existing top-down or bottom-up dynamic programming based query optimization framework, and a naive integration can increase the plan space complexity by a factor that is exponential to the number of relations in the join.
	\item We analyze the impact of bitvector filters on chain, star, and snowflake queries (Section~\ref{sec:bitvector_aware_qo}). We show that, with bitvector filters, the number of plans of the three classes of queries reduces from quadratic and exponential to linear in the plan space of right deep trees without cross products with $C_{out}$ cost function.
	\item We propose an algorithm to construct a join order for general decision support queries based on our analysis (Section~\ref{sec:general_qo}). We proposes additional heuristics to handle complexities in decision support queries beyond chain, star, and snowflake queries.
	\item We implement our algorithm in a commercial database \sqlserver and evaluate our techniques on standard and customer benchmarks (Section~\ref{sec:experiment}). Our evaluation show that the query plan produced by our algorithm reduces the CPU execution time of individual queries by up to two orders of magnitude, resulting in up to 64\% reduction in total CPU execution time at a workload level .
\end{itemize}
}

We discuss related work in Section~\ref{sec:related_work} and conclude the work in Section~\ref{sec:conclusion}.

	\section{Bitvector filter algorithm}
\label{sec:bitvector_algorithm}

In this section, we describe the details of bitvector filters creation and push-down algorithm following~\cite{graefe1993query}. 

\point{high level intuition: create one bitmap per hash join operator and push it down to the lowest possible level}

\changed{At a high level, each hash join operator creates a \emph{single} bitvector filter from the equi-join columns on the build side. This bitvector filter is then pushed down to the lowest possible level on the subtree rooted at the probe side so that it can eliminate tuples from that subtree as early as possible.}
	
\outline{Walk through the bitvector push down algorithm}

\begin{algorithm}[t]
\DontPrintSemicolon
\SetAlgoLined
\LinesNumbered

\caption{Push down bitvectors}
\label{algorithm:bitvector_pushdown}
\SetKwProg{FuncPushDownPlan}{}{:}{}
\FuncPushDownPlan{\textbf{PlanPushDown($plan$)}}{
	\KwIn{Query plan $plan$}
	\KwOut{New query plan $plan'$ with bitvectors}
	
	$root\leftarrow plan.GetRootOperator()$\;
	$plan'\leftarrow plan$\;
	$root'\leftarrow OpPushDown(op, \emptyset)$\;
	$plan'.SetRootOp(root')$\;
	\Return{$plan'$}	
}
\;

\vspace{-1em}

\SetKwProg{FuncPushDownOp}{}{:}{}

\FuncPushDownOp{\textbf{OpPushDown($op$, $B$)}}{
	\KwIn{Operator $op$, set of bitvectors $B$}
	\KwOut{New operator $op'$ with bitvectors}
	
	$residualSet \leftarrow \emptyset$\;
	$pushDownMap \leftarrow \emptyset$

	\If{$op$ is Hash Join}{
		$b \leftarrow$ bitvector created from $op.GetBuildChild()$\;
		$pushDownMap[op.GetProbeChild()]\leftarrow pushDownMap[op.GetProbeChild()]\cup b$
	}	

	
	\ForEach {bitvector $b$ in $B$}{
		$ops \leftarrow \emptyset$\;
		\ForEach {child $c$ of operator $op$}{
			\If{$b$ can be pushed down to $c$}{
				$ops \leftarrow ops \cup \{c\}$
			}
		}
		\lIf{$|ops| \neq 1$}{
			$residualSet \leftarrow residualSet \cup \{b\}$
		}
		\Else{
			$pushDownMap[c] \leftarrow pushDownMap[c] \cup \{b\}$
		}
	}
	
	$op' \leftarrow op$\;
	\If{$residualSet \neq \emptyset$}{
		$filterOp \leftarrow CreateFilterOp(op, residualSet)$\;
		$filterOp.AddChild(op)$\;
		$op' \leftarrow filterOp$
	}

	\ForEach{child $c$ of $op$}{
		$c' \leftarrow OpPushDown(c, pushDownMap[c])$\;
		$op.UpdateChild(c, c')$
	}
	
	\Return{$op'$}
}
\end{algorithm}

Algorithm~\ref{algorithm:bitvector_pushdown} shows how to push down bitvectors given a query plan. The algorithm takes a query plan as its input. Starting from the root of the query plan, the set of bitvector pushed down to the root is initialized to be empty (line 3) and each operator is then processed recursively in a pre-order traversal.
At each operator, it takes the set of bitvector filters pushed down to this operator as an input. If the operator is a hash join, a bitvector filter is created from the build side with the equi-join columns of this hash join as the keys of the bitvector filter and is added to the set of bitvector filters applied to the probe side of this hash join (line 8-10). Now consider every bitvector filter that is pushed down to this hash join operator. If one of the child operator of the join operator contains all the columns in the bitvector filter, the bitvector filter is added to the set of bitvector filters pushed down to this child operator; otherwise, the bitvector filter cannot be pushed down further, and it is added to the set of bitvector filters pushed down to this join operator (line 12 - 23). If the set of bitvector filters pushed down to this join operator is non-empty, add a filter operator on top of this join operator to apply the bitvector filters. In this case, update the root of this subplan to the filter operator (line 24-29). Recursively process the bitvector filters pushed down to the child operators and update the children accordingly (line 30 - 33). Finally, return the updated root operator of this subplan (line 34). An example of creating and pushing down bitvector filters with Algorithm~\ref{algorithm:bitvector_pushdown} is shown in Figure~\ref{fig:bitvector_pushdown}.

\cut{
\outline{Describe an example of the algorithm}
Figure~\ref{fig:bitvector_pushdown} shows an example of pushing down bitvectors with Algorithm~\ref{algorithm:bitvector_pushdown} for a query plan joining relations $A, B, C, D$. Figure~\ref{fig:bitvector_pushdown:join_graph} shows the join graph of the relations, and Figure~\ref{fig:bitvector_pushdown:plan} shows the query plan. Since $D$ joins with both $C$ and $A$, the bitvector filter created from $D$ is pushed down to the hash join $HJ_2$. Since $C$ joins only with $B$, the bitvector filter created from $C$ goes passed $HJ_3$ and is pushed down to $A$.
}

	\changed{\section{Overview and preliminaries}
\label{sec:bitvector_aware_qo}
}

\sloppy
\cut{
In this section, we formally define the queries and cost function discussed in this work. We then describe properties of bitvector filters.} 

\subsection{Overview}

\point{Overview of section 3, 4, 5, 6}

We start with the properties of bitvector filters and the cost function (Section~\ref{sec:bitvector_aware_qo}). We then show that, with bitvector filters, the number of candidate plans of the minimal cost is linear for star and snowflake queries with PKFK joins in the plan space of right deep trees without cross products (Section~\ref{sec:star} and Section~\ref{sec:snowflake}).
\changed{We finally describe the general bitvector-aware query optimization \emph{algorithm} for arbitrary decision support queries and how to integrate it with a Volcano / Cascades style optimizer (Section~\ref{sec:general_qo}).}
Table~\ref{table:notation} summarizes the notations.
Table~\ref{table:analysis_summary} summarizes the results of our analysis.

\begin{table}[]
\caption{List of notations}	
\label{table:notation}
\begin{tabular}{|m{2.7cm}|m{5.3cm}|}
\hline
Notation & Description     \\
\hline
$q$         & a query           \\
\hline
$R$         & a relation        \\
\hline
$\mathcal{R}$         & a set of relations        \\
\hline
$\mathcal{T}=T(R_1,\cdots, R_n)$         & a right deep tree with $R_1$ as the right most leaf and $R_n$ as the left most leaf \\
\hline
$S(R_1, \cdots, R_n,$ $B_1, \cdots, B_m)$ & join of relations $R_1, \cdots, R_n$ after applying bitvector filters created from $B_1, B_2, \cdots, B_m$, where $B_i$ is either a base relation or a join result. We omit $B_1, \cdots, B_m$ when they are clear from the context. We use the notation interchangeably with $\Join$ \\
\hline
$|R|$ & cardinality of a base relation or an intermediate join result after applying bitvector filters \\
\hline
$R_1/R_2$ & semi join of $R_1$ with $R_2$, where $R_1/R_2\subseteq R_1$ \\
\hline
$R_1/(R_2, \cdots, R_n)$ & semi join of $R_1$ with $R_2, \cdots, R_n$, where $R_1/(R_2, \cdots, R_n)\subseteq R_1$ \\
\hline
$R_1\to R_2$ & the join columns of $R_1$ and $R_2$ is a key in $R_2$. If the join columns form a primary key in $R_2$, then $R_1\to R_2$ is a primary-key-foreign-key join \\
\hline
$C_{out}$         & cost function (See Section~\ref{sec:cost_function})  \\
\hline
$\prod_{R_1}(R_2)$ & project out all the columns in $R_1$ from $R_2$, where the columns in $R_2$ is a superset of that in $R_1$. The resulting relation has the same number of rows as $R_2$ but less number of columns per row \\
\hline 
\end{tabular}
\end{table}

\begin{table*}[]
	\caption{\changed{Summary of the plan space complexity for star and snowflake queries with unique key joins}}
	\label{table:analysis_summary}
\begin{tabular}{|m{1.3cm}|m{1.9cm}|m{1.6cm}|m{3cm}|m{1.5cm}|m{6.3cm}|}
	\hline
join graph & graph size                     & \# of relations & original complexity & complexity w/ our analysis & candidate plans with minimal $C_{out}$ \\
\hline
star & $n$ dimension tables & $n+1$ & exponential to $n$ & $n+1$ & $T(R_0, R_1, \cdots, R_n)$, $\{T(R_k, R_0, R_1, R_2, \cdots, R_{k-1}, R_{k+1}, \cdots, R_n), \break 1\leq k \leq n \}$ \\
\hline
snowflake  & $m$ branches of lengths $n_i, 1\leq i\leq m$ & $n+1,n=\sum_{i=1}^m n_i$ & exponential to $n$ & $n+1$ & $T(R_0,R_{1,1}, \cdots, R_{1,n_1}, \cdots, R_{n,1}, \cdots, R_{n,n_m})$, $\{T(R_{i,a_1}, \cdots, R_{i,a_{n_1}},R_0, R_{1, n_1}, \cdots, R_{i-1,1}, \break \cdots, R_{i-1,n_{i-1}}, R_{i+1,1}, \cdots, R_{i+1,n_{i+1}},  \cdots, R_{n,1}, \break \cdots, R_{n,n_m}) \}$ (see Section~\ref{sec:snowflake} for $a_1, \cdots, a_{n_1}$) \\
\hline                                 
\end{tabular}
\end{table*}

\subsection{Properties of bitvector filters}
\label{sec:bitvector_properties}
We start with the properties of bitvector filters:

\vspace{-.5em}
\begin{property}\label{property:commutativity}
	\textbf{Commutativity}: $R/(R_1, R_2)=R/(R_2,R_1)$
\end{property}
\vspace{-1em}
\begin{property}\label{property:reduction}
	\textbf{Reduction}: $|R/R_1| \leq |R|$
\end{property}
\vspace{-1em}
\begin{property}\label{property:redundancy}
	\textbf{Redundancy}: $(R_1\bowtie R_2) / R_2 = R_1  \bowtie R_2$
\end{property}
\vspace{-1em}
\begin{property}\label{property:associativity}
	\textbf{Associativity}: $R/(R_1, R_2)=(R/R_1)/R_2$ \changed{if there are no false positives with the bitvector filters created from $(R_1, R_2), R_1,$ and $R_2$.}
\end{property}
\vspace{-.5em}

Now we prove the absorption rule of bitvector filters for PKFK joins. The absorption rule says that, if $R_1$ joins $R_2$ with a key in $R_2$, the result of joining $R_1$ and $R_2$ is \changed{a subset} of the result of semi-joining $R_1$ and $R_2$. Formally,

\vspace{-.5em}
\begin{lemma}\label{rule:absorption}
	\textbf{Absorption rule}: If $R_1\to R_2$, then \changed{$R_1/R_2\supseteq \prod_{R_1}(R_1\bowtie R_2)$ and $|R_1/R_2|\geq|R_1\bowtie R_2|$. The equality happens if the bitvector filter created from $R_2$ has no false positives.}
\end{lemma}
\vspace{-.5em}

\begin{proof}
	For every tuple $r$ in $R_1$, it can join with a tuple in $R_2$ if and only if the join columns in $r$ exist in $R_2$. Because $R_1\to R_2$, there is at most one such tuple in $R_2$. Thus, \changed{$R_1/R_2\subseteq\prod_{R_1}(R_1\bowtie R_2)$}.
\end{proof}
\vspace{-.5em}

\cut{
\subsection{Notation}
\label{sec:notation}
We first describe all the notations used in this paper as summarized in Table~\ref{table:notation}:

\begin{itemize}[leftmargin=*]
	\item $q$: a join query
	\item $R$: a base relation or a join result
	\item $\mathcal{R}=\{R_1, \cdots, R_n\}$: a set of relations
	\item $\mathcal{T}=T(R_1, \cdots, R_n)$: a query plan of a right deep tree, where $R_1$ is the right most leaf and $R_n$ is the left most leaf. Figure~\ref{fig:bitvector_plan:plan_0} shows an example of a right deep tree $T(B, A, C)$.
	\item $S(R_1, \cdots, R_n, B_1, \cdots, B_m)$: join result of $R_1, R_2, \cdots, R_n$ after applying bitvector filters created from $B_1, \cdots, B_m$, where $B_i$ is either a base relation or a join result. We omit $B_1, \cdots, B_m$ when they are clear from the context. We use the notation interchangeably with $\Join$.
	\item $|R|$: cardinality of a base relation or an intermediate join result after applying bitvector filters.
	\item $R_1 /R_2$: semi join of $R_1$ and $R_2$, where the resulting relation consists of the tuples from $R_1$ that qualifies the join condition.
	\item $R_1/ (R_2, R_3, \cdots, R_n)$: semi join of $R_1$ with $R_2, R_3, \cdots, R_n$, where the resulting relation consists of the tuples from $R_1$ that qualifies the join conditions with $R_2, R_3, \cdots, R_n$.
	\item $R_1\to R_2$: the join columns of $R_1$ and $R_2$ is a unique key in $R_2$. If the join columns form a primary key in $R_2$, then $R_1\to R_2$ is a primary-key foreign-key join.
	\item $C_{out}(\mathcal{T})$: cost of a query plan (See Section~\ref{sec:cost_function}).
	\item $\prod_{R_1}(R_2)$: project out all the columns in $R_1$ from $R_2$, where the columns in $R_2$ is a superset of that in $R_1$. The resulting relation has the same number of rows as $R_2$ but less number of columns per row.
\end{itemize}
}

\subsection{Cost function}
\label{sec:cost_function}

\outline{What is $C_{out}$ and why we use it}

\point{We use $C_{out}$ because it does not require physical information and is a good approximation for actual cost}
Since our analysis focuses on the quality of logical join ordering, we measure the intermediate result sizes (i.e., $C_{out}$) as our cost function similar to prior work on join order analysis~\cite{neumann2009query, neumann2013taking}. In practice, $C_{out}$ is a good approximation for comparing the actual execution cost of plans. \cut{Using the cost function with physical information is future work.}

\point{Formal definition of $C_{out}$}
$C_{out}$ measures the cost of a query plan by the sum of intermediate result sizes. Because bitvector filters also impact the cardinality of a base table, we adapt $C_{out}$ to include the base table cardinality as well. Formally, 
\begin{equation}\label{def:c_out}
C_{out}(T)=\begin{cases}
	|T| & \text{if $T$ is a base table} \\
	|T|+C_{out}(T_1)+C_{out}(T_2) & \text{if $T=T_1\bowtie T_2$}	
	\end{cases}
\end{equation}

\changed{Note that $|T|$ has reflected the impact of bitvector filters, where $|T|$ represents the cardinality after bitvector filters being applied} for both base tables and join results.

\cut{
\subsection{Decision support queries}
\label{sec:definition_query}

\outline{Formally define chain, star, and snowflake queries}
In this work, we analyze three common patterns in decision support queries: chain, star, and snowflake queries.

\point{Unique key join}

We first define a generalized concept of primary-key-foreign-key join called unique key join:
\vspace{-.5em}
\begin{definition}\label{def:unique_key_join}
	\textbf{Unique key join: } If $R_1$ joins with $R_2$, where the join columns are uniquely valued in $R_2$, we call $R_1$ joins with $R_2$ with a unique key join, and denote it as $R_1\to R_2$.
\end{definition}
\vspace{-.5em}

The unique key join is transitive. Formally,
\vspace{-.5em}
\begin{lemma}\label{lemma:unique_key_transitivity}
	\textbf{Transitivity of unique key join: }
	If $R_1 \to R_2$, then $(R_1 \Join R_3) \to R_2$.
\end{lemma}
\vspace{-.5em}

Now we define \changed{star and snowflake queries} as the following:

\point{Star query}
\vspace{-.5em}
\begin{definition}\label{def:star_query}
	\textbf{Star query: }
	Let $\mathcal{R}=\{R_0, R_1, \cdots, R_n\}$ be a set of relations and $q$ be a query joining relations in $\mathcal{R}$. The query $q$ is a star query if $R_0\to R_k$ for $1\leq k\leq n$. $R_0$ is called a fact table, and $R_k, 1\leq k\leq n$ is called a dimension table.
\end{definition}
\vspace{-.5em}

Figure~\ref{fig:star_query} shows an example of a star query, where $R_0$ is the fact table and $R_1, R_2, R_3$ are dimension tables.

\point{Snowflake query}
\vspace{-.5em}
\begin{definition}\label{def:snowflake_query}
	\textbf{Snowflake query}: Let $\mathcal{R}=\{R_0, R_{1,1}, \cdots, R_{1,n_1}, R_{2,1}, \cdots, R_{2,n_2},\cdots, R_{m,1},\cdots, R_{m,n_m}\}$ be a set of relations and $q$ be a query joining relations in $\mathcal{R}$. The query $q$ is a snowflake query if
	\begin{itemize}[leftmargin=*]
		\item $R_{0}\to R_{i,1}$ for $1\leq i\leq m$ and
		\item $R_{i,j-1}\to R_{i,j}$ for $1\leq i\leq m, 1< j\leq n_i$.
	\end{itemize}
	We call $R_0$ the fact table and $R_{i,1}, R_{i,2}, \cdots, R_{i,n_i}$ a branch. We denote $\{R_{i,1}, R_{i,2},\cdots, R_{i,n_i}\}$ on each chain as $\mathcal{R}_i$.
\end{definition}
\vspace{-.5em}

Figure~\ref{fig:snowflake_query} shows an example of a snowflake query, where $R_0$ is the fact table, and $\{R_{1,1}, R_{1,2}\}, \{R_{2,1}, R_{2,2}\}, \{R_3\}$ are three branches of dimension tables.

\cut{
\subsection{Scope}
\begin{itemize}
	\item Cost model: the class of cost models that satisfies $C_{out}$. Note that nested loop join does not satisfy $C_{out}$ unless it is a cross product.
	\item Query type: star query, chain query, snowflake query with unique key joins (see Section~\ref{sec:notation}). Note that key-foreign-key equi-join is a special case of unique key join.
	\item Plan space: right deep tree with cross products and its variants.
	\item Assume bitvector is perfect.
	\item Assume the build side will always create a bitvector unless the join is a cross product, i.e., nested loop join. Note that no bitvector will be created for merge join and nested loop join.
\end{itemize}
}
}

\cut{
\changed{
\subsection{Scope of our analysis}
\label{sec:assumption}
}

In the scope of this work, we analyze the first order impact of bitvector filters with the following assumptions:
\begin{itemize}[leftmargin=*]
	\item A bitvector filter is perfect, i.e., there is no false positive.
	\item The cost of creating and applying a bitvector filter is 0.
	\item The cost function is based on intermediate result sizes (See Section~\ref{sec:cost_function}). Physical information of operators is ignored.
\end{itemize}

\changed{
	In practice, bitvector filters can have collisions, which results in false positives.
	\cut{Since bitvector filters can compactly encode a large value domain, with moderate and sufficient amount of memory, the rate of false positives is often very low.} In addition, applying bitvector filters adds overhead and can be expensive for certain data types, e.g., string without dictionary encoding. We will relax the assumptions and discuss cost-based bitvector filters in Section~\ref{subsec:cost_based_bitvector}.
}
}

\cut{
\begin{lemma}\label{rule:decomposition}
	\textbf{Decomposition rule}: If $S(B_1,B_2)$ are cross products, $A/(B_1, B_2)=A/S(B_1, B_2)$.
\end{lemma}

\begin{proof}
	Let $r\in A$ and $\prod_{P(A, B_1, B_2)}r\in \prod_{P(A, B_1, B_2)}A/(B_1, B_2)$. We have $\prod_{P(A, B_1)}r\in \prod_{P(A, B_1)}A/B_1$. Because $S(B_1, B_2)$ is a cross product, $\prod_{P(A, B_1)}r\in \prod_{P(A, B_1)}S(B_1, B_2)$. Since $r\in A$, $\prod_{P(A, B_1)}r\in \prod_{P(A, B_1)}A/S(B_1, B_2)$.
	Conversely, if $r\in A$ and $\prod_{P(A, B_1, B_2)}r\in \prod_{P(A, B_1, B_2)}A/S(B_1, B_2)$, $\prod_{P(A, B_1)}r\in \prod_{P(A, B_1)} S(B_1, B_2)$ and $\prod_{P(A, B_2)}r\in \prod_{P(A, B_2)} S(B_1, B_2)$. Thus, $\prod_{P(A, B_1)}r\in \prod_{P(A, B_1)}B_1$ and $\prod_{P(A, B_2)r\in\prod_{P(A, B_2}}B_2$. Since $r\in A$, $\prod_{P(A, B_1, B_2)}r\in \prod_{P(A, B_1, B_2)}A/(B_1, B_2)$. 
	Therefore, $A/(B_1, B_2)=A/S(B_1, B_2)$.
\end{proof}	
}

\changed{\section{Analysis of Star Queries with PKFK joins}\label{sec:star}
}
\sloppy

\todo{Shrink this section}
\cut{
\point{Unique key join}

We first define a generalized concept of primary-key-foreign-key join called unique key join:
\vspace{-.5em}
\begin{definition}\label{def:unique_key_join}
	\textbf{Unique key join: } If $R_1$ joins with $R_2$, where the join columns are uniquely valued in $R_2$, we call $R_1$ joins with $R_2$ with a unique key join, and denote it as $R_1\to R_2$.
\end{definition}
\vspace{-.5em}

The unique key join is transitive. Formally,
\vspace{-.5em}
\begin{lemma}\label{lemma:unique_key_transitivity}
	\textbf{Transitivity of unique key join: }
	If $R_1 \to R_2$, then $(R_1 \Join R_3) \to R_2$.
\end{lemma}
\vspace{-.5em}
}

We define \changed{star queries with PKFK joins} as the following:

\point{Star query}
\begin{definition}\label{def:star_query}
	\changed{\textbf{Star query with PKFK joins: }}
	Let $\mathcal{R}=\{R_0, R_1, \cdots, R_n\}$ be a set of relations and $q$ be a query joining relations in $\mathcal{R}$. The query $q$ is a star query with PKFK joins if $R_0\to R_k$ for $1\leq k\leq n$. $R_0$ is called a fact table, and $R_k, 1\leq k\leq n,$ is called a dimension table.
\end{definition}
\vspace{-.5em}

\begin{figure}
	\centering
	\includegraphics[width=.4\linewidth]{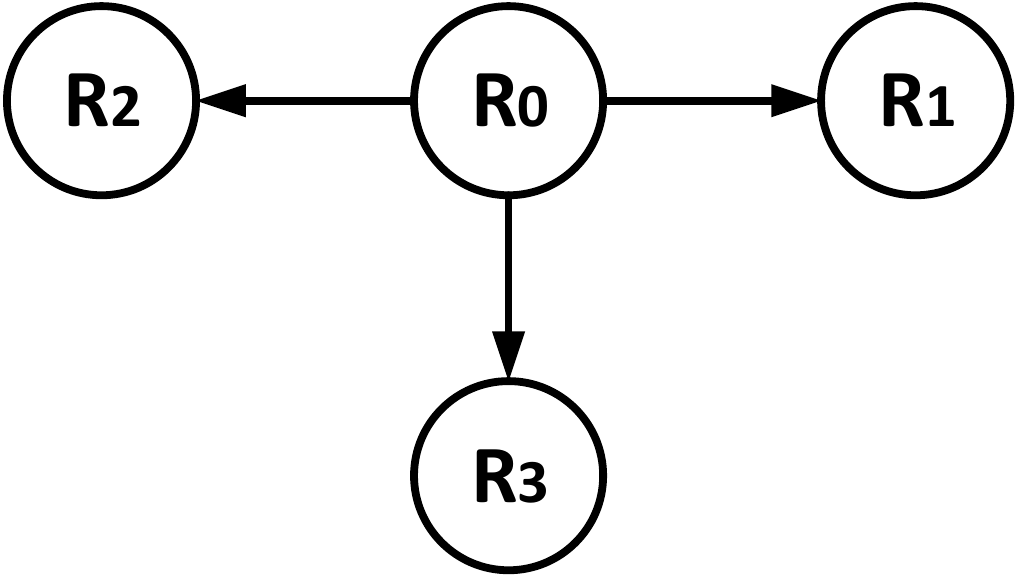}
	\caption{Star query graph \changed{with PKFK joins}, where the fact table is $R_0$ and dimension tables are $R_1, R_2, R_3$}
	\label{fig:star_query}
\end{figure}

Figure~\ref{fig:star_query} shows an example of a star query, where $R_0$ is the fact table and $R_1, R_2, R_3$ are dimension tables.

Now we analyze the plan space complexity for star queries with PKFK joins. We show that, in the plan space of right deep trees without cross products, we can find the query plan of the minimal cost (under the cost function from Section~\ref{sec:cost_function}) from $n+1$ plans with bitvector filters if the bitvector filters have no false positives, where $n+1$ is the number of relations in the query. In contrast, the original plan space complexity for star queries in this plan space is exponential to $n$~\cite{ono1990measuring}.

\point{key intuition of the proof}
\begin{figure}
	\begin{subfigure}[b]{.4\linewidth}
		\centering
		\includegraphics[width=\linewidth]{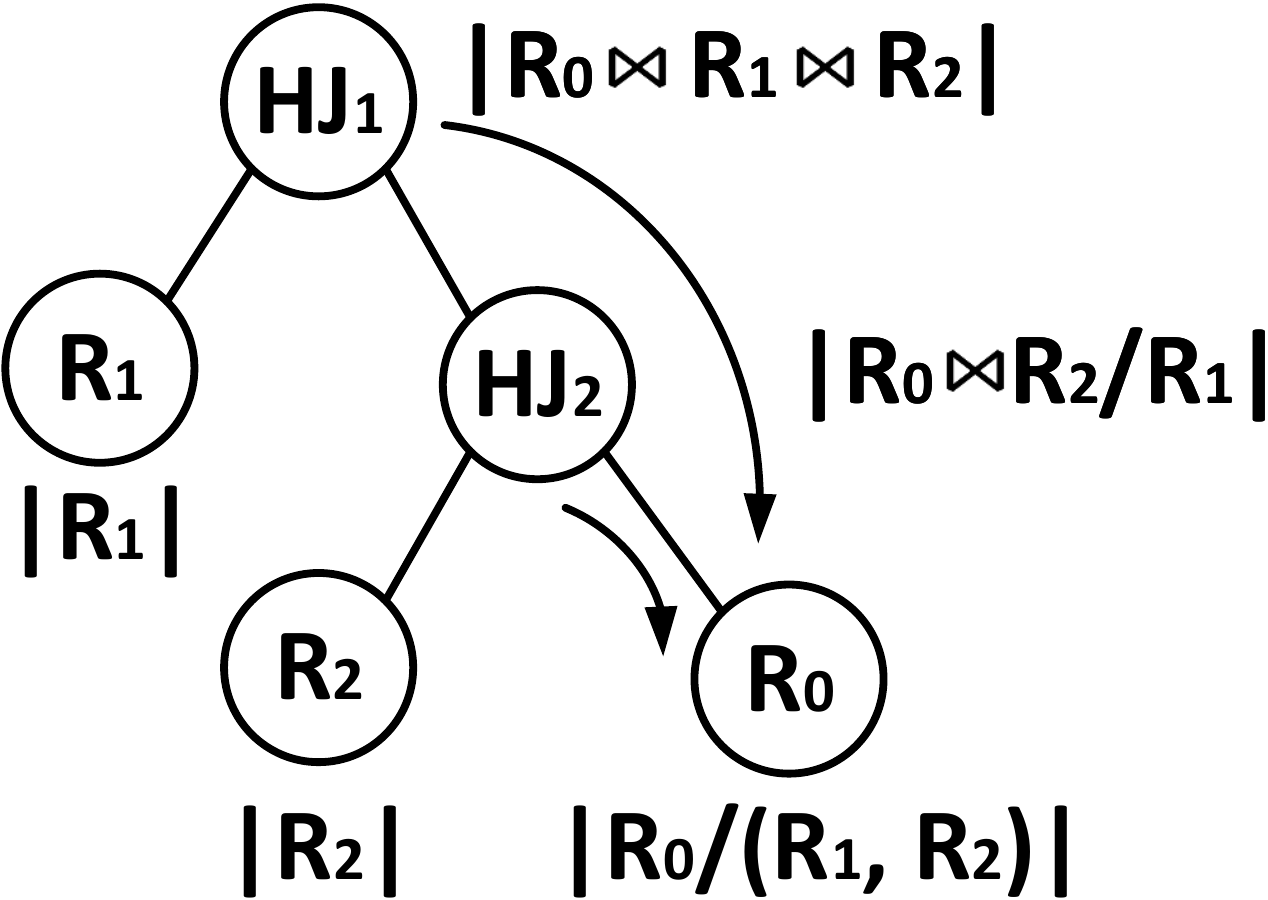}
		\caption{Plan $P_1$}
	\end{subfigure}
	\qquad
	\begin{subfigure}[b]{.4\linewidth}
		\centering
		\includegraphics[width=\linewidth]{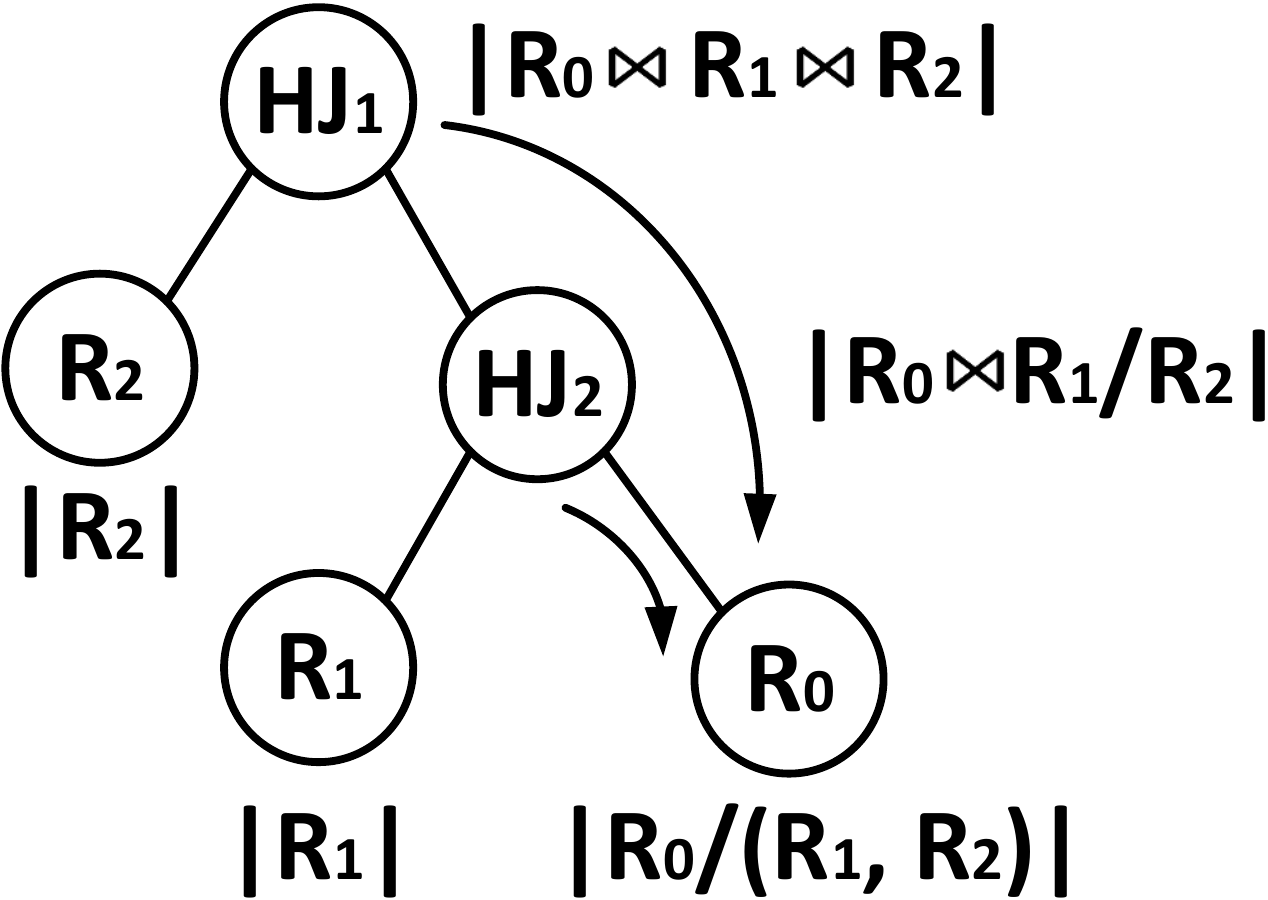}
		\caption{Plan $P_2$}
	\end{subfigure}
	\caption{Example of two plans of a star query $\{R_0, R_1, R_2\}$ with PKFK joins using bitvector filters. Each operator is annotated with the intermediate result size. Plan $P_1$ and $P_2$ have different join orders of dimension tables but the same cost.}
	\label{fig:proof:star}
\end{figure}

\textbf{Our key intuition} is that, in the plan space of right deep trees without cross products, the cost of plans of a star query with PKFK joins can be the same with different join orders of dimension tables. This is because all the bitvector filters for a star query will be pushed down to the fact table; and by Lemma~\ref{rule:absorption}, we can show the cost of many join orders is the same. Figure~\ref{fig:proof:star} shows an example of two plans of a star query with PKFK joins using different join orders of dimension tables but having the same cost.

Formally, \textbf{our key results} in this section are:

\vspace{-.5em}
\begin{theorem}\label{theorem:star:mincost_plan}
	\textbf{Minimal cost right deep trees for star query: }
	Let $\mathcal{R}$ be the set of relations of a star query as defined in Definition~\ref{def:star_query}. Let $\mathcal{A} =\{T(X_0, \cdots, X_n)\}$ be the set of right deep trees without cross products for $q$, where $X_0, \cdots, X_n$ is a permutation of $R_0, \cdots, R_n$. If $C_{min}=min\{C_{out}(\mathcal{T}), \mathcal{T}\in\mathcal{A} \}$, then there exists a plan $\mathcal{T}\in\mathcal{A}_{candidates}=\{T(R_0, R_1, \cdots, R_n)\} \cup \{T(R_k, R_0, R_1, \cdots, R_{k-1}, R_{k+1}, \cdots, R_n), 1\leq k \leq n \}$ such that $C_{out}(\mathcal{T})=C_{min}$.
\end{theorem}
\vspace{-.5em}

\vspace{-.5em}
\begin{theorem}\label{theorem:star_plan_space}
	\textbf{Plan space complexity for star query: }
	Let $\mathcal{R}$ be the set of $n+1$ relations of a star query as defined in Definition~\ref{def:star_query}. We can find the query plan with the minimal cost in the place space of right deep trees without cross products from $n+1$ candidate plans.
\end{theorem}
\vspace{-.5em}

\cut{
We omit most of the proofs due to space limit, and they can be found in our technical report~\cite{moreproofs}.
}

\cut{
\point{The steps that we prove the space complexity}
We will start by analyzing the plan space of right deep trees without cross products for star queries. we will divide the plan space into two subspaces: whether the fact table is the right most leaf or not. We will then analyze the plans in each subspace and show that many of them are equivalent in cost. }

\changed{We start the analysis by understanding} the plan space of right deep trees without cross products for star queries:
\cut{
\vspace{-.5em}
\begin{lemma}\label{lemma:star:trees}
	\textbf{Right deep trees for star query: } 
	Let $\mathcal{R}$ be the set of relations of a star query as defined in Definition~\ref{def:star_query}. Let $\mathcal{T} = T(X_0, X_1, X_2, \cdots, X_n)$ be a query plan, where $X_0, \cdots, X_n$ is a permutation of $\{R_0, R_1, R_2, \cdots, R_n\}$. Then $\mathcal{T}$ is a right deep tree without cross products if and only if $X_0=R_0$ or $X_1=R_0$.
\end{lemma}
\vspace{-.5em}
}

\vspace{-.5em}
\begin{restatable}{lemma}{lemmastartrees}\label{lemma:star:trees}
	\textbf{Right deep trees for star query: } 
	Let $\mathcal{R}$ be the set of relations of a star query as defined in Definition~\ref{def:star_query}. Let $\mathcal{T} = T(X_0, X_1, X_2, \cdots, X_n)$ be a query plan, where $X_0, \cdots, X_n$ is a permutation of $\{R_0, R_1, R_2, \cdots, R_n\}$. Then $\mathcal{T}$ is a right deep tree without cross products if and only if $X_0=R_0$ or $X_1=R_0$.
\end{restatable}
\vspace{-.5em}

\cut{
\begin{proof}
	Assume $X_0=R_i, X_1=R_j, i\neq 1, j\neq 1$. Then $R_i$ and $R_j$ do not have a join condition based on Definition~\ref{def:star_query}. Thus, $T(X_0, X_1)$ has a cross product, which is a contradiction.
	
	If $X_0=R_0$ or $X_1=R_0$, since $R_0$ joins with $R_1, \cdots, R_n$, then $\mathcal{T}=T(X_0, X_1, \cdots, X_n)$ does not contain any cross product.
	
\cut{	Thus, $\mathcal{T}$ is a right deep tree without cross products if and only if $X_0=R_0$ or $X_1=R_0$.}
\end{proof}
\vspace{-.5em}
}

The proof can be found in Appendix~\ref{sec:moreproofs}.

\changed{By Lemma~\ref{lemma:star:trees},} we divide the plans into two cases: whether $R_0$ is the right most leaf or not.\cut{ The key insight to reduce the plan space complexity to linear is to derive plans that are equivalent in cost based on the properties of bitvector filters.}

\point{Additional rule needed}
\changed{We first generalize} Lemma~\ref{rule:absorption} to multiple relations:
\vspace{-.5em}
\cut{
\begin{lemma}\label{lemma:star_absorption}
	\textbf{Star query absorption rule}: Let $\mathcal{R}$ be a star query as defined in Definition~\ref{def:star_query}, then $R_0/(R_1, R_2, \cdots, R_n)$\changed{$\supseteq$} $\prod_{R_0} (R_0 \Join R_1 \Join \cdots \Join R_n)$ and $|R_0/(R_1, R_2, \cdots, R_n)|$\changed{$\geq$}$|R_0 \Join R_1 \Join \cdots \Join R_n|$. \changed{The equality happens when the bitvector filters created from $(R_1, R_2, \cdots, R_n)$ has no false positives.}
\end{lemma}}

\begin{restatable}{lemma}{lemmastarabsorption}\label{lemma:star_absorption}
	\textbf{Star query absorption rule}: Let $\mathcal{R}$ be a star query as defined in Definition~\ref{def:star_query}, then $R_0/(R_1, R_2, \cdots, R_n)$\changed{$\supseteq$} $\prod_{R_0} (R_0 \Join R_1 \Join \cdots \Join R_n)$ and $|R_0/(R_1, R_2, \cdots, R_n)|$\changed{$\geq$}$|R_0 \Join R_1 \Join \cdots \Join R_n|$. \changed{The equality happens when the bitvector filters created from $(R_1, R_2, \cdots, R_n)$ has no false positives.}
\end{restatable}

\vspace{-.5em}

The proof can be found in Appendix~\ref{sec:moreproofs}.

\cut{
\changed{This can be proved by Property~\ref{property:associativity} and Lemma~\ref{rule:absorption}.}
}
\cut{
\begin{proof}
	By Property~\ref{property:associativity}, $R_0/(R_1,R_2)=(R_0/R_1)/R_2$.
	Since $R_0\to R_1$, by Lemma~\ref{rule:absorption}, $R_0/R_1=\prod_{R_0} (R_0 \Join R_1)$. 
	Since $R_0\to R_2$, $R_0 \Join R_1\to R_2$.
	By applying Lemma~\ref{rule:absorption} again, we have $(R_0 \Join R_1) / R_2=\prod_{R_0}(R_0 \Join R_1 \Join R_2)$.
	Thus, $R_0/(R_1,R_2)=\prod_{R_0} (R_0 \Join R_1 \Join R_2)$.
	
	By induction, we can prove $R_0/(R_1,R_2,\cdots, R_n)=\prod_{R_0} (R_0 \Join R_1 \Join \cdots \Join R_n)$.	
\end{proof}
\vspace{-.5em}
}

\cut{
The star query redundancy rule generalizes the \redundancy to multiple relations:
\begin{lemma}\label{rule:starredundancy}
	\textbf{Star query redundancy rule: }
	If the join graph of $A, B_1, \cdots, B_n$ is a star join and $B_1\to A, B_2\to A, \cdots, B_n\to A$, then $S(A/(B_1, B_2, \cdots, B_n), B_i)=S(A/(B_1,B_2,\cdots, B_n))$.
\end{lemma}

\begin{proof}
	By~\ref{property:commutativity}, $A/(B_1, B_2,\cdots, B_n)=(A/B_i)/(B_1,B_2,\cdots,B_{i-1}, B_{i+1},\cdots, B_n)$. By~\ref{rule:redundancy}, $S((A/B_i)/(B_1,B_2,\cdots,B_{i-1}, B_{i+1},\cdots, B_n), B_i)=S((A/B_i)/(B_1,B_2,\cdots,B_{i-1}, B_{i+1},\cdots, B_n))$. By~\ref{property:commutativity} again, we have $S((A/B_i)/(B_1,B_2,\cdots,B_{i-1}, B_{i+1},\cdots, B_n))=S(A/(B_1,B_2,\cdots, B_n))$. Thus, $S(A/(B_1, B_2, \cdots, B_n), B_i)=S(A/(B_1,B_2,\cdots, B_n))$.
\end{proof}
}

\cut{
\subsection{$R_0$ is the right most leaf}
}

\point{Describe what the plans look like for star query}

\point{Prove plans equivalent in cost for the two cases}

\changed{We now show that, } all the plans in this plan space where the right most leaf is $R_0$ has the same cost $C_{out}$ \changed{if bitvector filters have no false positives.} Formally,

\vspace{-.5em}
\begin{lemma}\label{lemma:star:mincost_rightmost1}
\textbf{Minimal cost right deep tree for star query with right most leaf $R_0$: }
	Let $\mathcal{R}$ be the set of relations of a star query as defined in Definition~\ref{def:star_query}. The cost of the right deep tree $C_{out}(T(R_0, X_1, X_2,\cdots, X_n))$ is the same for every permutation $X_1,X_2,\cdots,X_n$ of $R_1,R_2,\cdots,R_n$.
\end{lemma}

\vspace{-.5em}
\begin{proof}
Because $R_1, R_2, \cdots, R_n$ only connects to $R_0$, and $R_0$ is the right most leaf, based on Algorithm~\ref{algorithm:bitvector_pushdown}, all the bitvector filters created from $R_1, R_2, \cdots, R_n$ will be pushed down to $R_0$. Thus, $C_{out}(X_k)=|X_k|$ for $1\leq k\leq n$ and $C_{out}(R_0)=|R_0/(X_1, X_2, \cdots, X_n)|$. By Lemma~\ref{lemma:star_absorption}, $C_{out}(R_0)=|R_0/(R_1, R_2, \cdots, R_n)|$.

Now consider the intermediate join result for $S(R_0,X_1,\break  \cdots, X_k)$, where $1\leq k\leq n$. By Lemma~\ref{lemma:star_absorption}, $|S(R_0, X_1, \cdots,X_k)|\break =|S(R_0/(R_1, \cdots, R_n), X_1, \cdots, X_k)|=|S(R_0, R_1, \cdots, R_n)|$. Thus, $C_{out}(S(R_0, X_1, \cdots, X_k))=C_{out}(S(R_0, R_1, \cdots, R_n))$ for all $1\leq k \leq n$.

Since the total cost of the plan is $C_{out}(T(R_0, X_1, \cdots, X_{n-1}))\break =\sum_{i=1}^{n}|R_i|+n\cdot |S(R_0, R_1, \cdots, R_0)|$, every permutation $X_1, \cdots, X_n$ of $R_1, \cdots, R_n$ has the same cost.
\end{proof}
\vspace{-.5em}

\cut{
\changed{
\vspace{-.5em}
\begin{proof}
Because $R_1, R_2, \cdots, R_n$ only connects to $R_0$, and $R_0$ is the right most leaf, based on Algorithm~\ref{algorithm:bitvector_pushdown}, all the bitvector filters created from $R_1, R_2, \cdots, R_n$ will be pushed down to $R_0$.
	By Lemma~\ref{lemma:star_absorption}, $C_{out}(R_0)=|R_0/(R_1, R_2, \cdots, R_n)|$.

By Lemma~\ref{lemma:star_absorption}, the intermediate join size $|S(R_0, X_1, \cdots, X_k)|$ $=|S(R_0/(R_1, R_2, \cdots, R_n), X_1, \cdots, X_k)|=|S(R_0, R_1, \cdots, R_n)|$.

	Thus, the total cost of the plan is $C_{out}(T(R_0, X_1, \break \cdots, X_{n-1}))$ $=\sum_{i=1}^{n}|R_i|+n\cdot |S(R_0, R_1, \cdots, R_0)|$. So every permutation $X_1, \cdots, X_n$ has the same cost.
\end{proof}
\vspace{-.5em}
}
}

\cut{\subsection{$R_0$ is not the right most leaf}}

Now consider the other case where $R_0$ is not the right most leaf, and $X_1=R_0$. Let $X_1=R_k, 1\leq k\leq n$, similarly, we show that the cost of the plans in the form of $T(R_k, R_0, X_1, X_2, \cdots, X_{n-1})$ is the same for every permutation of $R_1, R_2, \cdots, R_{k-1}, R_{k+1}, \cdots, R_n$ \changed{if bitvector filters have no false positives.} Formally,

\vspace{-.5em}
\cut{
\begin{lemma}\label{lemma:star:mincost_rightmost2}
	\textbf{Minimal cost right deep tree for star query with right most leaf $R_k$: }
		Let $\mathcal{R}$ be the set of relations of a star query as defined in Definition~\ref{def:star_query}. The cost of the right deep tree $C_{out}(T(R_k, R_0, X_1, X_2, \cdots, X_{n-1})$ is the same for every permutation $X_1, X_2, \cdots, X_{n-1}$ of $R_2, R_3, \cdots, R_{k-1}, R_{k+1}, \cdots, R_n$.
\end{lemma}
}

\begin{restatable}{lemma}{lemmastarmincostrightmosttwo}\label{lemma:star:mincost_rightmost2}
	\textbf{Minimal cost right deep tree for star query with right most leaf $R_k$: }
	Let $\mathcal{R}$ be the set of relations of a star query as defined in Definition~\ref{def:star_query}. The cost of the right deep tree $C_{out}(T(R_k, R_0, X_1, X_2, \cdots, X_{n-1})$ is the same for every permutation $X_1, X_2, \cdots, X_{n-1}$ of $R_2, R_3, \cdots, R_{k-1}, R_{k+1}, \cdots, R_n$.
\end{restatable}

The proof can be found in Appendix~\ref{sec:moreproofs}.

\cut{
The proof is similar to that of Lemma~\ref{lemma:star:mincost_rightmost1}.
}
\cut{
\vspace{-.5em}
\begin{proof}
Because $R_1, \cdots, R_n$ only connects to $R_0$, the bitvector filters created from $R_1, \cdots, R_{k-1}, R_{k+1}, \cdots, R_n$ will be pushed down to $R_0$, and the bitvector created from $R_0$ will be pushed down to $R_k$. Thus, $C_{out}(R_0)=|R_0/(R_1, \cdots, R_{k-1}, R_{k+1}, \cdots, R_n)|$. Let $R_0'=R_0/(R_1, \cdots, R_{k-1}, R_{k+1}, \cdots, R_n)$, then $C_{out}(R_k)=|R_k/R_0'|$.

By Lemma~\ref{lemma:star_absorption} and Property~\ref{property:redundancy}, $|S(R_k/R_0', R_0', X_1, X_2, \cdots, \break X_k)|=|S(R_0, R_1, \cdots, R_n)|$. Thus, the total cost of the plan is $C_{out}(T(R_k, R_0, X_1, X_2, \cdots, X_{n-1}))=\sum_{i=1, i\neq k}^{n}|R_i|+C_{out}(R_0)+C_{out}(R_k)+(n-1)\cdot |S(R_0, R_1, \cdots, R-n)|$. Thus, $C_{out}(T(R_k, R_0, X_1, X_2, \cdots, X_{n-1}))$ is the same for every permutation $X_1, X_2, \cdots, X_{n-1}$ of $R_1, R_2, \cdots, R_{k-1}, R_{k+1}, \cdots, R_n$.

\end{proof}
\vspace{-.5em}
}

\cut{
\changed{The proof leverages Lemma~\ref{lemma:star_absorption} and Property~\ref{property:redundancy}, and it can be found in~\cite{moreproofs}.}
}

\cut{
\changed{
\vspace{-.5em}
\begin{proof}
	Because $R_1, R_2, \cdots, R_n$ only connects to $R_0$, the bitvector filters created from $R_1, R_2, \cdots, R_{k-1}, R_{k+1}, \cdots, R_n$ will be pushed down to $R_0$, and the bitvector created from $R_0$ will be pushed down to $R_k$. Thus, $C_{out}(R_0)=|R_0/(R_1, \cdots, R_{k-1}, R_{k+1}, \cdots, R_n)|$. Let $R_0'=R_0/(R_1, \cdots, R_{k-1}, R_{k+1}, \cdots, R_n)$, then $C_{out}(R_k)=|R_k/R_0'|$.
	
	By Lemma~\ref{lemma:star_absorption} and Property~\ref{property:redundancy}, the intermediate join size $|S(R_k/R_0', R_0', X_1, \cdots, X_k)|=|S(R_0, R_1, \cdots, R_n)|$. Thus, the total cost of the plan is $C_{out}(T(R_k, R_0, X_1, \cdots, X_{n-1}))=\sum_{i=1, i\neq k}^{n}|R_i|+C_{out}(R_0)+$ $C_{out}(R_k)+(n-1)\cdot |S(R_0, R_1, \cdots, R-n)|$, which is the same for every permutation $X_1, \cdots, X_{n-1}$.
	
\end{proof}
\vspace{-.5em}
}
}

\cut{
\subsection{Plan space complexity}
}
\point{Show what candidate plans of min cost look like and the number of candidate plans}

By combining Lemma~\ref{lemma:star:mincost_rightmost1} and Lemma~\ref{lemma:star:mincost_rightmost2}, we can prove Theorem~\ref{theorem:star:mincost_plan} and Theorem~\ref{theorem:star_plan_space}.

\cut{
Combine Theorem~\ref{theorem:star:mincost_rightmost1} and Theorem~\ref{lemma:star:mincost_rightmost2}, we have

\vspace{-.5em}
\begin{theorem}\label{theorem:star:mincost_plan}
	\textbf{Minimal cost right deep trees for star query: }
		Let $\mathcal{R}$ be the set of relations of a star query as defined in Definition~\ref{def:star_query}. Let $\mathcal{A} =\{T(X_0, \cdots, X_n)\}$ be the set of right deep trees without cross products for $q$, where $X_0, \cdots, X_n$ is a permutation of $R_0, \cdots, R_n$. If $C_{min}=min\{C_{out}(\mathcal{T}), \mathcal{T}\in\mathcal{A} \}$, then there exists a plan $\mathcal{T}\in\mathcal{A}_{candidates}=\{T(R_0, R_1, \cdots, R_n)\} \cup \{T(R_k, R_0, R_1, \cdots, R_{k-1}, R_{k+1}, \cdots, R_n), 1\leq k \leq n \}$ such that $C_{out}(\mathcal{T})=C_{min}$.
\end{theorem}
\vspace{-.5em}

Directly follow Theorem~\ref{theorem:star:mincost_plan}, we have

\vspace{-.5em}
\begin{theorem}\label{theorem:star_plan_space}
	\textbf{plan space complexity for star query: }
		Let $\mathcal{R}$ be the set of $n+1$ relations of a star query as defined in Definition~\ref{def:star_query}. We can find the query plan with the minimal cost in the place space of right deep trees without cross products from $n+1$ candidate plans.
\end{theorem}
\vspace{-.5em}
}
\cut{

\subsection{Right deep tree with cross products for star query}

Now let's consider the right deep tree with cross products.

Let $R_1, R_2, \cdots, R_n$ be relations of a star query as defined in Definition~\ref{def:star_query}. Let $\mathcal{T}=T(X_1, X_2, \cdots, X_n)$ be a right deep tree, where $X_1, X_2, \cdots, X_n$ is a permutation of $R_0, R_1, \cdots, R_n$. Let $X_k=R_0, k<n$, then $\mathcal{T_{sub}}=T(X_{k+1}, X_{k+2}, \cdots, X_n)$ is a subplan of $\mathcal{T}$ with all cross products. Because $T(X_{k+1}, X_{k+2}, \cdots, X_n)$ are all cross products, no bitvector will be created.

\begin{lemma}\label{lemma:star_mincost_cp}
	\textbf{Minimal cost cross product plan}: Let $R_1, R_2, \cdots, R_m$ be relations without any join condition, and $Y_1, Y_2, \cdots, Y_m$ be a permutation of $R_1, R_2, \cdots, R_m$. Let $\mathcal{T}=T(Y_m, Y_{m-1}, \cdots, Y_1)$ be a right deep tree, then $C_{out}(\mathcal{T})$ is minimal if $|Y_1|\leq |Y_2|\leq \cdots \leq |Y_m|$.
\end{lemma}

\begin{proof}
Because $R_1, R_2, \cdots, R_m$ have no join conditions, all the joins are cross products and no bitvector filters will be created. Thus, $S(Y_i)=|Y_i|, 1\leq i\leq m$ and $S(Y_1, Y_2, \cdots, Y_i)=|Y_1|\times |Y_2|\times \cdots \times |Y_i|$.
Thus, $C_{out}(T(Y_1, Y_2, \cdots, Y_m))=\sum_{i=1}^n|Y_i|+\sum_{i=2}^m\prod_{j=1}^i |Y_j|$. 

Now we prove $C_{out}(T(Y_m, Y_{m-1}, \cdots, Y_1))$ is minimal if $|Y_1|\leq |Y_2|\leq \cdots \leq |Y_m|$. Assume there exists $Y_k$, $Y_l$, where $k<l$ and $|Y_k|>|Y_l|$. Consider a new permutation $Z_1, Z_2, \cdots, Z_m$, where $Z_i=Y_i$ if $i\neq k$ and $i\neq l$, $Z_k=Y_l$ and $Z_l=Y_k$. Then
\begin{align*}
C_{out}&(T(Y_m, Y_{m-1}, \cdots, Y_1))-C_{out}(T(Z_m, Z_{m-1},\cdots, Z_1))\\
&=\sum_{i=2}^m\prod_{j=1}^i|Y_j|-\sum_{i=2}^m\prod_{j=1}^i|Z_j| \\
&=\sum_{i=k}^{l-1}\prod_{j=1}^i|Y_j|-\sum_{i=k}^{l-1}\prod_{j=1}^i|Z_j| \\
&=(|Y_k|-|Z_k|)\sum_{i=k}^{l-1}\prod_{j=1,j\neq k}^i|Y_j| \\
&=(|Y_k|-|Y_l|)\sum_{i=k}^{l-1}\prod_{j=1,j\neq k}^i|Y_j|>0
\end{align*}

Thus, $T(Z_m, Z_{m-1},\cdots, Z_1)$ has smaller cost than $T(Y_m, Y_{m-1}, \cdots, Y_1)$, which contradicts to the assumption that $C_{out}(\mathcal{T})$ is minimal. Therefore, $|Y_1|\leq |Y_2|\leq \cdots \leq |Y_m|$.
\end{proof}

\todo{Plans with minimal cost for right deep tree with cross products}
\begin{lemma}\label{lemma:star_mincost_plan_cp}
	\textbf{Minimal cost right deep tree with cross products for star query}: Let $R_0, R_1, R_2, \cdots, R_{n-1}$ as defined in Definition~\ref{def:star_query}. Let $X_1, X_2, \cdots, X_n$ be a permutation of $\{R_1, R_2, \cdots, R_n\}$. Let $\mathcal{A}$ be the set of right deep trees with cross products. Then $\mathcal{T}=T(X_1, X_2, \cdots, X_n) \in \mathcal{A}$ has minimal cost if
	\begin{enumerate}
		\item $|X_1|\leq |X_2|\leq \cdots \leq |X_k|$ and
		\item $|X_i|\leq |X_j|$ for $1\leq i\leq k$ and $k+1\leq j\leq n-1$
	\end{enumerate}
\end{lemma}

\begin{proof}
	By Lemma~\ref{lemma:mincostleftcp}, we have $|U_1|\leq |U_2|\leq \cdots \leq |U_k|$ for a given set of $\{U_1, U_2, \cdots, U_k\}$.

Let $T_{left}=T_{left}(U_1, U_2, \cdots, U_k)$.	Now assume $T=T_{right}(R_1, T_{left}, U_{k+1}, \cdots, U_{n-1})$ has the minimal cost among all quasi right deep tree with cross products of $k$ relations, and there exists $i, j$ such that $|U_i|>|U_j|$ and $1\leq i\leq k, k+1\leq j\leq n-1$. Consider an alternative permutation $U_1, U_2, \cdots, U_{i-1}, U_j, U_{i+1}, \cdots, U_{n-1}$. Let $T'_{left}=T_{left}(U_1, U_2, \cdots, U_{i-1}, U_j, U_{i+1}, \cdots, U_k)$, the corresponding quasi right deep tree is $T'=T_{right}(R_1, T'_{left}, U_{k+1}, \cdots, U_{j-1}, U_i, U_{j+1}, \cdots, U_{n-1})$. 
	
 	First, it is obvious that $C_{out}(T_{left})>C_{out}(T'_{left})$.
	Consider the set of bitvectors created at $T_1$. $T_1$ has created a bitvector from each $U_l, k+1\leq l\leq n-1$ and one bitvector from $S(U_1, \cdots, U_k)$. Because $S(U_1, \cdots, U_k)$ are cross products, by~\ref{rule:decomposition}, $R_1/S(U_1, \cdots, U_k)=R_1/(U_1, \cdots, U_k)$. By~\ref{property:commutativity}, $R_1/(S(U_1, \cdots, U_k), U_{k+1}, \cdots, U_{n-1})=R_1/(U_1, \cdots, U_{n-1})$. Similarly, we can show the bitvectors created at $T_2$ for $R_1$ is also equivalent to $R_1/(U_1, \cdots, U_n)$.
	
	Now consider cost of $T_1$ and $T_2$:
	\begin{align*}
	C_{out}(T)&=C_{out}(T_{left})+\sum_{i=k+1}^{n-1} |U_i|\\
	&+(n-k)\times |S(R_1/(U_1,\cdots, U_{n-1}))| \\
	C_{out}(T')&=C_{out}(T'_{left})+\sum_{i=k+1}^{n-1} |U_i|\\
	&+(n-k)\times |S(R_1/(U_1,\cdots, U_{n-1}))| \\
	C_{out}(T_{left})&>C_{out}(T'_{left}) \rightarrow C_{out}(T)>C_{out}(T')
	\end{align*}
	This contradicts to the assumption that $C_{out}(T)$ is minimal. Thus, there does not exist such $i, j$ and $|U_i|\leq |U_j|$ for $1\leq i\leq k$ and $k+1\leq j\leq n-1$.
\end{proof}	

\begin{theorem}\label{theorem:mincoststarquasirightcp}
	\textbf{Minimal cost quasi right deep tree for star query}: Let $R_1, R_2, \cdots, R_n$ be relations of a star join, where $R_1$ is the fact table and $R_i, 2\leq i\leq n$ are dimension tables. WOLG, assume $|R_2|\leq |R_3|\leq \cdots |R_n|$. Let $T_k=T(R_1, T_{left}(U^k_1, U^k_2, \cdots, U_k), U^k_{k+1}, \cdots, U^k_{n-1})$ be the minimal cost quasi right deep tree with cross products of $k$ relations among all permutations $U^k_1, U^k_2, \cdots, U^k_{n-1}$ of $\{R_2, \cdots, R_n\}$. Then $T=T_k$ is the minimal cost quasi right deep tree if
	\begin{enumerate}
		\item $U^k_i=R_i$ for $1\leq i\leq k$ and
		\item $\prod_{i=2}^k |R_i|\leq |S(R_1/(R_2, \cdots, R_n))|$ and
		\item $\prod_{i=2}^{k+1} |R_i|>|S(R_1/(R_2, \cdots, R_n))|$
	\end{enumerate}
\end{theorem}

\begin{proof}
	Since $|R_2|\leq |R_3|\leq \cdots \leq |R_n|$, by Lemma~\ref{lemma:mincoststarquasirightcpk}, we have $T_k$ is a minimal cost quasi right deep tree with cross products of $k$ relations if $U_i=R_i$ for $1\leq i\leq k$.
	
	Now consider $T_{k}$ and $T_{k+1}$, where $T_k=T(R_1, T_{left}(R_2, R_3, \cdots, R_k), U^k_{k+1}, \cdots, U^k_{n-1})$ and $T_{k+1}=T(R_1, T_{left}(R_2, R_3, \cdots, R_k, R_{k+1}), U^{k+1}_{k+2}, \cdots, U^{k+1}_{n-1})$. We have
	\begin{align*}
	C_{out}(T_k)-C_{out}(T_{k+1})=|S(R_1/(R_2, \cdots, R_n))|-\prod_{i=2}^{k+1}|R_i|
	\end{align*}
	Since $|S(R_1/(R_2, \cdots, R_n))|$ is a constant and $\prod_{i=2}^{k+1}|R_i|$ monotonically increasing with larger $k$, $C_{out}(T_k)$ is minimal if $\prod_{i=2}^k |R_i|\leq |S(R_1/(R_2, \cdots, R_n))|$ and $\prod_{i=2}^{k+1} |R_i|>|S(R_1/(R_2, \cdots, R_n))|$.
\end{proof}	

\todo{optimization: replace the right deep tree of cross products with left deep trees}
Thus, it is better to replace this right deep tree $T_{right}$ with a left deep tree $T_{left}(R_2, R_3, \cdots, R_k)$ since $T_{left}$ has the same cost than $T_{right}$ but lower memory consumption.

In addition, because the fact table can be potentially benefit from index seeks with cross products of dimension tables, the join with the fact table and the cross products will treat the cross products as the build side and the fact table as the probe side.

\subsection{Algorithm}
\begin{algorithm}
\DontPrintSemicolon
\SetAlgoLined
\LinesNumbered

\caption{Construct minimal cost quasi right deep tree query plan for star query}
\label{algorithm:quasi_star}

\SetKw{Continue}{continue}
\SetKwProg{FuncQuasiStar}{}{:}{}
\FuncQuasiStar{\textbf{QuasiStarPlan($G$, $f$, $B$)}}{
	\KwIn{Join graph $G$, fact table $f$, set of branches $B$}
	\KwOut{Query plan $plan$}
	$sortedBranches \leftarrow SortByCardAsc(B)$\;
	$finalCard\leftarrow JoinAll(f, B).Card()$\;
	$plan \leftarrow \emptyset$\;
	$crossProduct \leftarrow \emptyset$\;
	$crossProductMode\leftarrow True$\;
	\ForEach{$b \in sortedBranches$}{
		\If{$crossProductMode$}{
			\If{$crossProduct$ is empty}{
				$crossProduct\leftarrow b$\;
				\Continue
			}
			$newCP = Join(crossProduct, b)$\;
			\lIf{$finalCard \geq newCP.Card()$}{
				$crossProduct\leftarrow newCP$
			}
			\Else{
				$plan\leftarrow Join(crossProduct, f)$\;
				$plan\leftarrow HashJoin(b, plan)$\;
				$crossProductMode\leftarrow False$
			}
		}
		\lElse{
			$plan\leftarrow HashJoin(b, plan)$
		}
	}
	\lIf{$crossProductMode$}{
		$plan\leftarrow Join(crossProduct, f)$
	}
	\Return $plan$
}

\end{algorithm}

Algorithm~\ref{algorithm:quasi_star} shows the process to construct the minimal cost plan with quasi right deep tree for star queries.
\todo{Describe the algorithm}
\todo{(Optional) Add an example of the algorithm}
}
\cut{\section{Chain Query}
\label{sec:chain}
\sloppy

\done{This section is ready}

We continue our analysis with the chain query as defined in Definition~\ref{def:chain_query}. We want to show that, in the plan space of right deep trees without cross products, we can find the optimal query plan (subject to $C_{out}$) from $n$ query plans with bitvector filters, where $n$ is the number of relations in the chain query. The original plan space complexity for chain queries in this plan space is $(n-1)^2$~\cite{ono1990measuring}.

We divide the plans in this plan space into two cases: whether $R_n$ is the right most leaf or not. The key insight is to derive that the plans where $R_n, R_{n-1}, \cdots, R_k$ joins consecutively have equal or less cost than other plans based on \associativity, \commutativity and \reduction of the bitvector filters. For example, if the chain query is $R_0 \to R_1 \to R_2 \to R_3$, we try to prove that $C_{out}(R_2, R_3, X, Y)\leq C_{out}(R_2, X, R_3, Y)$.

\subsection{$R_n$ is the right most leaf}
\point{Plans where $R_n$ is the right most leaf}

Let's first look at the query plans where $R_n$ is the right most leaf. Formally,
\vspace{-.5em}

\begin{lemma}\label{lemma:chain_tree_nocp}
Let $\mathcal{R}$ be the set of relations of a chain query as described in Definition~\ref{def:chain_query}. There exists only one right deep tree without cross products such that $R_n$ is the right most leaf, that is, $T(R_n, R_{n-1}, \cdots, R_0)$.
\end{lemma}
\vspace{-.5em}

\begin{proof}
If $R_n$ is the right most leaf and there is no cross product in the query plan, then $R_n$ can only join with $R_{n-1}$. Thus, the right most subplan with two relations is $T_2=T(R_n, R_{n-1})$. Similarly, if the right most subplan is $T_k=T(R_n, R_{n-1}, \cdots, R_{n-k+1})$ and there is no cross product, then $T_k$ can only join with $R_{n-k}$. By induction, $T(R_n, R_{n-1}, \cdots, R_0)$ is the only right deep tree without cross products where $R_n$ is the right most leaf.
\end{proof}
\vspace{-.5em}

\subsection{$R_n$ is not the right most leaf}

\point{Plans where $R_n$ is not the right most leaf. First prove we can reduce the cost by pushing down $R_n, R_{n-1}, \cdots, R_k$}
	
Now we look at the query plans where $R_n$ is not the right most leaf. Let $T(X_0, \cdots, X_n)$ be a right deep tree without cross products where $X_0, \cdots, X_n$ is a permutation of $R_0, \cdots, R_n$. We will show that the plans that join $R_n, R_{n-1}, \cdots, R_k$ consecutively have equal or less cost than other plans.
\vspace{-.5em}

\begin{lemma}\label{lemma:chain_pushdown1}
	\textbf{Cost reduction by pushing down $R_n$: }
	Let $\mathcal{R}$ be the set of relations of a chain query as described in Definition~\ref{def:chain_query}. Let $\mathcal{T}=T(X_0, X_1, \cdots, X_n)$ be a right deep tree without cross products for $R_0, R_1, \cdots, R_n$. Assume $X_k=R_n$ for some $1 \leq k \leq n$. If $X_{k-1}\neq R_{n-1}$, then $\mathcal{T'}=T(X_0, X_1, \cdots, X_k, X_{k-1}, X_{k+1}, X_{k+2}, \cdots, X_n)$ is a right deep tree without cross products and $C_{cout}(\mathcal{T'}) \leq  C_{out}(\mathcal{T})$.
\end{lemma}

\vspace{-.5em}
\begin{proof}
Since there is no cross product in $\mathcal{T}=T(X_0, X_1,\cdots, X_{k-1}, R_n, X_{k+1}, \cdots, X_n)$, one relation in $\mathcal{A}=\{X_0, X_1, \cdots, X_{k-1}\}$ must connect to $R_n$. Since $R_{n-1}$ is the only relation that connects to $R_n$ in the join graph, $R_{n-1}\in \{X_0, X_1, \cdots, X_{k-1}\}$. By induction, we can show that $ R_{n-2}, R_{n-3},\cdots, R_{n-k} \in \mathcal{A}$. Thus, $\mathcal{A}=\{R_{n-k}, R_{n-k+1}, \cdots, R_{n-1}\}$.

If $X_{k-1}\neq R_{n-1}$, then $X_{k-1}=R_{n-k}$; otherwise, the join graph of $X_0, X_1, \cdots, X_{k-2}$ is not connected and the subplan $T(X_0, X_1, \cdots, X_{k-2})$ has cross products.

Now consider the relations $\{X_{k+1}, X_{k+2}, \cdots, X_n\}$. Because $X_{k+1}$ joins with $\{X_0, X_1,\cdots, X_k\}=\{R_n, R_{n-1}, \cdots, R_{n-k}\}$, $X_{k+1}=R_{n-k-1}$. Similarly, we can show that $X_i=R_{n-i}$ for $k<i\leq n$.

If we swap $R_n$ and $R_{n-k}$, we get a new plan $\mathcal{T'}=T(X_0, X_1, \cdots, X_{k-2}, R_n, R_{n-k}, X_{k+1}, \cdots, X_n)$. Because $\{X_0, X_1, \cdots, X_{k-2}\}=\mathcal{A}\setminus \{R_{n-k}\}=\{R_{n-k+1}, R_{n-k+2}, \cdots, R_{n-1}\}$. Thus, $\mathcal{T'}$ has no cross product.

Now we prove $C_{out}(\mathcal{T'})\leq C_{out}(\mathcal{T})$.

\cut{
First, consider $X_i$ for $k< i\leq n$. Because there is no change from the root to $X_{k+1}$, the bitvector filters created from $X_{k+1}, \cdots, X_n$ and the bitvector filters pushed down to $X_{k+1}, \cdots, X_n$ are the same. Thus, $C_{out}(X_i)$ is the same for $\mathcal{T}$ and $\mathcal{T'}$.
}

First, consider $X_i$ for $k< i\leq n$. Since there is no change in bitvector filters, it is easy to see that $C_{out}(X_i)$ is the same for $\mathcal{T}$ and $\mathcal{T'}$.

\cut{
Next, consider $X_i$ for $0\leq i < k$. Since $\mathcal{B}=\{X_0, X_1, \cdots, X_{k-1}\}=\{R_{n-k+1}, \cdots, R_{n-1}\}$, only $R_{n-k}$ and $R_n$ will create bitvector filters that can be pushed down to subplans of $\mathcal{B}$. Because $R_{n-1}\in \mathcal{B}$, no bitvector filter will be pushed down to $R_n$. Thus, the bitvector filter created from $R_n$ is the same for $\mathcal{T}$ and $\mathcal{T'}$, and the same bitvector filter will be pushed down to $R_{n-1}$ the same way in $\mathcal{T}$ and $\mathcal{T'}$. Similarly, the bitvector filters pushed down to and created from $R_{n-k}$ are the same in $\mathcal{T}$ and $\mathcal{T'}$, and the bitvector filters created from $R_{n-k}$ will be pushed down to $R_{n-k+1}$ the same way in $\mathcal{T}$ and $\mathcal{T'}$.
}

Next, consider $X_i$ for $0\leq i < k$. Since $\mathcal{B}=\{X_0, X_1, \cdots, X_{k-1}\}=\{R_{n-k+1}, \cdots, R_{n-1}\}$, only $R_{n-k}$ and $R_n$ will create bitvector filters that can be pushed down to subplans of $\mathcal{B}$. Because $R_{n-1}\in \mathcal{B}$, no bitvector filter will be pushed down to $R_n$. Thus, the bitvector filter created from $R_n$ is the same for $\mathcal{T}$ and $\mathcal{T'}$, and the same bitvector filter will be pushed down to $R_{n-1}$ the same way in $\mathcal{T}$ and $\mathcal{T'}$. Similarly, the bitvector filters created from and pushed down to $R_{n-k}$ and $R_{n-k+1}$ are the same in $\mathcal{T}$ and $\mathcal{T'}$.

Thus, we have proved $C_{out}(X_i)$ is the same for $0\leq i\leq n$.

Next, we show that the intermediate join sizes in $\mathcal{T'}$ is equal to or smaller than these in $\mathcal{T}$.

Since $\mathcal{T}$ and $\mathcal{T'}$ share the same subplan $\mathcal{T}_{j}=T(X_0, X_1, \cdots, X_j), 0\leq j\leq k-2$, and we have shown the bitvector filters pushed down to $\mathcal{T}_{j}$ is the same in $\mathcal{T}$ and $\mathcal{T'}$, the intermediate join sizes are the same in $\mathcal{T}_{j}$ for both plans.

Consider the cardinalities of the join $S(T_{k-2}, R_{n-k})$ in $\mathcal{T}$ and $S(T_{k-2}, R_n)$ in $\mathcal{T'}$. Since $R_\{n-1\} \in\mathcal{A}$, $S(T_{k-2}, R_n)$ is a unique key join. By \absorption,  $|S(T_{k-2}, R_n)|=|T_{k-2}|$. Since $R_{n-k-1}=X_{k+1} \notin \mathcal{A}$, $S(T_{k-2}, R_{n-k})$ is not a unique key join. By \reduction, $|T_{k-2} \Join R_{n-k}|\geq |T_{k-2}|$. Thus, $|T_{k-2} \Join R_n|=|T_{k-2}|\leq |T_{k-2} \Join R_{n-k}|$.

Now consider the cardinalities for $S(T_{k-2}, R_{n-k}, R_n)$ in $\mathcal{T}$ and $S(T_{k-2}, R_n, R_{n-k})$ in $\mathcal{T'}$. Since the set of bitvector filters from $B$ pushed down to $S(T_{k-2}, R_{n-k}, R_n)$ is the same as those pushed down to $S(T_{k-2}, R_n, R_{n-k})$ and the join relations are the same, $|S(T_{k-2}, R_{n-k}, R_n)|=|S(T_{k-2}, R_n, R_{n-k})|$. Similarly, we can show that $|S(T_{k-2}, R_{n-k}, R_n, X_{k+1}, \cdots, X_i)| =|S(T_{k-2}, R_{n}, R_{n-k}, X_{k+1}, \cdots, X_i)|$ for $k+1\leq i\leq n$.

Thus, $C_{out}(\mathcal{T})=\sum_{i=1}^n C_{out}(X_i)+\sum_{i=0, i\neq k-1, k}^n |S(X_0, \cdots, X_i)|+|S(T_{k-2}, R_{n-k})| +|S(T_{k-2}, R_{n-k}, R_n)| \geq \sum_{i=1}^n C_{out}(X_i)+\sum_{i=0, i\neq k-1, k}^n |S(X_0, \cdots, X_i)|+|S(T_{k-2}, R_n)| +|S(T_{k-2}, R_n, R_{n-k})|=C_{out}(\mathcal{T'})$.

\end{proof}
\vspace{-.5em}

\begin{lemma}\label{lemma:chain_pushdown2}
	\textbf{Cost reduction by pushing down $R_n, R_{n-1}, \cdots, R_{n-m}$: }
		Let $\mathcal{R}$ be the set of relations of a chain query as described in Definition~\ref{def:chain_query}. Let $\mathcal{T}=T(X_0, X_1, \cdots, X_n)$ be a right deep tree without cross products for $R_0, R_1, \cdots, R_n$. Let $X_k=R_n, X_{k-1}=R_{n-1}, \cdots, X_{k-m}=R_{n-m}$ for some $m\leq k\leq n$. If $X_{k-m-1}\neq R_{n-m-1}$, then $\mathcal{T'} =T(X_0, X_1, \cdots, X_{k-m-2}, X_{k-m}, X_{k-m+1}, \cdots, X_k, X_{k-m-1}, \break X_{k+1},\cdots, X_n)$ is a right deep tree without cross products and $C_{cout}(\mathcal{T'}) \leq C_{out}(\mathcal{T})$.
\end{lemma}
\vspace{-.5em}

\cut{
\begin{proof}
Similar to the proof of Lemma~\ref{lemma:chain_pushdown1}, we can show that $X_{k-m-1}=R_{n-k}$ if $X_{k-m-1}\neq R_{n-m-1}$, $\mathcal{A}=\{X_{0}, X_{1}, \cdots, X_{k-m-1}\}=\{R_{n-k}, R_{n-k+1}, \cdots, R_{n-m-1}\}$, and $X_i=R_{n-i}$ for $k<i\leq n$.

Now consider swapping $R_n, R_{n-1}, \cdots, R_{n-m}$ with $R_{n-k}$, the resulting plan is $\mathcal{T'}$. Similar to the proof of Lemma~\ref{lemma:chain_pushdown1}, we can show that $\mathcal{T'}$ has no cross product.

Consider $C_{out}$ for $X_0, X_1, \cdots, X_n$. Similar to the proof of Lemma~\ref{lemma:chain_pushdown1}, we can show that $C_{out}(X_i)$ is the same for $X_0, X_1, \cdots, X_n$ in $\mathcal{T}$ and $\mathcal{T'}$.

Next, consider the intermediate join sizes. Since both $\mathcal{T}$ and $\mathcal{T'}$ share the same subplan $\mathcal{T}_{j}(X_0, X_1, \cdots, X_j), 0\leq j\leq k-m-2$, similar to the proof of  Lemma~\ref{lemma:chain_pushdown1}, we can show that $\mathcal{T}_{j}, 0\leq j\leq k-m-2$ is the same for $\mathcal{T}$ and $\mathcal{T'}$.

Now consider the cardinality of joins $S(\mathcal{T}_{k-m-2}, R_{k-m-1})$ and $S(\mathcal{T}_{k-m-2}, R_{n-k})$, similar to Lemma~\ref{lemma:chain_pushdown1}, we can show $|S(\mathcal{T}_{k-m-2}, R_{k-m-1})| \leq S(\mathcal{T}_{k-m-2}, R_{n-k})$.
 
Now consider the cardinality of joins $S(\mathcal{T}_{k-m-2}, R_{k-m-1}, R_{k-m})$ and $S(\mathcal{T}_{k-m-2}, R_{n-k}, R_{k-m-1})$. Since $R_{k-m-2}$ is a unique key join with $S(\mathcal{T}_{k-m-2}, R_{k-m-1})$, $|S(\mathcal{T}_{k-m-2}, R_{k-m-1}, R_{k-m})|=S(\mathcal{T}_{k-m-2}, R_{k-m-1}$. Similarly, since $R_{k-m-1}$ is a unique key join with $S(\mathcal{T}_{k-m-2}, R_{n-k}, R_{k-m-1})$, we have $|S(\mathcal{T}_{k-m-2}, R_{n-k}, R_{k-m-1})|=|S(\mathcal{T}_{k-m-2}, R_{n-k})|$. Thus, $|S(\mathcal{T}_{k-m-2}, R_{k-m-1}, R_{k-m})|\leq S(\mathcal{T}_{k-m-2}, R_{n-k}, R_{k-m-1})$.

By similar reasoning, we can show that $|S(\mathcal{T}_{k-m-2}, R_{k-m-1}, R_{k-m}, \cdots, R_j)|\leq |S(\mathcal{T}_{k-m-2}, R_{n-k}, R_{k-m-1}, R_{k-m}, \cdots, R_{j-1})|, k-m-1\leq j\leq n$.
		
Finally, we can show that $|S(\mathcal{T}_{k-m-2}, R_{k-m-1}, R_{k-m}, \cdots, R_n, R_{n-k})|=|S(\mathcal{T}_{k-m-2}, R_{n-k}, R_{k-m-1}, R_{k-m}, \cdots, R_n)|$ and $|S(\mathcal{T}_{k-m-2}, R_{k-m-1}, R_{k-m}, \cdots, R_n, R_{n-k}, R_{n-k-1}, \cdots, R_j)|=|S(\mathcal{T}_{k-m-2}, R_{n-k}, R_{k-m-1}, R_{k-m}, \cdots, R_n, R_{n-k-1}, \cdots, R_j)|, 0\leq j\leq n-k-1$ as in Lemma~\ref{lemma:chain_pushdown1}.

By summing up everything together, we have $C_{out}(\mathcal{T})\geq C_{out}(\mathcal{T'})$.

\end{proof}
}

The proof is similar to Lemma~\ref{lemma:chain_pushdown1} and is omitted due to space limit.

\subsection{Plan space complexity}
\point{Show what candidate plans of min cost look like and the number of candidate plans}

Combine Lemma~\ref{lemma:chain_tree_nocp} and Lemma~\ref{lemma:chain_pushdown2}, we have
\vspace{-.5em}

\begin{theorem}\label{theorem:chain_mincost_plan}
	\textbf{Minimal cost right deep trees for chain query: }
			Let $\mathcal{R}$ be the set of relations of a chain query as described in Definition~\ref{def:chain_query}. Let $\mathcal{A} =\{T(X_0, X_1, \cdots, X_n)\}$ be the set of right deep trees without cross products for $q$, where $X_0, X_1, \cdots, X_n$ is a permutation of $R_0, R_1, \cdots, R_n$. If $C_{min}=min\{C_{out}(T(X_0, X_1, \cdots, X_n))\}$, then there exists a plan $\mathcal{T} \in \mathcal{A_{candidates}}=\{T(R_n, R_{n-1}, \cdots, R_0)\} \cup \{T(R_k, R_{k+1}, \cdots, R_n, R_{k-1}, R_{k-2}, \cdots, R_0), 0\leq k \leq n-1 \}$ such that $C_{out}(\mathcal{T})=C_{min}$.
\end{theorem}
\vspace{-.5em}

Directly follow Theorem~\ref{theorem:chain_mincost_plan}, we have

\vspace{-.5em}
\begin{theorem}\label{theorem:chain_plan_space}
	\textbf{Plan space complexity for chain query: }
		Let $\mathcal{R}$ be the set of $n+1$ relations of a chain query as described in Definition~\ref{def:chain_query}. We can find the query plan with the minimal cost in the place space of right deep trees without cross products from $n+1$ candidate plans.
\end{theorem}

\vspace{-.5em}
\begin{proof}
By Theorem~\ref{theorem:chain_mincost_plan}, there exists a plan in $\mathcal{A_{candidates}}=\{T(R_n, R_{n-1}, \cdots, R_1)\} \cup \{T(R_1, R_2, \cdots, R_k, R_n, R_{n-1}, \cdots, R_{k+1}), 1\leq k \leq n-1 \}$ such that the cost of the plan is minimal, we can find the plan with the minimal cost from $n$ candidate query plans.
\end{proof}
\vspace{-.5em}

\cut{
\subsection{Right deep tree with cross products}
\todo{Prove right deep tree with cross products is suboptimal}
\begin{theorem}\label{theorem:suboptchainwithcp}
	If a right deep tree plan $p$ contains a cross product for a chain query $q$, then there exists a right deep tree plan $p'$, such that $cost(p')\leq cost(p)$.
\end{theorem}

\todo{Seems this is not true???}

In other words, we don't need to consider right deep tree query plans with cross products for chain queries.
}
}
\changed{\section{Analysis of Snowflake queries with PKFK joins}\label{sec:snowflake}
}
\sloppy

\point{Snowflake query}
We define snowflake queries with PKFK joins as below:
\begin{definition}\label{def:snowflake_query}
	\changed{\textbf{Snowflake query with PKFK joins}}: Let $\break \mathcal{R}=\{R_0, R_{1,1}, \cdots, R_{1,n_1}, R_{2,1}, \cdots, R_{2,n_2},\cdots, R_{m,1},\cdots, R_{m,n_m}\}$ be a set of relations and $q$ be a query joining relations in $\mathcal{R}$. The query $q$ is a snowflake query with PKFK joins if
	\begin{itemize}[leftmargin=*]
		\item $R_{0}\to R_{i,1}$ for $1\leq i\leq m$ and
		\item $R_{i,j-1}\to R_{i,j}$ for $1\leq i\leq m, 1< j\leq n_i$.
	\end{itemize}
	We call $R_0$ the fact table and $R_{i,1}, R_{i,2}, \cdots, R_{i,n_i}$ a branch. We denote the branch $\{R_{i,1}, R_{i,2},\cdots, R_{i,n_i}\}$ as $\mathcal{R}_i$.
\end{definition}
\vspace{-.5em}

Figure~\ref{fig:snowflake_query} shows an example of a snowflake query, where $R_0$ is the fact table, and $\{R_{1,1}\}, \{R_{2,1}, R_{2,2}\}, \{R_{3,1}, R_{3,2}\}$ are three branches of dimension tables.

\begin{figure}
	\centering
	\includegraphics[width=.5\linewidth]{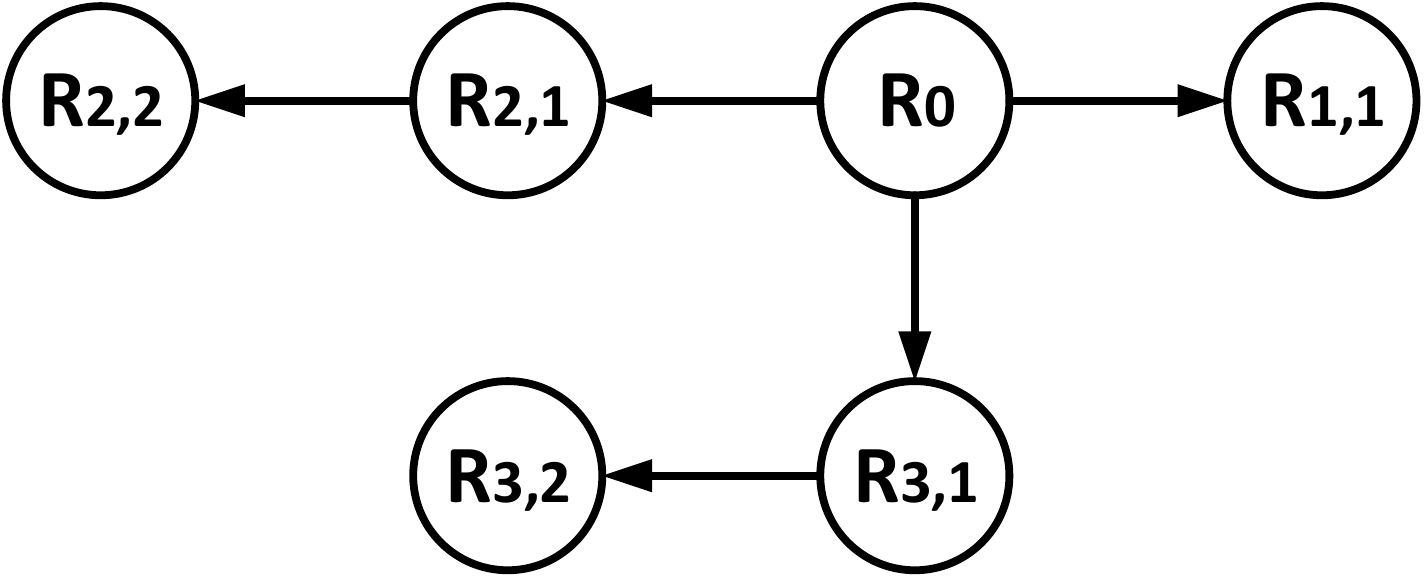}
	\caption{Snowflake query with \changed{PKFK joins}, where the fact table is $R_0$ and the branches are $\{R_{1,1}\}, \{R_{2,1}, R_{2,2}\}, \{R_{3,1}, R_{3,2}\}$}
	\label{fig:snowflake_query}
\end{figure}

\outline{The case where $R_0$ is the right most leaf}

Now we analyze the plan space complexity for the snowflake query (Definition~\ref{def:snowflake_query}). We will show that, in the plan space of right deep trees without cross products, we can find the query plan of the minimal cost (under the cost function from Section~\ref{sec:cost_function}) from $n+1$ query plans with bitvector filters if \changed{the bitvector filters have no false positives}, where $n+1$ is the number of relations in the snowflake query. In contrast, the original plan space complexity for snowflake queries in this plan space is exponential to $n$.

We divide the plans into two cases: whether $R_0$ is the right most leaf or not. \changed{We start with the case where $R_0$ is the right most leaf. Then we analyze a subproblem of the plan space for a branch in a snowflake query. We finally analyze the case where $R_0$ is not the right most leaf.}

 \cut{The key insight to reduce the plan space complexity is to derive plans that are equivalent or less in cost and form the final set of candidate plans that can be minimal in cost.}

Formally, \textbf{our key results} in this section are:

\vspace{-.5em}
\begin{theorem}\label{theorem:snowflake:min_cost_plan}
	\textbf{Minimal cost right deep trees for snowflake query: }
	Let $\mathcal{R}$ be the set of relations of a snowflake query $q$ as described in Definition~\ref{def:snowflake_query}. Let $C_{min}=min\{C_{out}(T(X_0, X_1, \cdots, X_n))\}$, where $X_0, X_1, \cdots, \break X_n$ is a permutation of $\mathcal{R}$, and $T(X_1, X_2, \cdots, X_n)$ is a right deep tree without cross products for $q$. Then there exists a right deep tree $\mathcal{T'}\in \{T(R_{i,a_1}, R_{i,a_2}, \cdots,\break R_{i,a_{n_i}},R_0, R_{1,1}, \cdots, R_{1, n_1}, \cdots, R_{i-1,1}, \cdots, R_{i-1,n_{i-1}}, R_{i+1,1}, \break \cdots, R_{i+1,n_{i+1}} \cdots, R_{n,1}, \cdots,  R_{n,n_m})\} \cup \{T(R_0,R_{1,1}, R_{1,2}, \cdots, \break R_{n,1}, \cdots, R_{n,n_m})\}$, where $a_1, a_2, \cdots, a_{n_i}$ is a permutation of $1,2, \cdots, n_i$, such that $C_{out}(\mathcal{T'})=C_{min}$.
\end{theorem}
\vspace{-.5em}

\cut{
Directly follow \changed{Theorem~\ref{theorem:chain_plan_space} and} Theorem~\ref{theorem:snowflake:min_cost_plan}, we have
}
\vspace{-.5em}
\begin{theorem}\label{theorem:snowflake:plan_space}
	\textbf{Plan space complexity for snowflake query: }
	Let $\mathcal{R}$ be the set of $n+1$ relations of a snowflake query $q$ as described in Definition~\ref{def:snowflake_query}. We can find the query plan with the minimal cost in the place space of right deep trees without cross products from $n+1$ candidate plans.
\end{theorem}
\vspace{-.5em}

\cut{
We omit most of the proofs due to space limit, and they can be found in our technical report~\cite{moreproofs}.
}

\subsection{$R_0$ is the right most leaf}

Let's first look at the right deep trees where $R_0$ is the right most leaf. \textbf{Our key insight} is to extend our analysis on star queries and show that all the trees in this plan space have the same $C_{out}$.

We define a class of right deep trees where a relation with a PKFK join condition only appears on the right side of the relations it joins with in a snowflake query. Formally,
\begin{definition}\label{def:snowflake:partially_ordered_tree}
	\textbf{Partially-ordered right deep tree: }
	Let $\mathcal{R}$ be the set of relations of a snowflake query $q$ as described in Definition~\ref{def:snowflake_query}. 
	Let $\mathcal{T}=T(R_0, X_1, \cdots, X_n)$ be a plan for $q$, where $X_1, \cdots, X_n$ is a permutation of $\mathcal{R}-\{R_0\}$. If for any $X_i, 1\leq i\leq n$, either $X_i=R_{p,1}$ or there exists $X_j, 1\leq j <i$ such that $X_j\to X_i$, we call $\mathcal{T}$ a partially-ordered right deep tree.
\end{definition}

\cut{

Similarly, we define a class of right deep trees where a right subtree of $k$ joins is partially ordered:
\begin{definition}\label{def:partiallyorderedrighttreek}

\textbf{$k$-partially-ordered right deep tree}: Let $T_k=T(R_0, U_1, U_2, \cdots, U_k)$, where $U_1, U_2, \cdots, U_n$ is a permutation of $\mathcal{R}-\{R_0\}$. If for any $U_i, 1\leq i\leq k$, either $U_i=R_{p,1}$ or there exists $U_j, 1\leq j <i$ such that $U_i\to U_j$, we call $T$ a $k$-partially ordered right deep tree.
\end{definition}
}

Now we show that the plans in the space of right deep trees without cross products are partially-ordered trees if $R_0$ is the right most leaf. Formally,
\vspace{-.5em}

\begin{lemma}\label{lemma:snowflake:right_deep_tree_nocp}
	\textbf{Right deep tree without cross products for snowflake query: }
	Let $\mathcal{R}$ be the set of relations of a snowflake query $q$ as described in Definition~\ref{def:snowflake_query}. If $\mathcal{T}=T(R_0, X_1, X_2, \cdots, X_n)$ is a right deep tree without cross products for $q$, then $\mathcal{T}$ is a partially-ordered right deep tree. 
\end{lemma}

\vspace{-.5em}
\begin{proof}
If $\mathcal{T}$ is not partially ordered, then there exists $X_i$ such that $X_i\notin \{R_{1,1}, R_{2,1}, \cdots, R_{n,1}\}$ and there does not exist $X_j, i<j\leq n$ such that $X_j\to X_i$. Then $X_i$ does not join with $R_0, X_n, X_{n-1}, \cdots, X_{i+1}$. So there exists a cross product.
\end{proof}

\vspace{-.5em}
Now we show all the partially-ordered right deep trees have the same cost if $R_0$ is the right most leaf. 

Follow Lemma~\ref{lemma:snowflake:right_deep_tree_nocp} and Algorithm~\ref{algorithm:bitvector_pushdown}, we have 
\vspace{-.5em}
\cut{
\begin{lemma}\label{lemma:snowflake:bitvector_pushdown}
	\textbf{Bitvector filters in partially-ordered right deep tree: }
	Let $\mathcal{R}$ be the set of relations of a snowflake query $q$ as described in Definition~\ref{def:snowflake_query}. If $\mathcal{T}=T(R_0, X_1, X_2, \cdots, X_n)$ is a right deep tree without cross products for $q$, then the bitvector filter created from $R_{i,j}$ will be pushed down to $R_{i,j-1}$ if $j>1$ or $R_0$ if $j=1$.
\end{lemma}
}

\begin{restatable}{lemma}{lemmasnowflakebitvectorpushdown}\label{lemma:snowflake:bitvector_pushdown}
	\textbf{Bitvector filters in partially-ordered right deep tree: }
	Let $\mathcal{R}$ be the set of relations of a snowflake query $q$ as described in Definition~\ref{def:snowflake_query}. If $\mathcal{T}=T(R_0, X_1, X_2, \cdots, X_n)$ is a right deep tree without cross products for $q$, then the bitvector filter created from $R_{i,j}$ will be pushed down to $R_{i,j-1}$ if $j>1$ or $R_0$ if $j=1$.
\end{restatable}

\cut{
\vspace{-.5em}
\begin{proof}
Because $\mathcal{T}$ is partially ordered, for every relation $X_k=R_{i,j}, j>1$, there exists one and only one relation $X_p, p<k$ such that $X_k$ connects to $X_p$. Thus, the bitvector filter created from $X_k$ will be pushed down to $X_p$. If $X_k=R_{i,1}$, it only connects to $R_0$. Thus, the bitvector filter created from $X_k$ will be pushed down to $R_0$.
\end{proof}
}

The proof can be found in Appendix~\ref{sec:moreproofs}.

Follow Lemma~\ref{lemma:snowflake:bitvector_pushdown}, we have
\vspace{-.5em}

\cut{
\begin{lemma}\label{lemma:snowflake:equal_cost_right_deep_tree}
	\textbf{Equal cost for partially-ordered right deep tree: }
	Let $\mathcal{R}$ be the set of relations of a snowflake query $q$ as described in Definition~\ref{def:snowflake_query}. Let $\mathcal{T}=T(R_0, X_1, X_2, \cdots, X_n)$ and $\mathcal{T'}=T(R_0, Y_1, Y_2, \cdots, Y_n)$ be two partially ordered right deep trees of $q$. Then $C_{out}(\mathcal{T})=C_{out}(\mathcal{T'})$.
\end{lemma}
}

\begin{restatable}{lemma}{lemmasnowflakequalcostrightdeeptree}\label{lemma:snowflake:equal_cost_right_deep_tree}
	\textbf{Equal cost for partially-ordered right deep tree: }
	Let $\mathcal{R}$ be the set of relations of a snowflake query $q$ as described in Definition~\ref{def:snowflake_query}. Let $\mathcal{T}=T(R_0, X_1, X_2, \cdots, X_n)$ and $\mathcal{T'}=T(R_0, Y_1, Y_2, \cdots, Y_n)$ be two partially ordered right deep trees of $q$. Then $C_{out}(\mathcal{T})=C_{out}(\mathcal{T'})$.
\end{restatable}

The proof can be found in Appendix~\ref{sec:moreproofs}.

\cut{
\changed{The proof leverages Lemma~\ref{lemma:snowflake:bitvector_pushdown} and Lemma~\ref{lemma:star_absorption}, and it can be found in~\cite{moreproofs}.}
}

\cut{
The proof can be found in~\cite{moreproofs}.
}

\cut{
\vspace{-.5em}
\begin{proof}
Consider the bitvector filters created in both $\mathcal{T}$ and $\mathcal{T'}$. BY Lemma~\ref{lemma:snowflake:bitvector_pushdown}, the bitvector filters created from $\mathcal{T}$ and $\mathcal{T'}$ from the same relation $R_{i, j}$ will be pushed down to the same relation $R_{i,j-1}$ if $j>1$ or $R_0$ if $j=1$. 

Since, $S(R_{i,n_i})$ is the same in $\mathcal{T}$ and $\mathcal{T'}$. By induction, we can show that $S(R_{i,j})$ is the same in $\mathcal{T}$ and $\mathcal{T'}$. Since, $S(R_0)=R_0/(S(R_{1,1}), S(R_{2,1}), \cdots, S(R_{n,1}))$, $S(R_0)$ is the same in $\mathcal{T}$ and $\mathcal{T'}$.

Now consider the join cardinality in $\mathcal{T}$ and $\mathcal{T'}$. By Lemma~\ref{lemma:star_absorption}, $S(R_{i,j})=S(R_{i,j}, R_{i,j+1}, \cdots, R_{i, n_i})$. Thus, $S(R_0)=R_0/(S(R_{1,1}), S(R_{2,1}), \cdots, S(R_{n,1}))=S(R_0,R_{1,1}, R_{1,2}, \cdots, R{1,n_1}, R{2,1}, \cdots, R_{m,1}, R_{m,2}, \cdots, R_{m,n_m})$. Thus, $S(R_0, X_1, X_2, \cdots, X_u)=S(R_0), 1\leq u\leq n$ and $S(R_0, Y_1, Y_2, \cdots, Y_v)=S(R_0), 1\leq v\leq n$. Thus, $C_{out}(S(R_0, X_1, \cdots, X_u))=C_{out}(S(R_0, Y_1, \cdots, Y_v)), 1\leq u, v\leq n$.

Thus, $C_{out}(\mathcal{T})=C_{out}(\mathcal{T'})$.

\end{proof}
}

\cut{
	\subsection{Chain subquery}
}
\changed{
	\subsection{Branch of a snowflake query}
	\label{sec:branch}
}
\todo{Consolidate this section}
\changed{
Before diving into the case where $R_0$ is not the right most leaf, we first analyze a subproblem of a branch in a snowflake query in the plan space of right deep trees without cross products. We show that the plan space complexity is linear in the number of relations in the branch. Formally, we define a branch as the following:
}

\point{Chain query}
\changed{
	\begin{definition}\label{def:chain_query}
		\textbf{Branch of a snowflake query: } Let $\mathcal{R}=\{R_0, R_1, \cdots, R_n\}$  be a set of relations and $q$ be a query joining relations in $\mathcal{R}$. The query $q$ is a branch if $R_{k-1} \to R_k$ for all $1\leq k\leq n$.
	\end{definition}
	\vspace{-.5em}
	Figure~\ref{fig:snowflake_query} shows an example of a snowflake query with three branches.
}

\cut{
	We divide the plans in this plan space into two cases: whether $R_n$ is the right most leaf or not. The key insight is to derive that the plans where $R_n, R_{n-1}, \cdots, R_k$ joins consecutively have equal or less cost than other plans based on \associativity, \commutativity and \reduction of the bitvector filters. For example, if the chain query is $R_0 \to R_1 \to R_2 \to R_3$, we try to prove that $C_{out}(R_2, R_3, X, Y)\leq C_{out}(R_2, X, R_3, Y)$.
}

\point{Key results}
We show that, in the plan space of right deep trees without cross products, we can find the query plan with minimal $C_{out}$ from $n+1$ plans with bitvector filters if the bitvector filters have no false positives, where $n+1$ is the number of relations in the query. In contrast, the original plan space complexity for a branch is $n^2$~\cite{ono1990measuring}. 

\point{key intuition of the proof}
\begin{figure}
	\begin{subfigure}[b]{.45\linewidth}
		\centering
		\includegraphics[width=\linewidth]{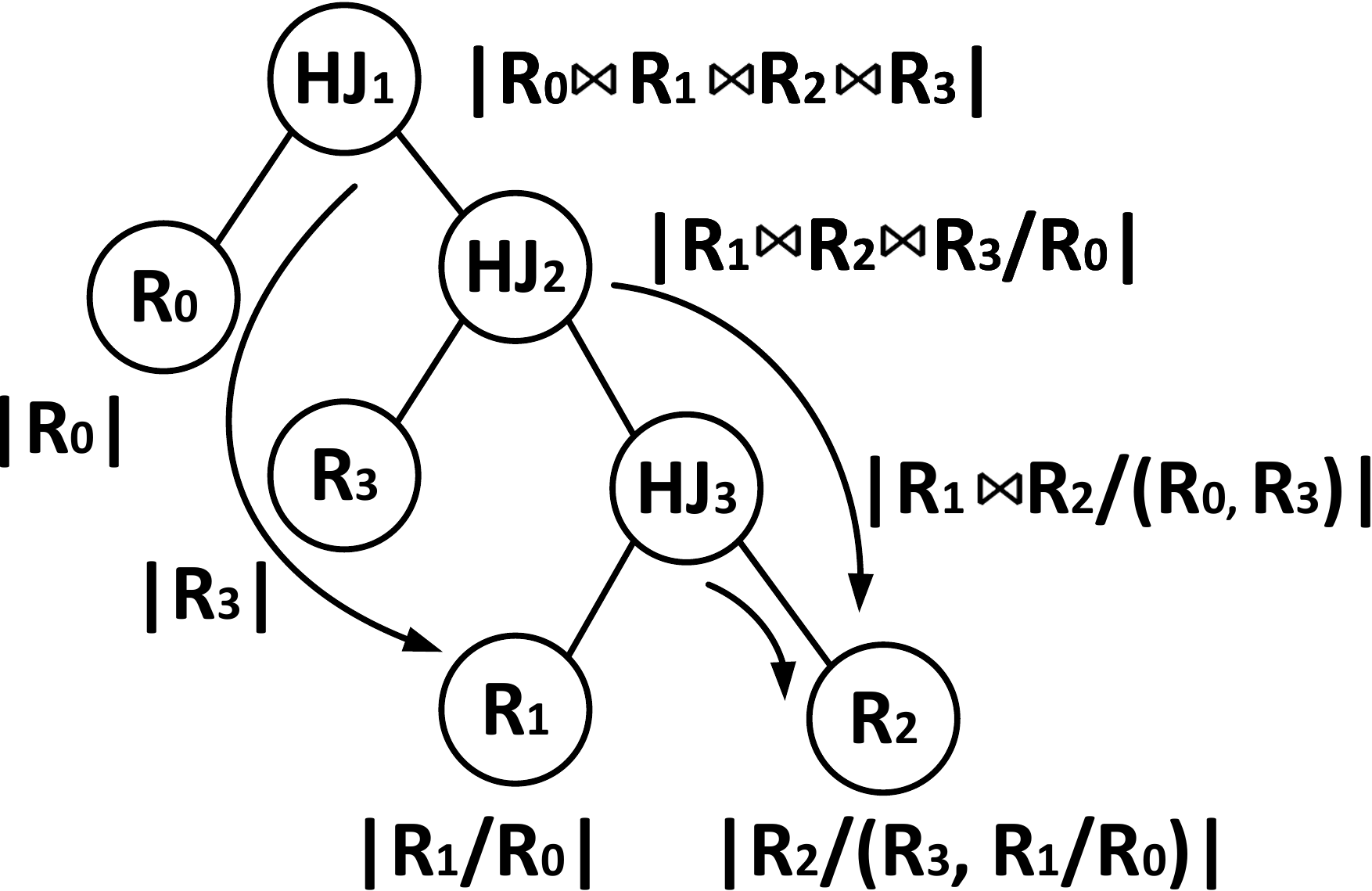}
		\caption{Plan $P_1$}
			\label{fig:proof_chain:plan_1}
	\end{subfigure}
	\qquad
	\begin{subfigure}[b]{.45\linewidth}
		\centering
		\includegraphics[width=\linewidth]{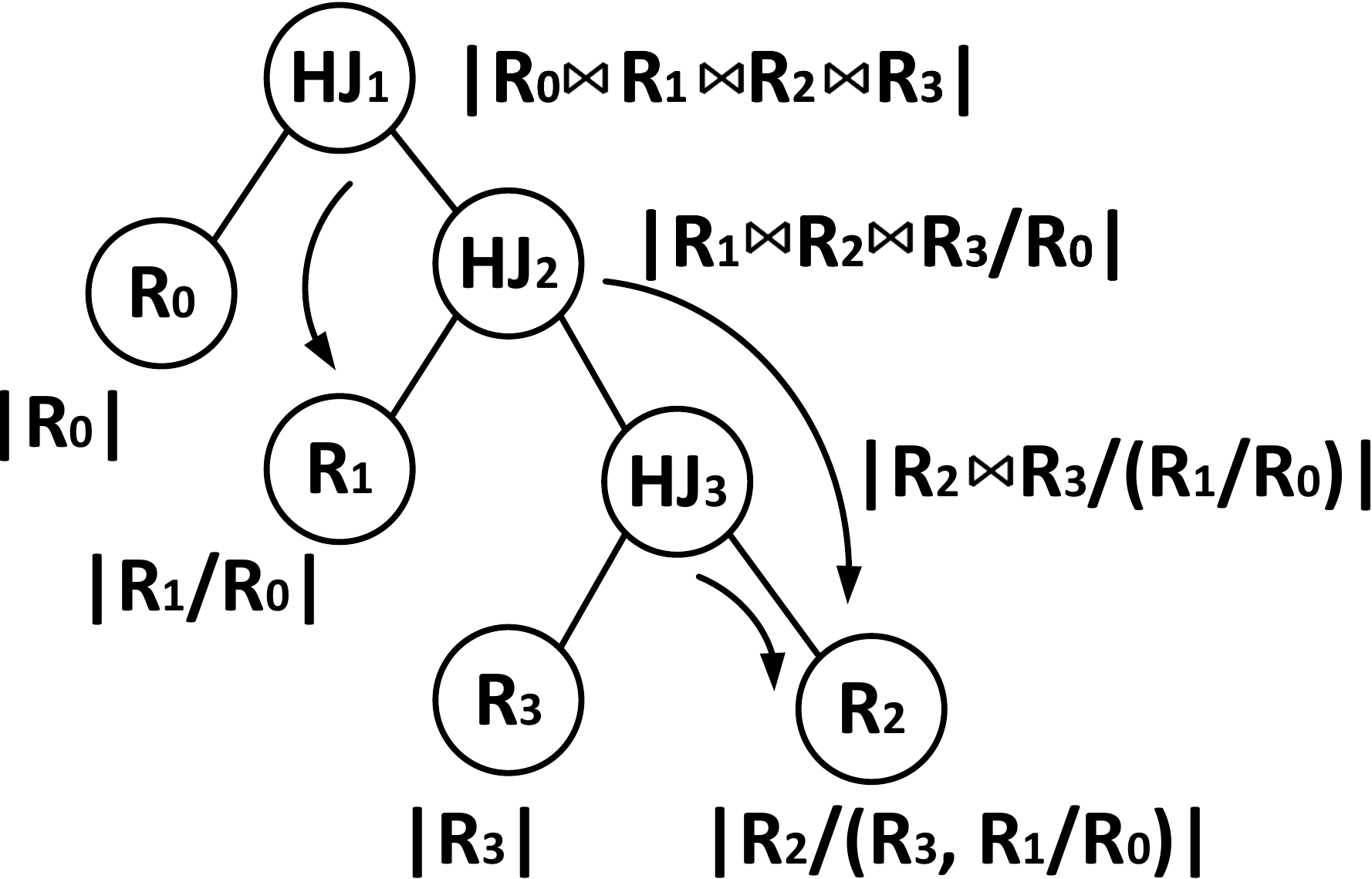}
		\caption{Plan $P_2$}
			\label{fig:proof_chain:plan_2}
	\end{subfigure}
	\caption{Example of two plans for a branch $\{R_0, R_1, R_2, R_3\}$ of a snowflake query with PKFK joins using bitvector filters. Each operator is annotated with the intermediate result size. Plan $P_1$ does not join $R_2$ and $R_3$ consecutively in its right subtree. Pushing down $R_3$ to join with $R_2$ consecutively results in plan $P_2$ with reduced cost.}
	\label{fig:proof:chain}
\end{figure}

\textbf{Our key insight} is that, for a plan with right most leaf $R_k$, where $1\leq k \leq n$, if the plan has minimal cost, it must join $R_k, R_{k+1}, \cdots, R_n$ consecutively in its right subtree. Otherwise, we can reduce the plan cost by altering the join order and 'pushing down' the relations $R_n, R_{n-1}, \cdots, R_{m+1}$ into the right subtree. Figure~\ref{fig:proof:chain} shows an example of how the plan cost can be reduced by 'pushing down' the relations.

Formally, \textbf{our key results} in this subsection are:

\begin{theorem}\label{theorem:chain_mincost_plan}
	\textbf{Minimal cost right deep trees for a branch: }
	Let $\mathcal{R}$ be the set of relations of a \changed{branch} as described in Definition~\ref{def:chain_query}. Let $\mathcal{A} =\{T(X_0, X_1, \cdots, X_n)\}$ be the set of right deep trees without cross products for $q$, where $X_0, X_1, \cdots, X_n$ is a permutation of $R_0, R_1, \cdots, R_n$. If $C_{min}=min\{C_{out}(T(X_0, X_1, \cdots, X_n))\}$, then there exists a plan $\mathcal{T} \in \break \mathcal{A}_{candidates}=\{T(R_n, R_{n-1}, \cdots, R_0)\} \cup \{T(R_k, R_{k+1}, \cdots, \break R_n, R_{k-1}, R_{k-2}, \cdots, R_0), 0\leq k \leq n-1 \}$ such that $C_{out}(\mathcal{T})=C_{min}$.
\end{theorem}
\vspace{-.5em}

\cut{
Directly follow \changed{Lemma~\ref{lemma:chain_tree_nocp} and} Theorem~\ref{theorem:chain_mincost_plan}, we have
}

\vspace{-.5em}
\begin{theorem}\label{theorem:chain_plan_space}
	\textbf{Plan space complexity for a branch: }
	Let $\mathcal{R}$ be the set of $n+1$ relations of a \changed{branch} as described in Definition~\ref{def:chain_query}. We can find the query plan with the minimal cost in the place space of right deep trees without cross products from $n+1$ candidate plans.
\end{theorem}

\changed{Consider the query plan for a branch $\{R_0, R_1, \cdots, R_n\}$ of the snowflake in the plan space of right deep trees without cross products. Let's first look at the query plans where $R_n$ is the right most leaf. Formally,}

\cut{
	\point{Chain subquery}
	We define the chain subquery as the following:
	\vspace{-.5em}
	\begin{definition}\label{def:chain_subquery}
		\textbf{Chain query: } Let $\mathcal{R}=\{R_0, R_2, \cdots, R_n\}$  be a set of relations and $q$ be a query joining relations in $\mathcal{R}$. The query $q$ is a chain query if $R_{k-1} \to R_k$ for all $1\leq k\leq n$.
	\end{definition}
	\vspace{-.5em}
}

\cut{
	\subsubsection{$R_n$ is the right most leaf}
	\point{Plans where $R_n$ is the right most leaf}
}

\cut{
	Let's first look at the query plans where $R_n$ is the right most leaf. Formally,
}
\vspace{-.5em}

\cut{
\begin{lemma}\label{lemma:chain_tree_nocp}
	Let $\mathcal{R}$ be the set of relations of a branch as described in Definition~\ref{def:chain_query}. There exists only one right deep tree without cross products such that $R_n$ is the right most leaf, that is, $T(R_n, R_{n-1}, \cdots, R_0)$.
\end{lemma}
}

\begin{restatable}{lemma}{lemmachaintreenocp}\label{lemma:chain_tree_nocp}
	Let $\mathcal{R}$ be the set of relations of a branch as described in Definition~\ref{def:chain_query}. There exists only one right deep tree without cross products such that $R_n$ is the right most leaf, that is, $T(R_n, R_{n-1}, \cdots, R_0)$.
\end{restatable}

\vspace{-.5em}

\cut{This can be derived from the join graph of a branch.}
The proof can be found in Appendix~\ref{sec:moreproofs}.

\cut{
\vspace{-.5em}

\begin{proof}
	If $R_n$ is the right most leaf and there is no cross product in the query plan, then $R_n$ can only join with $R_{n-1}$. Thus, the right most subplan with two relations is $T_2=T(R_n, R_{n-1})$. Similarly, if the right most subplan is $T_k=T(R_n, R_{n-1}, \cdots, R_{n-k+1})$ and there is no cross product, then $T_k$ can only join with $R_{n-k}$. By induction, $T(R_n, R_{n-1}, \cdots, R_0)$ is the only right deep tree without cross products where $R_n$ is the right most leaf.
\end{proof}
\vspace{-.5em}
}

\cut{
	\subsubsection{$R_n$ is not the right most leaf}
}
\point{Plans where $R_n$ is not the right most leaf. First prove we can reduce the cost by pushing down $R_n, R_{n-1}, \cdots, R_k$}

Now we look at the query plans where $R_n$ is not the right most leaf. Let $T(X_0, \cdots, X_n)$ be a right deep tree without cross products where $X_0, \cdots, X_n$ is a permutation of $R_0, \cdots, R_n$. We show that, without joining $R_n,R_{n-1}, \cdots, R_k$ consecutively, a plan cannot have the minimal cost. Formally,
\vspace{-.5em}

\cut{
\begin{lemma}\label{lemma:chain_pushdown1}
	\textbf{Cost reduction by pushing down $R_n$: }
	Let $\mathcal{R}$ be the set of relations of a \changed{branch} as described in Definition~\ref{def:chain_query}. Let $\mathcal{T}=T(X_0, X_1, \cdots, X_n)$ be a right deep tree without cross products for $R_0, R_1, \cdots, R_n$. Assume $X_k=R_n$ for some $1 \leq k \leq n$. If $X_{k-1}\neq R_{n-1}$, then $\mathcal{T'}=T(X_0, X_1, \cdots, X_k, X_{k-1}, X_{k+1}, X_{k+2}, \cdots, X_n)$ is a right deep tree without cross products and $C_{cout}(\mathcal{T'}) \leq  C_{out}(\mathcal{T})$.
\end{lemma}
}

\begin{restatable}{lemma}{lemmachainpushdownone}\label{lemma:chain_pushdown1}
	\textbf{Cost reduction by pushing down $R_n$: }
	Let $\mathcal{R}$ be the set of relations of a \changed{branch} as described in Definition~\ref{def:chain_query}. Let $\mathcal{T}=T(X_0, X_1, \cdots, X_n)$ be a right deep tree without cross products for $R_0, R_1, \cdots, R_n$. Assume $X_k=R_n$ for some $1 \leq k \leq n$. If $X_{k-1}\neq R_{n-1}$, then $\mathcal{T'}=T(X_0, X_1, \cdots, X_k, X_{k-1}, X_{k+1}, X_{k+2}, \cdots, X_n)$ is a right deep tree without cross products and $C_{cout}(\mathcal{T'}) \leq  C_{out}(\mathcal{T})$.
\end{restatable}

The proof can be found in Appendix~\ref{sec:moreproofs}.

\cut{\changed{The proof can be found in~\cite{moreproofs}.}
	
	We further generalize Lemma~\ref{lemma:chain_pushdown1} as the following:}
\cut{
\changed{See Appendix~\ref{sec:additional_proofs} for the proof. We further generalize Lemma~\ref{lemma:chain_pushdown1} as the following:}
}

\cut{
\vspace{-.5em}
\begin{proof}
	Since there is no cross product in $\mathcal{T}=T(X_0, X_1,\cdots, X_{k-1}, R_n, X_{k+1}, \cdots, X_n)$, one relation in $\mathcal{A}=\{X_0, X_1, \cdots, X_{k-1}\}$ must connect to $R_n$. Since $R_{n-1}$ is the only relation that connects to $R_n$ in the join graph, $R_{n-1}\in \{X_0, X_1, \cdots, X_{k-1}\}$. By induction, we can show that $ R_{n-2}, R_{n-3},\cdots, R_{n-k} \in \mathcal{A}$. Thus, $\mathcal{A}=\{R_{n-k}, R_{n-k+1}, \cdots, R_{n-1}\}$.
	
	If $X_{k-1}\neq R_{n-1}$, then $X_{k-1}=R_{n-k}$; otherwise, the join graph of $X_0, X_1, \cdots, X_{k-2}$ is not connected and the subplan $T(X_0, X_1, \cdots, X_{k-2})$ has cross products.
	
	Now consider the relations $\{X_{k+1}, X_{k+2}, \cdots, X_n\}$. Because $X_{k+1}$ joins with $\{X_0, X_1,\cdots, X_k\}=\{R_n, R_{n-1}, \cdots, R_{n-k}\}$, $X_{k+1}=R_{n-k-1}$. Similarly, we can show that $X_i=R_{n-i}$ for $k<i\leq n$.
	
	If we swap $R_n$ and $R_{n-k}$, we get a new plan $\mathcal{T'}=T(X_0, X_1, \cdots, X_{k-2}, R_n, R_{n-k}, X_{k+1}, \cdots, X_n)$. Because $\{X_0, X_1, \cdots, X_{k-2}\}=\mathcal{A}\setminus \{R_{n-k}\}=\{R_{n-k+1}, R_{n-k+2}, \cdots, R_{n-1}\}$. Thus, $\mathcal{T'}$ has no cross product.
	
	Now we prove $C_{out}(\mathcal{T'})\leq C_{out}(\mathcal{T})$.
	
	\cut{
		First, consider $X_i$ for $k< i\leq n$. Because there is no change from the root to $X_{k+1}$, the bitvector filters created from $X_{k+1}, \cdots, X_n$ and the bitvector filters pushed down to $X_{k+1}, \cdots, X_n$ are the same. Thus, $C_{out}(X_i)$ is the same for $\mathcal{T}$ and $\mathcal{T'}$.
	}
	
	First, consider $X_i$ for $k< i\leq n$. Since there is no change in bitvector filters, it is easy to see that $C_{out}(X_i)$ is the same for $\mathcal{T}$ and $\mathcal{T'}$.
	
	\cut{
		Next, consider $X_i$ for $0\leq i < k$. Since $\mathcal{B}=\{X_0, X_1, \cdots, X_{k-1}\}=\{R_{n-k+1}, \cdots, R_{n-1}\}$, only $R_{n-k}$ and $R_n$ will create bitvector filters that can be pushed down to subplans of $\mathcal{B}$. Because $R_{n-1}\in \mathcal{B}$, no bitvector filter will be pushed down to $R_n$. Thus, the bitvector filter created from $R_n$ is the same for $\mathcal{T}$ and $\mathcal{T'}$, and the same bitvector filter will be pushed down to $R_{n-1}$ the same way in $\mathcal{T}$ and $\mathcal{T'}$. Similarly, the bitvector filters pushed down to and created from $R_{n-k}$ are the same in $\mathcal{T}$ and $\mathcal{T'}$, and the bitvector filters created from $R_{n-k}$ will be pushed down to $R_{n-k+1}$ the same way in $\mathcal{T}$ and $\mathcal{T'}$.
	}
	
	Next, consider $X_i$ for $0\leq i < k$. Since $\mathcal{B}=\{X_0, X_1, \cdots, X_{k-1}\}=\{R_{n-k+1}, \cdots, R_{n-1}\}$, only $R_{n-k}$ and $R_n$ will create bitvector filters that can be pushed down to subplans of $\mathcal{B}$. Because $R_{n-1}\in \mathcal{B}$, no bitvector filter will be pushed down to $R_n$. Thus, the bitvector filter created from $R_n$ is the same for $\mathcal{T}$ and $\mathcal{T'}$, and the same bitvector filter will be pushed down to $R_{n-1}$ the same way in $\mathcal{T}$ and $\mathcal{T'}$. Similarly, the bitvector filters created from and pushed down to $R_{n-k}$ and $R_{n-k+1}$ are the same in $\mathcal{T}$ and $\mathcal{T'}$.
	
	Thus, we have proved $C_{out}(X_i)$ is the same for $0\leq i\leq n$.
	
	Next, we show that the intermediate join sizes in $\mathcal{T'}$ is equal to or smaller than these in $\mathcal{T}$.
	
	Since $\mathcal{T}$ and $\mathcal{T'}$ share the same subplan $\mathcal{T}_{j}=T(X_0, X_1, \cdots, X_j), 0\leq j\leq k-2$, and we have shown the bitvector filters pushed down to $\mathcal{T}_{j}$ is the same in $\mathcal{T}$ and $\mathcal{T'}$, the intermediate join sizes are the same in $\mathcal{T}_{j}$ for both plans.
	
	Consider the cardinalities of the join $S(T_{k-2}, R_{n-k})$ in $\mathcal{T}$ and $S(T_{k-2}, R_n)$ in $\mathcal{T'}$. Since $R_\{n-1\} \in\mathcal{A}$, $S(T_{k-2}, R_n)$ is a PKFK join. By \absorption,  $|S(T_{k-2}, R_n)|=|T_{k-2}|$. Since $R_{n-k-1}=X_{k+1} \notin \mathcal{A}$, $S(T_{k-2}, R_{n-k})$ is not a PKFK join. By \reduction, $|T_{k-2} \Join R_{n-k}|\geq |T_{k-2}|$. Thus, $|T_{k-2} \Join R_n|=|T_{k-2}|\leq |T_{k-2} \Join R_{n-k}|$.
	
	Now consider the cardinalities for $S(T_{k-2}, R_{n-k}, R_n)$ in $\mathcal{T}$ and $S(T_{k-2}, R_n, R_{n-k})$ in $\mathcal{T'}$. Since the set of bitvector filters from $B$ pushed down to $S(T_{k-2}, R_{n-k}, R_n)$ is the same as those pushed down to $S(T_{k-2}, R_n, R_{n-k})$ and the join relations are the same, $|S(T_{k-2}, R_{n-k}, R_n)|=|S(T_{k-2}, R_n, R_{n-k})|$. Similarly, we can show that $|S(T_{k-2}, R_{n-k}, R_n, X_{k+1}, \cdots, X_i)| =|S(T_{k-2}, R_{n}, R_{n-k}, X_{k+1}, \cdots, X_i)|$ for $k+1\leq i\leq n$.
	
	Thus, $C_{out}(\mathcal{T})=\sum_{i=1}^n C_{out}(X_i)+\sum_{i=0, i\neq k-1, k}^n |S(X_0, \cdots, X_i)|+|S(T_{k-2}, R_{n-k})| +|S(T_{k-2}, R_{n-k}, R_n)| \geq \sum_{i=1}^n C_{out}(X_i)+\sum_{i=0, i\neq k-1, k}^n |S(X_0, \cdots, X_i)|+|S(T_{k-2}, R_n)| +|S(T_{k-2}, R_n, R_{n-k})|=C_{out}(\mathcal{T'})$.
	
\end{proof}
\vspace{-.5em}
}

\cut{
\begin{lemma}\label{lemma:chain_pushdown2}
	\textbf{Cost reduction by pushing down $R_n, R_{n-1}, \cdots, R_{n-m}$: }
	Let $\mathcal{R}$ be the set of relations of a \changed{branch} as described in Definition~\ref{def:chain_query}. Let $\mathcal{T}=T(X_0, X_1, \cdots, X_n)$ be a right deep tree without cross products for $R_0, R_1, \cdots, R_n$. Let $X_k=R_n, X_{k-1}=R_{n-1}, \cdots, X_{k-m}=R_{n-m}$ for some $m\leq k\leq n$. If $X_{k-m-1}\neq R_{n-m-1}$, then $\mathcal{T'} =T(X_0, X_1, \cdots, X_{k-m-2}, X_{k-m}, X_{k-m+1}, \cdots, X_k, X_{k-m-1}, \break X_{k+1},\cdots, X_n)$ is a right deep tree without cross products and $C_{cout}(\mathcal{T'}) \leq C_{out}(\mathcal{T})$.
\end{lemma}
}

\begin{restatable}{lemma}{lemmachainpushdowntwo}\label{lemma:chain_pushdown2}
	\textbf{Cost reduction by pushing down $R_n, R_{n-1}, \cdots, R_{n-m}$: }
	Let $\mathcal{R}$ be the set of relations of a \changed{branch} as described in Definition~\ref{def:chain_query}. Let $\mathcal{T}=T(X_0, X_1, \cdots, X_n)$ be a right deep tree without cross products for $R_0, R_1, \cdots, R_n$. Let $X_k=R_n, X_{k-1}=R_{n-1}, \cdots, X_{k-m}=R_{n-m}$ for some $m\leq k\leq n$. If $X_{k-m-1}\neq R_{n-m-1}$, then $\mathcal{T'} =T(X_0, X_1, \cdots, X_{k-m-2}, X_{k-m}, X_{k-m+1}, \cdots, X_k, X_{k-m-1}, \break X_{k+1},\cdots, X_n)$ is a right deep tree without cross products and $C_{cout}(\mathcal{T'}) \leq C_{out}(\mathcal{T})$.
\end{restatable}

\vspace{-.5em}

The proof can be found in Appendix~\ref{sec:moreproofs}.

\cut{
	\begin{proof}
		Similar to the proof of Lemma~\ref{lemma:chain_pushdown1}, we can show that $X_{k-m-1}=R_{n-k}$ if $X_{k-m-1}\neq R_{n-m-1}$, $\mathcal{A}=\{X_{0}, X_{1}, \cdots, X_{k-m-1}\}=\{R_{n-k}, R_{n-k+1}, \cdots, R_{n-m-1}\}$, and $X_i=R_{n-i}$ for $k<i\leq n$.
		
		Now consider swapping $R_n, R_{n-1}, \cdots, R_{n-m}$ with $R_{n-k}$, the resulting plan is $\mathcal{T'}$. Similar to the proof of Lemma~\ref{lemma:chain_pushdown1}, we can show that $\mathcal{T'}$ has no cross product.
		
		Consider $C_{out}$ for $X_0, X_1, \cdots, X_n$. Similar to the proof of Lemma~\ref{lemma:chain_pushdown1}, we can show that $C_{out}(X_i)$ is the same for $X_0, X_1, \cdots, X_n$ in $\mathcal{T}$ and $\mathcal{T'}$.
		
		Next, consider the intermediate join sizes. Since both $\mathcal{T}$ and $\mathcal{T'}$ share the same subplan $\mathcal{T}_{j}(X_0, X_1, \cdots, X_j), 0\leq j\leq k-m-2$, similar to the proof of  Lemma~\ref{lemma:chain_pushdown1}, we can show that $\mathcal{T}_{j}, 0\leq j\leq k-m-2$ is the same for $\mathcal{T}$ and $\mathcal{T'}$.
		
		Now consider the cardinality of joins $S(\mathcal{T}_{k-m-2}, R_{k-m-1})$ and $S(\mathcal{T}_{k-m-2}, R_{n-k})$, similar to Lemma~\ref{lemma:chain_pushdown1}, we can show $|S(\mathcal{T}_{k-m-2}, R_{k-m-1})| \leq S(\mathcal{T}_{k-m-2}, R_{n-k})$.
		
		Now consider the cardinality of joins $S(\mathcal{T}_{k-m-2}, R_{k-m-1}, R_{k-m})$ and $S(\mathcal{T}_{k-m-2}, R_{n-k}, R_{k-m-1})$. Since $R_{k-m-2}$ is a PKFK join with $S(\mathcal{T}_{k-m-2}, R_{k-m-1})$, $|S(\mathcal{T}_{k-m-2}, R_{k-m-1}, R_{k-m})|=S(\mathcal{T}_{k-m-2}, R_{k-m-1}$. Similarly, since $R_{k-m-1}$ is a PKFK join with $S(\mathcal{T}_{k-m-2}, R_{n-k}, R_{k-m-1})$, we have $|S(\mathcal{T}_{k-m-2}, R_{n-k}, R_{k-m-1})|=|S(\mathcal{T}_{k-m-2}, R_{n-k})|$. Thus, $|S(\mathcal{T}_{k-m-2}, R_{k-m-1}, R_{k-m})|\leq S(\mathcal{T}_{k-m-2}, R_{n-k}, R_{k-m-1})$.
		
		By similar reasoning, we can show that $|S(\mathcal{T}_{k-m-2}, R_{k-m-1}, R_{k-m}, \cdots, R_j)|\leq |S(\mathcal{T}_{k-m-2}, R_{n-k}, R_{k-m-1}, R_{k-m}, \cdots, R_{j-1})|, k-m-1\leq j\leq n$.
		
		Finally, we can show that $|S(\mathcal{T}_{k-m-2}, R_{k-m-1}, R_{k-m}, \cdots, R_n, R_{n-k})|=|S(\mathcal{T}_{k-m-2}, R_{n-k}, R_{k-m-1}, R_{k-m}, \cdots, R_n)|$ and $|S(\mathcal{T}_{k-m-2}, R_{k-m-1}, R_{k-m}, \cdots, R_n, R_{n-k}, R_{n-k-1}, \cdots, R_j)|=|S(\mathcal{T}_{k-m-2}, R_{n-k}, R_{k-m-1}, R_{k-m}, \cdots, R_n, R_{n-k-1}, \cdots, R_j)|, 0\leq j\leq n-k-1$ as in Lemma~\ref{lemma:chain_pushdown1}.
		
		By summing up everything together, we have $C_{out}(\mathcal{T})\geq C_{out}(\mathcal{T'})$.
		
	\end{proof}
}

\cut{The proof is similar to Lemma~\ref{lemma:chain_pushdown1}, and it can be found in~\cite{moreproofs}.}
\cut{
\changed{The proof can be found in~\cite{moreproofs}.}
}

\cut{
	\subsubsection{Plan space complexity}
}

\cut{
The proofs can be found in our technical report~\cite{moreproofs}.
}

\point{Show what candidate plans of min cost look like and the number of candidate plans}

By combining Lemma~\ref{lemma:chain_tree_nocp} and Lemma~\ref{lemma:chain_pushdown2}, we can prove Theorem~\ref{theorem:chain_mincost_plan} and Theorem~\ref{theorem:chain_plan_space}.

\cut{
Combine Lemma~\ref{lemma:chain_tree_nocp} and Lemma~\ref{lemma:chain_pushdown2}, we have
\vspace{-.5em}

\begin{theorem}\label{theorem:chain_mincost_plan}
	\textbf{Minimal cost right deep trees for a branch: }
	Let $\mathcal{R}$ be the set of relations of a \changed{branch} as described in Definition~\ref{def:chain_query}. Let $\mathcal{A} =\{T(X_0, X_1, \cdots, X_n)\}$ be the set of right deep trees without cross products for $q$, where $X_0, X_1, \cdots, X_n$ is a permutation of $R_0, R_1, \cdots, R_n$. If $C_{min}=min\{C_{out}(T(X_0, X_1, \cdots, X_n))\}$, then there exists a plan $\mathcal{T} \in \mathcal{A}_{candidates}=\{T(R_n, R_{n-1}, \cdots, R_0)\} \cup \{T(R_k, R_{k+1}, \cdots, R_n, R_{k-1}, R_{k-2}, \cdots, R_0), 0\leq k \leq n-1 \}$ such that $C_{out}(\mathcal{T})=C_{min}$.
\end{theorem}
\vspace{-.5em}

Directly follow \changed{Lemma~\ref{lemma:chain_tree_nocp} and} Theorem~\ref{theorem:chain_mincost_plan}, we have

\vspace{-.5em}
\begin{theorem}\label{theorem:chain_plan_space}
	\textbf{Plan space complexity for a branch: }
	Let $\mathcal{R}$ be the set of $n+1$ relations of a \changed{branch} as described in Definition~\ref{def:chain_query}. We can find the query plan with the minimal cost in the place space of right deep trees without cross products from $n+1$ candidate plans.
\end{theorem}
}

\cut{
	\vspace{-.5em}
	\begin{proof}
		By Theorem~\ref{theorem:chain_mincost_plan}, there exists a plan in $\mathcal{A}_{candidates}=\{T(R_n, R_{n-1}, \cdots, R_1)\} \cup \{T(R_1, R_2, \cdots, R_k, R_n, R_{n-1}, \cdots, R_{k+1}),$ $1\leq k \leq n-1 \}$ such that the cost of the plan is minimal, we can find the plan with the minimal cost from $n$ candidate query plans.
	\end{proof}
	\vspace{-.5em}
}

\cut{
	\vspace{-.5em}
	\begin{proof}
		By Theorem~\ref{theorem:chain_mincost_plan}, there exists a plan in $\mathcal{A}=\{T(R_n, \cdots, R_1)\} \cup \{T(R_1, \cdots, R_k, R_n, R_{n-1}, \cdots, R_{k+1}),$ $1\leq k \leq n-1 \}$ such that the cost of the plan is minimal, we can find the plan with the minimal cost from $n$ candidate query plans.
	\end{proof}
	\vspace{-.5em}
}

\cut{
	\subsection{Right deep tree with cross products}
	\todo{Prove right deep tree with cross products is suboptimal}
	\begin{theorem}\label{theorem:suboptchainwithcp}
		If a right deep tree plan $p$ contains a cross product for a branch $q$, then there exists a right deep tree plan $p'$, such that $cost(p')\leq cost(p)$.
	\end{theorem}
	
	\todo{Seems this is not true???}
	
	In other words, we don't need to consider right deep tree query plans with cross products for chain queries.
}

\vspace{-.5em}
\subsection{$R_0$ is not the right most leaf}
\outline{The case where $R_0$ is not the right most leaf}

Now let's look at the right deep trees where $R_0$ is not the right most leaf for a snowflake query with PKFK joins.

We first show that the relations appear on the left side of $R_0$ can only come from a single branch given the join graph of a snowflake query. Formally,

\vspace{-.5em}

\cut{
\begin{lemma}\label{lemma:snowflake:single_branch}
	\textbf{Single branch in right most leaves: }Let $\mathcal{R}$ be the set of relations of a snowflake query $q$ as described in Definition~\ref{def:snowflake_query}. Let $\mathcal{T}=T(X_0, X_1, \cdots, X_n)$ be a right deep tree without cross products for $q$, where $X_0, X_1, \cdots, X_n$ is a permutation of $\mathcal{R}$. If $X_k=R_0$, then $X_{0}, X_{1}, \cdots, X_{k-1}$ is a permutation of $R_{i, 1}, R_{i, 2}, \cdots, R_{i, k}$ for some $1\leq i\leq m$.
\end{lemma}
}

\begin{restatable}{lemma}{lemmasnowflakesinglebranch}\label{lemma:snowflake:single_branch}
	\textbf{Single branch in right most leaves: }Let $\mathcal{R}$ be the set of relations of a snowflake query $q$ as described in Definition~\ref{def:snowflake_query}. Let $\mathcal{T}=T(X_0, X_1, \cdots, X_n)$ be a right deep tree without cross products for $q$, where $X_0, X_1, \cdots, X_n$ is a permutation of $\mathcal{R}$. If $X_k=R_0$, then $X_{0}, X_{1}, \cdots, X_{k-1}$ is a permutation of $R_{i, 1}, R_{i, 2}, \cdots, R_{i, k}$ for some $1\leq i\leq m$.
\end{restatable}

The proof can be found in Appendix~\ref{sec:moreproofs}.

\cut{
\vspace{-.5em}
\begin{proof}
	Assume there exists $X_u=R_{i_1,j_1}$ and $X_v=R_{i_2,j_2}$ such that $0\leq u, v\leq k-1$ and $i_1\neq i_2$. Because $X_k=R_0$, $X_u$ does not connect to $X_v$ by joining with $X_0, X_1, \cdots, X_{k-1}$. Thus, there must be a cross product. This contradicts that $\mathcal{T}$ has no cross product. So $\{X_0, X_1, \cdots, X_{k-1}\}\subseteq \mathcal{R_i}$ for some $1\leq i\leq m$.
	
	Since $X_0, X_1, \cdots, X_{k-1}$ has a join condition with $R_0$, $R_{i,1}\in \{X_0, X_1, \cdots, X_{k-1}\}$. Because $T(X_0, X_1, \cdots, X_{k-1})$ has no cross product, $\{X_0, X_1, \cdots, X_{k-1}\}=\{R_{i,1}, R_{i,2}, \cdots, R_{i,k}\}$. Thus, $X_{0}, X_{1}, \cdots, X_{k-1}$ is a permutation of $R_{i, 1}, R_{i, 2}, \cdots, R_{i, k}$. 
\end{proof}
\vspace{-.5em}
}

\cut{
\vspace{-.5em}
\begin{proof}
	Assume there exists $X_u=R_{i_1,j_1}$ and $X_v=R_{i_2,j_2}$ such that $0\leq u, v\leq k-1$ and $i_1\neq i_2$. Because $X_k=R_0$, $X_u$ does not connect to $X_v$ by joining with $X_0, X_1, \cdots, X_{k-1}$. Thus, there must be a cross product, \changed{which is a contradiction.}\cut{ This contradicts that $\mathcal{T}$ has no cross product. So $\{X_0, X_1, \cdots, X_{k-1}\}\subseteq \mathcal{R_i}$ for some $1\leq i\leq m$.}
	
	Since $X_0, X_1, \cdots, X_{k-1}$ has a join condition with $R_0$, $R_{i,1}\in \{X_0, X_1, \cdots, X_{k-1}\}$. Because $T(X_0, X_1, \cdots, X_{k-1})$ has no cross product, $\{X_0, X_1, \cdots, X_{k-1}\}=\{R_{i,1}, R_{i,2}, \cdots, R_{i,k}\}$. Thus, $X_{0}, X_{1}, \cdots, X_{k-1}$ is a permutation of $R_{i, 1}, R_{i, 2}, \cdots, R_{i, k}$. 
\end{proof}
\vspace{-.5em}
}

Now we show that the relations on the left side of $R_0$ are partially ordered. Formally,

\vspace{-.5em}

\cut{
\begin{lemma}\label{lemma:snowflake:partially_ordered_tree2}
	\textbf{Partially-ordered subtree: }Let $\mathcal{R}$ be the set of relations of a snowflake query $q$ as described in Definition~\ref{def:snowflake_query}. Let $\mathcal{T}=T(X_0, X_1, \cdots, X_n)$ be a right deep tree without cross products for $q$, where $X_0, X_1, \cdots, X_n$ is a permutation of $\mathcal{R}$. If $X_k=R_0$, then $X_{k+1}, X_{k+2}, \cdots, X_{n}$ is a partially ordered right deep tree of the new relation $R_0'=X_0 \Join X_1 \Join \cdots \Join X_k$.
\end{lemma}
}

\begin{restatable}{lemma}{lemmasnowflakepartiallyorderedtreetwo}\label{lemma:snowflake:partially_ordered_tree2}
	\textbf{Partially-ordered subtree: }Let $\mathcal{R}$ be the set of relations of a snowflake query $q$ as described in Definition~\ref{def:snowflake_query}. Let $\mathcal{T}=T(X_0, X_1, \cdots, X_n)$ be a right deep tree without cross products for $q$, where $X_0, X_1, \cdots, X_n$ is a permutation of $\mathcal{R}$. If $X_k=R_0$, then $X_{k+1}, X_{k+2}, \cdots, X_{n}$ is a partially ordered right deep tree of the new relation $R_0'=X_0 \Join X_1 \Join \cdots \Join X_k$.
\end{restatable}

\vspace{-.5em}

\cut{
\changed{The proof leverages Lemma~\ref{lemma:snowflake:single_branch} and Theorem~\ref{lemma:snowflake:right_deep_tree_nocp}, and it can be found in~\cite{moreproofs}.}
}

\cut{
The proof can be found in~\cite{moreproofs}.
}

The proof can be found in Appendix~\ref{sec:moreproofs}.

\cut{
\begin{proof}
\cut{We first show that the join graph of $\{R_0', X_{k+1}, X_{k+2}, \cdots, X_n\}$ is a snowflake as defined in Definition~\ref{def:snowflake_query}.
}
By Lemma~\ref{lemma:snowflake:single_branch}, $X_{0}, X_{1}, \cdots, X_{k-1}$ is a permutation of $R_{i, 1}, R_{i, 2}, \cdots, R_{i, k}$ for some $1\leq i\leq m$. Let's create a new relation $R_0'=Join(X_0, X_1, \cdots, X_{k-1}, R_0)$. For $X_j, k<j\leq n$, if $X_j=R_{i,k+1}$, $R_{i,k}\to X_j$ and thus $R_0' \to X_j$; if $X_j=R_{u, 1}$, $R_0 \to X_j$ and thus $X_j\to R_0'$; if $X_j=R_{u, v}, v>1$, then there exists $R_{u, v-1}\in \{X_{k+1}, X_{k+2}, \cdots, X_n\}$ such that $R_{u,v-1}\to X_j$. Thus, $\{R_0', X_{k+1}, X_{k+2}, \cdots, X_n\}$ is a snowflake query. By Lemma~\ref{lemma:snowflake:right_deep_tree_nocp}, and $X_{k+1}, X_{k+2}, \cdots, X_n$ is a partially ordered right deep tree of the new snowflake query.
\end{proof}
\vspace{-.5em}
}

Now we show that if a subset of relations of a single branch $\mathcal{R}_i$ is on the right side of $R_0$, there exists a query plan with lower cost where all the relations in $\mathcal{R}_i$ are on the right side of $R_0$. Formally,

\vspace{-.5em}

\cut{
\begin{lemma}\label{lemma:snowflake:cost_reduction_push_down}
		\textbf{Cost reduction by consolidating a single branch: } 
	Let $\mathcal{R}$ be the set of relations of a snowflake query $q$ as described in Definition~\ref{def:snowflake_query}. Let $\mathcal{T}=T(X_0, X_1, \cdots, X_{k-1}, R_0, X_{k+1}, \cdots, X_n)$ be a right deep tree without cross products for $q$, where $X_0, X_1, \cdots, X_{k-1}$ is a permutation of $R_{i,1}, R_{i,2}, \cdots, R_{i, k}$ for some $1\leq i\leq m, 1\leq k\leq n_i-1$. Then there exists a right deep tree without cross products $\mathcal{T'}=T(X_0, X_1, \cdots, X_{k-1}, R_{i,k+1}, R_{i,k+2}, \cdots, R_{i, n_i}, R_0, Y_1, Y_2, \cdots, \break Y_{n-n_i-1})$ for $q$ such that $C_{out}(\mathcal{T'})\leq C_{out}(\mathcal{T})$.
\end{lemma}
}

\begin{restatable}{lemma}{lemmasnowflakecostreductionpushdown}\label{lemma:snowflake:cost_reduction_push_down}
	\textbf{Cost reduction by consolidating a single branch: } 
	Let $\mathcal{R}$ be the set of relations of a snowflake query $q$ as described in Definition~\ref{def:snowflake_query}. Let $\mathcal{T}=T(X_0, X_1, \cdots, X_{k-1}, R_0, X_{k+1}, \cdots, X_n)$ be a right deep tree without cross products for $q$, where $X_0, X_1, \cdots, X_{k-1}$ is a permutation of $R_{i,1}, R_{i,2}, \cdots, R_{i, k}$ for some $1\leq i\leq m, 1\leq k\leq n_i-1$. Then there exists a right deep tree without cross products $\mathcal{T'}=T(X_0, X_1, \cdots, X_{k-1}, R_{i,k+1}, R_{i,k+2}, \cdots, R_{i, n_i}, R_0, Y_1, Y_2, \cdots, \break Y_{n-n_i-1})$ for $q$ such that $C_{out}(\mathcal{T'})\leq C_{out}(\mathcal{T})$.
\end{restatable}

\vspace{-.5em}

The proof can be found in Appendix~\ref{sec:moreproofs}.

\cut{
\begin{proof}
	By Lemma~\ref{lemma:snowflake:partially_ordered_tree2}, $\mathcal{T}$ is a partially-ordered subtree. Let $\mathcal{T}_p=T(X_0, X_1, \cdots, X_{k-1}, R_0, R_{i,k+1}, R_{i,k+2}, \cdots, R_{i,n_i}, Y_1, Y_2, \break Y_{n-n_i-1})$, where $Y_1, Y_2, \cdots, Y_{n-n_i-1}$ is a permutation of $\mathcal{A}=\{X_{k+1}, X_{k+2}, \cdots, X_n\}\setminus \{R_{i,k+1}, R_{i,k+2},\cdots, R_{i,n_i}\}$, and $Y_1, Y_2, \cdots, Y_{n-n_i-1}$ is partially ordered. By Theorem~\ref{lemma:snowflake:equal_cost_right_deep_tree}, $C_{out}(\mathcal{T}_p)=C_{out}(\mathcal{T})$.
	
	Now consider $\mathcal{T'}= T(X_0, X_1, \cdots,  X_{k-1}, R_{i,k+1}, R_{i,k+2},\break \cdots, R_{i,n_i}, R_0, Y_1, Y_2, Y_{n-n_i-1})$. Let $R_0'=R_0/(Y_1, Y_2, \cdots, Y_{n-n_i-1})$. Since $X_0, X_1, \cdots, X_{k-1}$ is a permutation of $\{R_{i,1}, R_{i,2}, \cdots, R_{i,k}\}$, joining $\{X_0, X_1, \cdots, X_{k-1},\break  R_{i,k+1}, R_{i,k+2}, \cdots, R_{i,n_i},  R_0'\}$ is a branch. By Lemma~\ref{lemma:chain_pushdown2}, $C_{cout}(T(X_0, X_1, \cdots, X_{k-1}, R_{i,k+1}, R_{i,k+2},\cdots, R_{i,n_i}, R_0'))\leq C_{cout}(T(X_0, X_1, \cdots, X_{k-1}, R_0', R_{i,k+1}, R_{i,k+2}, \cdots, R_{i,n_i}))$.
	
	Consider $\mathcal{T'}$ and $\mathcal{T}_p$. Because $\mathcal{T}_p$ is a partially-ordered subtree, $C_{out}(\mathcal{T}_p)=C_{out}(T(X_0, X_1, \cdots, X_{k-1}, R_0', R_{i,k+1}, R_{i,k+2}),\cdots, R_{i,n_i})+\sum_{j=1}^{n-n_i-1}C_{out}(Y_j)+(n-n_i-1)\cdot |S(R_0, R_{i,1}, R_{i,2}, \cdots, R_{i, n_i}, Y_1, Y_2, \cdots, Y_{n-n_i-1})|$. Thus, $C_{out}(\mathcal{T}_p)\geq C_{out}(T(X_1, X_2, \cdots, X_{k-1}, R_{i,k+1}, R_{i,k+2}, ,\cdots,\break R_{i,n_i}, R_0'))+(n-n_i-1)\cdot |S(R_0, R_{i,1}, R_{i,2}, \cdots, R_{i, n_i}, Y_1, Y_2, \cdots, \break Y_{n-n_i-1})|=C_{out}(\mathcal{T'})$.
	
	Thus, $C_{out}(\mathcal{T'})\leq C_{out}(\mathcal{T}_p)=C_{out}(\mathcal{T})$.
\end{proof}
\vspace{-.5em}
}

\cut{
\changed{The proof leverages Lemma~\ref{lemma:snowflake:partially_ordered_tree2} and Lemma~\ref{lemma:chain_pushdown2}, and it can be found in~\cite{moreproofs}.}
}

By combining Lemma~\ref{lemma:snowflake:equal_cost_right_deep_tree} and Lemma~\ref{lemma:snowflake:cost_reduction_push_down}, we can prove Theorem~\ref{theorem:snowflake:min_cost_plan}, and Theorem~\ref{theorem:snowflake:plan_space} directly follows from Theorem~\ref{theorem:chain_plan_space} and Theorem~\ref{theorem:snowflake:min_cost_plan}.

\cut{
\begin{proof}
	By Lemma~\ref{lemma:snowflake:partially_ordered_tree2}, $\mathcal{T}$ is a partially-ordered subtree. Let $\mathcal{T}_p=T(X_0, X_1, \cdots, X_{k-1}, R_0, R_{i,k+1}, R_{i,k+2}, \cdots, R_{i,n_i}, Y_1, Y_2, \break\cdots, Y_{n-n_i-1})$, where $Y_1, Y_2,\cdots, Y_{n-n_i-1}$ is a permutation of $\mathcal{A}=\{X_{k+1}, X_{k+2}, \cdots, X_n\}\setminus \{R_{i,k+1}, R_{i,k+2},\cdots, R_{i,n_i}\}$, and $Y_1, Y_2, \cdots, Y_{n-n_i-1}$ is partially ordered. By Theorem~\ref{lemma:snowflake:equal_cost_right_deep_tree}, $C_{out}(\mathcal{T}_p)=C_{out}(\mathcal{T})$.
	
	Now consider $\mathcal{T'}= T(X_0, X_1, \cdots,  X_{k-1}, R_{i,k+1}, R_{i,k+2},\break \cdots, R_{i,n_i}, R_0, Y_1, Y_2, Y_{n-n_i-1})$. Let $R_0'=R_0/(Y_1, Y_2, \cdots, \break Y_{n-n_i-1})$. Since $X_0, X_1, \cdots, X_{k-1}$ is a permutation of $\{R_{i,1}, R_{i,2},\cdots, R_{i,k}\}$, joining $\{X_0, X_1, \cdots, X_{k-1}, R_{i,k+1},\break  R_{i,k+2}, \cdots, R_{i,n_i},  R_0'\}$ is a branch of a snowflake. By Lemma~\ref{lemma:chain_pushdown2}, $C_{cout}(T(X_0, X_1, \cdots, X_{k-1}, R_{i,k+1}, R_{i,k+2},\cdots, R_{i,n_i}, R_0'))\leq C_{cout}(T(X_0, X_1, \cdots, X_{k-1}, R_0', R_{i,k+1}, R_{i,k+2}, \cdots, R_{i,n_i}))$.
	
	Consider $\mathcal{T'}$ and $\mathcal{T}_p$. Because $\mathcal{T}_p$ is a partially-ordered subtree, $C_{out}(\mathcal{T}_p)=C_{out}(T(X_0, X_1, \cdots, X_{k-1}, \break R_0', R_{i,k+1}, R_{i,k+2}),\cdots, R_{i,n_i})+\sum_{j=1}^{n-n_i-1}C_{out}(Y_j)+(n-n_i-1)\cdot |S(R_0, R_{i,1}, R_{i,2}, \cdots, R_{i, n_i}, Y_1, Y_2, \cdots, Y_{n-n_i-1})|$. Thus, $C_{out}(\mathcal{T}_p)\geq C_{out}(T(X_1, X_2, \cdots, X_{k-1}, R_{i,k+1}, R_{i,k+2} ,\cdots,\break R_{i,n_i}, R_0'))+(n-n_i-1)\cdot |S(R_0, R_{i,1}, R_{i,2}, \cdots, R_{i, n_i}, Y_1, Y_2, \cdots, \break Y_{n-n_i-1})|=C_{out}(\mathcal{T'})$.
	
	Thus, $C_{out}(\mathcal{T'})\leq C_{out}(\mathcal{T}_p)=C_{out}(\mathcal{T})$.
\end{proof}
\vspace{-.5em}
}

\cut{
\subsection{Plan space complexity}

\outline{Summarize the two cases for min cost plans and the number of plans to consider}
Combine Lemma~\ref{lemma:snowflake:equal_cost_right_deep_tree} and Lemma~\ref{lemma:snowflake:cost_reduction_push_down}, we have 

\vspace{-.5em}
\begin{theorem}\label{theorem:snowflake:min_cost_plan}
	\textbf{Minimal cost right deep trees for snowflake query: }
Let $\mathcal{R}$ be the set of relations of a snowflake query $q$ as described in Definition~\ref{def:snowflake_query}. Let $C_{min}=min\{C_{out}(T(X_0, X_1, \cdots, X_n))\}$, where $X_0, X_1, \cdots, X_n$ is a permutation of $\mathcal{R}$, and $T(X_1, X_2, \cdots, X_n)$ is a right deep tree without cross products for $q$. Then there exists a right deep tree $\mathcal{T'}\in \{T(R_{i,a_1}, R_{i,a_2}, \cdots, R_{i,a_{n_i}},R_0, R_{1,1}, \cdots, R_{1, n_1}, \break \cdots, R_{i-1,1}, \cdots, R_{i-1,n_{i-1}}, R_{i+1,1}, \cdots, R_{i+1,n_{i+1}} \cdots, R_{n,1}, \break \cdots, R_{n,n_m})\} \cup \{T(R_0,R_{1,1}, R_{1,2}, \cdots, R_{n,1}, \cdots, R_{n,n_m})\}$, where $a_1, a_2, \cdots, a_{n_i}$ is a permutation of $1,2, \cdots, n_i$, such that $C_{out}{T'}=C_{min}$.
\end{theorem}
\vspace{-.5em}

Directly follow \changed{Theorem~\ref{theorem:chain_plan_space} and} Theorem~\ref{theorem:snowflake:min_cost_plan}, we have

\vspace{-.5em}
\begin{theorem}\label{theorem:snowflake:plan_space}
	\textbf{Plan space complexity for snowflake query: }
	Let $\mathcal{R}$ be the set of $n+1$ relations of a snowflake query $q$ as described in Definition~\ref{def:snowflake_query}. We can find the query plan with the minimal cost in the place space of right deep trees without cross products from $n+1$ candidate plans.
\end{theorem}
\vspace{-.5em}
}

\cut{
\begin{proof}
By Theorem~\ref{theorem:snowflake:min_cost_plan}, we can find the plan with minimal $C_{out}$ from $\{T(R_{i,a_1}, \cdots, R_{i,a_{n_1}},R_0, R_{1, n_1}, \cdots, R_{i-1,1}, \cdots, R_{i-1,n_{i-1}}, R_{i+1,1}, \break \cdots, R_{i+1,n_{i+1}},  \cdots, R_{n,1}, \cdots, R_{n,n_m})\} \cup \{T(R_0,R_{1,1}, \cdots, R_{1,n_1}, \break \cdots, R_{n,1}, \cdots, R_{n,n_m})\}$, where $a_1, a_2, \cdots, a_{n_i}$ is a permutation of $1,2, \cdots, n_i$.

Let $R_0'=R_0/(\{R_{1,1}, \cdots, R_{m, n_m}\}\setminus\{R_{i,1}, \cdots, R_{i, n_i} \} )$. By Theorem~\ref{theorem:chain_plan_space}, we can find the minimal cost plan of right deep tree without cross products for the branch $R_{i,a_1} /R_0', R_{i, a_2}, \cdots, R_{i, a_{n_1}}$ from $n_i$ plans. Thus, we can find the minimal cost plan for $\mathcal{R}$ with $\sum_{i=1}^{n}n_i=n-1$ plans. Adding the final candidate $T(R_0,R_{1,1}, R_{1,2}, \cdots, R_{n,1}, \cdots, R_{n,n_m})$ results in $n$ plans.
\end{proof}
\vspace{-.5em}
}
\cut{
\begin{theorem}\label{theorem:snowflake:min_cost_plan}
	\textbf{Minimal cost right deep tree for snowflake query}: Let $\mathcal{R}=\{R_0, R_{1,1}, \cdots, R_{1,n_1}, R_{2,1}, \cdots, R_{2,n_2},\cdots, R_{m,1},\cdots, R_{m,n_m}\}$ be relations of a snowflake query as described in Definition~\ref{def:snowflake_query}. Let $\mathcal{T}=T(U_1, U_2, \cdots, U_n, R_0)$ be a right deep tree, where $U_1, U_2, \cdots, U_n$ is a permutation of $\mathcal{R}-\{R_0\}$. $C_{out}(\mathcal{T})$ is minimal if $T$ is a partially-ordered right deep tree.
\end{theorem}
}

\cut{
These proofs are for right deep tree with cross products	
	
We first prove the following lemmas:

\begin{lemma}\label{lemma:snowflake:single_connection}
\textbf{Bitvector filter push-down with no in-between connection: }	Let $T(U_1, U_2, \cdots, U_i, \cdots, U_j, \cdots, U_n)$ be a right deep tree and $B$ be the bitvector created from $U_j$ and pushed down to $U_i$. Then there does not exist any relation $U_k, 1\leq k\leq j-1, k\neq i$ such that $U_k$ connects to $U_j$ in the join graph.
\end{lemma}
\begin{proof}
	This follows directly from Algorithm~\ref{algorithm:bitvector_pushdown}. Assume there exists relations in $\{U_{i+1}, U_{i+2}, \cdots, U_{j-1}\}$ that connect to $U_j$ in the join graph. Let $U_w$ be the relation with the maximal $w$ among such relations. Then $B$ will be pushed down to the join result of $U_1, U_2, \cdots, U_w$. If such a relation $U_w$ does not exist, but if there exists a relation $U_v \in \{U_1, U_2, \cdots, U_{i-1}\}$ such that $U_v$ and $U_j$ are connected in the join graph, then $B$ will be pushed down to the join result of $U_1, U_2, \cdots, U_i$.
	
	Because $B$ is pushed down to $U_i$, there does not exist any relation that connects to $U_j$ in the join graph in $\{U_1, U_2, \cdots, U_{j-1}\}$ except for $U_i$.
\end{proof}

\begin{lemma}\label{lemma:snowflake:relation_move_down}
	\textbf{Change in base relation bitvector filtering with moving down a relation: }
	Let $T(U_1, U_2, \cdots, U_i, \cdots, U_j, \cdots, U_n)$ and $T'(U_1, U_2, \cdots, U_{i-1}, U_j, U_{i}, U_{i+1}, \cdots, U_{j-1}, U_{j+1}, \cdots, U_n)$ be right deep trees. Let $S(U)$ be the base relation $U$ filtered by bitvectors in $T$ and $S'(U)$ be that in $T'$. Then $S'(U_j)\subseteq S(U_j)$ and $S'(U_k)\supseteq S(U_k)$ for $i\leq k\leq j-1$.
\end{lemma}

\begin{proof}
	Let the bitvectors pushed down to $U_j$ in $T$ be $\mathcal{G}_j=\{B_1, B_2, \cdots, B_m\}$ and these pushed down to $U_j$ in $T'$ be $\mathcal{G'}_j$. Then $B_1, B_2,\cdots, B_m$ are created from relation $U_{j+1}, U_{j+2}, \cdots, U_n$ in $T$.
		
	Assume $B_k$ is created from $U_l, j+1\leq l\leq n$. By Lemma~\ref{lemma:snowflake:single_connection}, there does not exist any relation connected to $U_l$ in $U_1, U_2, \cdots, U_{l-1}$ except for $U_j$. Thus, $B_k$ will be pushed down to $U_j$ in $T'$.
	
	Since for every $B_k\in \mathcal{G}_j$, we have $B_k\in \mathcal{G'}_j$, then $\mathcal{G}_j\subseteq \mathcal{G'}_j$.	
	Thus, $S'(U_j)=S(U_j/\mathcal{G'}_j)\subseteq S(U_j/\mathcal{G}_j)=S(U_j)$.
	
	Similarly, let the bitvectors pushed down to $U_k$ in $T$ be $\mathcal{G}_k$ and these pushed down to $U_k$ in $T'$ be $\mathcal{G'}_k$ for $i\leq k\leq j-1$. We can show $\mathcal{G'}_k\subseteq \mathcal{G}_k$ and thus $S'(U_k)\supseteq S(U_k)$.
\end{proof}

\begin{corollary}\label{corollary:snowflake:nobitvector_relation_move_down}
	\textbf{Unchanged base relation with moving down a relation}
	Let $T(U_1, U_2, \cdots, U_i, \cdots, U_j, \cdots, U_n)$ and $T'(U_1, U_2, \cdots, U_{i-1}, U_j, U_{i}, U_{i+1}, \cdots, U_{j-1}, U_{j+1}, \cdots, U_n)$ be right deep trees. Let $S(U)$ be the base relation $U$ filtered by bitvectors in $T$ and $S'(U)$ be that in $T'$. If $U_j$ does not create a bitvector pushed down to $U_k$, where $i\leq k\leq j-1$, then $S'(U_k)= S(U_k)$.
\end{corollary}

\begin{proof}
Let the bitvectors pushed down to $U_k$ in $T$ be $\mathcal{G}_k$ and these pushed down to $U_k$ in $T'$ be $\mathcal{G'}_k$ for $i\leq k\leq j-1$.

Let $B\in \mathcal{G}$, then $B$ must be created from relations $U_w, j\leq w\leq n$. Since $U_j$ does not create a bitvector that is pushed down to $U_k, i\leq k\leq j-1$, $B$ must be created from relations $U_w, j+1\leq w\leq n$. 

By Lemma~\ref{lemma:snowflake:single_connection}, $U_w$ does not connect with any relation in $U_1, \cdots, U_j$ excepts for $U_k$. Thus, the bitvector created from $U_w$ will be pushed down to $U_j$ in $T'$. Since $B\in \mathcal{G'}$ for every $B\in \mathcal{G'}$, we have $\mathcal{G}\subseteq \mathcal{G'}$.

By Lemma~\ref{lemma:snowflake:relation_move_down}, we have $\mathcal{G'}\subseteq \mathcal{G}$. Thus, $\mathcal{G}=\mathcal{G'}$ for all $U_k$ where $i\leq k\leq j-1$. Thus, $S'(U_k)=S(U_k)$.

\end{proof}

\begin{lemma}\label{lemma:snowflake:min_cost_plan1}
	\textbf{Minimal cost right deep tree is $1$-partially-ordered: }
	Let $\mathcal{R}=\{R_0, R_{1,1}, \cdots, R_{1,n_1}, R_{2,1}, \cdots, R_{2,n_2},\cdots, R_{m,1},\cdots, R_{m,n_m}\}$ be relations of a snowflake query and $\mathcal{T} =T(R_0, U_1, U_2, \cdots, U_n)$ be a right deep tree, where $U_1, U_2, \cdots, U_n$ is a permutation of $\mathcal{R}-\{R_0\}$. If $C_{out}(\mathcal{T})$ is minimal, then there exists a $1$-partially-ordered right deep tree $\mathcal{T'}$ such that $C_{out}(\mathcal{T'})=C_{out}(\mathcal{T})$.
\end{lemma}

\begin{proof}
Assume $T=T(R_0, U_1, U_2, \cdots, U_n)$ is not a partially-ordered right deep tree of size $1$, then $U_1\neq R_{i, 1}$ for all $1\leq i\leq m$. Let $k$ be the smallest number where $U_k=R_{i,1}$ for some $1\leq i\leq m$. Consider an alternative right deep tree $T'=T(R_0, U_k, U_1, \cdots, U_{k-1}, U_{k+1}, \cdots, U_n)$.

Let $S(A)$ be the base relation filtered by the bitvectors pushed down to a relation $A$ in $T$ and $S'(A)$ be that for $A$ in $T'$. Let $S(A_1, A_2, \cdots, A_i)$ be the result of the join of right deep subtree $T(A_1, A_2, \cdots, A_i)$ in $T$ and $S'(A_1, A_2, \cdots, A_i)$ be that in $T'$.

For $U_j, k<j\leq n$, both the cardinality of the base relation filtered by the bitvectors and the join result size will not change. Thus, $S(U_j)=S'(U_j)$ and $S(R_0, U_1, \cdots, U_j)=S'(R_0, U_k, U_1, \cdots, U_{k-1}, U_{k+1}, \cdots, U_j)$ will be the same in $T$ and $T'$.

For $U_k$, by Lemma~\ref{lemma:snowflake:relation_move_down}, $S'(U_k)\subseteq S(U_k)$. Now consider the relations $U_1, \cdots, U_{k-1}$. If $U_k$ creates a bitvector that is pushed down to a relation $U_j$ where $1\leq j\leq k-1$, by Lemma~\ref{lemma:snowflake:single_connection}, $U_k$ must connect to and only to $U_j$ in the join graph among the relations $R_0, U_1, \cdots, U_{k-1}$. Since $U_k=R_{i,1}$ joins with $R_0$, the bitvector can only be pushed down to $R_0$. then $U_k$ cannot create a bitvector that is pushed down to $U_j$. By Corollary~\ref{corollary:snowflake:nobitvector_relation_move_down}, $S(U_j)=S'(U_j)$.

For $R_0$, let the bitvectors pushed down to $R_0$ be $\mathcal{G}_0$ in $T$ and these in $T'$ be $\mathcal{G'}_0$. Let the bitvector created from relation $U_i$ be $B_i$ in $T$ and that in $T'$ be $B'_i$. First, since $S(U_j)=S'(U_j)$ for all $1\leq j\leq n, j\neq k$, $B_i=B'_i$ for all $1\leq j\leq n, j\neq k$. By Lemma~\ref{lemma:snowflake:single_connection}, if $B_j\in \mathcal{G}$, $U_j\in \{R_{1, 1}, \cdots, R_{m,1}\}$. Because $U_k$ is the first relation that is in $\{R_{1,1}, \cdots, R_{m,1}\}$, the bitvectors created from $U_1,\cdots, U_{k-1}$ will be pushed down to $R_0$. Thus, $B_1, \cdots, B_{k-1} \notin \mathcal{G}$.

If $B_j\in \mathcal{G}_0, k<j\leq n$, by Lemma~\ref{lemma:snowflake:single_connection}, $U_j$ connects to $R_0$ but it is not connected to $U_1, U_2, \cdots, U_k$ in the join graph. Thus, $B'_j\in\mathcal{G'}_0$. Since $B'_j=B_j$, $B_j\in \mathcal{G'}_0$. For relation $U_k$, $B'_k$ is created from $S'(U_k)$ and is pushed down to $R_0$. Since $S'(U_k)\subseteq S(U_k)$, $R_0/B'_k\subseteq R_0/B_k$. Since $\mathcal{G}_0-{B_k}\subseteq \mathcal{G'}_0-{B'_k}$, $R_0/\mathcal{G}_0\supseteq R_0/(B_k, \mathcal{G}_0)\supseteq R_0/(B'_k, \mathcal{G'}_0) = R_0/\mathcal{G'}_0$. Thus, $S'(R_0)\subseteq S(R_0)$.

Consider the intermediate join result $S(R_0, U_1, U_2, \cdots, U_j)$ and $S'(R_0, U_k, U_1, U_2, \cdots, U_j)$, where $j<k$. Since $U_k\to R_0$, $S(R_0, U_k)\subseteq S(R_0)$. If a bitvector $B_w$ is created from $U_w$ and pushed down to $S(R_0, U_1, U_2, \cdots, U_j)$, \todo{How about bitvectors that are pushed down to the join results?}

\end{proof}
}

\cut{
\subsection{Additional optimizations}
\paragraph{Cross product for star query}
}
	\changed{\section{Bitvector-aware QO for general snowflake queries}
\label{sec:general_qo}
}

\point{Chain, star, snowflakes are common, but general decision support queries are more complicated}
\changed{
While \changed{star and snowflake queries with PKFK joins are important patterns in decision support queries, in practice, such queries can have more complicated join graphs.}\cut{the join graphs of decision support queries can be more complicated than standard snowflakes.} For example, a decision support query can join multiple fact tables, where the joins may not be PKFK joins. In addition, there can be join conditions between the dimension tables or branches, where the bitvector filters created from the dimension tables may not be pushed down to the fact table. Finally, there can be dimension tables or branches that are larger than the fact table after predicate filters, where the fact table should be on the build side in the plan space of right deep trees.\cut{. In such cases, the plan space of right deep trees will be suboptimal if the dimension tables are on the build side of the query plan.}

In this section, we first propose an algorithm to extend bitvector-aware query optimization to an arbitrary snowflake query with a single fact table. We then generalize it to arbitrary decision support queries with multiple fact tables. \changed{\cut{While our algorithm is optimal for a class of decision support queries as analyzed in Section~\ref{sec:snowflake}}Our algorithm applies to queries with arbitrary join graphs.}
We further optimize our algorithm with cost-based bitvector filters. We also discuss options to integrate our algorithm into a Volcano / Cascades query optimizer.
}

\changed{
	\subsection{Queries with a single fact table}	
}	
	\point{Chain, star, snowflakes are common, but general decision support queries are more complicated}
	\cut{
	While \changed{star and snowflake are important patterns in decision support queries, in practice, such queries can have more complicated join graphs.}\cut{the join graphs of decision support queries can be more complicated than standard snowflakes.} For example, a decision support query can join multiple fact tables, where the joins may not be PKFK joins. In addition, there can be join conditions between the dimension tables or branches, where the bitvector filters created from the dimension tables may not be pushed down to the fact table. Finally, there can be dimension tables or branches that are larger than the fact table after predicate filters, where the fact table should be on the build side in the plan space of right deep trees.\cut{. In such cases, the plan space of right deep trees will be suboptimal if the dimension tables are on the build side of the query plan.}
}
	
	\point{Algorithm uses heuristics to leverage the plans with equal cost in star and snowflake queries}
	We propose an algorithm \changed{(Algorithm~\ref{algorithm:general_snowflake})} with simple heuristics to construct the join order for an arbitrary snowflake query with a single fact table.
	The key insight is to \changed{leverage the candidate plans of minimal cost analyzed in \cut{intuition that different permutations of branches or dimension tables will result in the same cost based on our analysis} Section~\ref{sec:snowflake}}. 
\changed{Algorithm~\ref{algorithm:general_snowflake} shows how to construct the join order for a decision support query \changed{with a single fact table.}}

We first assign priorities to the branches based on their violations of the snowflake pattern as \changed{defined in Definition~\ref{def:snowflake_query}. We then sort the branches in descending order by their priorities (line 1)}. Intuitively, if the bitvector filters created from dimension tables are all pushed down to the fact table except for one, where the corresponding dimension table either joins with another dimension table or is not on the build side. Since this dimension table does not create a bitvector filter that is pushed down to the fact table, joining this dimension table early with the fact table can eliminate the unnecessary tuples that do not qualify the join condition early in the plan.

Specifically, we assign priorities to branches for snowflake queries with the following heuristics:
\begin{itemize}[leftmargin=*]
	\item Group P0: Relations that do not have join condition or PKFK joins with the fact table \changed{(line 23)}. This can happen when joining multiple fact tables. \changed{As a heuristic, we join these branches by descending selectivity on the fact table (line 23).}
	\item Group P1: Branches that do not join with any other branches and have smaller cardinality than the fact table \changed{(line 24)}. \changed{These branches are joined with the fact table before joining the branches in group P0.}
	\item Group P2: Branches joining with other branches \changed{(line 21)}. Such branches should be joined consecutively in the right deep tree to allow pushing down bitvector filters created by these branches. \changed{As a heuristic, within a set of connected branches, we join these branches with descending selectivity on the fact table (line 31); across sets of connected branches, we prioritize the sets of larger numbers of connected branches (line 21).}
	\item Group P3: Branches that are larger than the fact table \changed{(line 25)}. \changed{Since it is clearly suboptimal to put these branches on the build side, we reorder the build and probe sides for them (line 12-13). Joining these branches early allows pushing down the bitvector filters created from the fact table. As a heuristic, we order the branches in this group with descending selectivity on the fact table (line 31).} 
\end{itemize}

\cut{
We first assign priorities to the branches (line 1) based on their violations of the snowflake pattern as defined in Definition~\ref{def:snowflake_query} and sort them in descending order by their priorities using \changed{Algorithm \emph{SortBranches}(line 17-35)}.
}

\begin{algorithm}
\DontPrintSemicolon
\SetAlgoLined
\LinesNumbered

\caption{Construct a join order for a snowflake query \changed{with a single fact table}}
\label{algorithm:general_snowflake}

\SetKwProg{FuncOptSnowflake}{}{:}{}
\FuncOptSnowflake{\textbf{OptimizeSnowflake($G$)}}{
	\KwIn{Join graph $G$}
	\KwOut{Query plan $plan$}
	$B \leftarrow SortedBranches(G.Branches)$\;
	$best \to JoinBranches(B, G.Fact, \emptyset)$\;
	\ForEach {branch $b$ in $B$}{
		$p \leftarrow Join(OptimizeChain(b, G.Fact), G.Fact)$\;
		$p \leftarrow JoinBranches(B\setminus b, G.Fact, p)$\;
		\lIf{$best.Cost>p.Cost$}{
			$best \leftarrow p$
		}
	}
	\Return{$best$}
}

\cut{
\begin{algorithm}
	\DontPrintSemicolon
	\SetAlgoLined
	\LinesNumbered
	
	\caption{Group connected branches}
	\label{algorithm:group_branch}

\SetKwProg{FuncGroupBranches}{}{:}{}
\FuncGroupBranches{\textbf{GroupBranches($G$, $B$)}}{
	\KwIn{Join graph $G$, set of branches $B$}
	\KwOut{Groups of branches $groups$}
	\tcc{Group connected dimension tables}
	$groups \leftarrow \emptyset$\;
	$visited \leftarrow \emptyset$\;
	\ForEach {dimension table $d$ in $D$}{		
		\If{$d$ is not in $visited$}{
			$group\leftarrow \{d\}$\;
			$queue \leftarrow [d]$\;
			\While{$queue$ is not empty}{
				$d1 \leftarrow queue.Pop()$\;
				\ForEach {dimension table $d2$ in $D$}{
					\If{$d2\notin visited$ and $G.HasEdge(d1, d2)$}{
						$visited\leftarrow visited\cup\{d2\}$\;
						$group\leftarrow group\cup \{d2\}$\;
						$queue.Enqueue(d2)$
					}
				}
			}
		}
	}	
	\Return{$groups$}	
}
\end{algorithm}
}

\cut{
\begin{algorithm}
	\DontPrintSemicolon
	\SetAlgoLined
	\LinesNumbered
	
	\caption{Complete the rest of the join with remaining branches}
	\label{algorithm:join_branches}
}
	
	\SetKwProg{FuncJoinBranches}{}{:}{}
	\FuncJoinBranches{\textbf{JoinBranches($B$, $f$, $p$)}}{
		\KwIn{A set of branches $B$, fact table $f$, a plan $p$}
		\KwOut{A query plan $p'$}
		$p'\leftarrow p$\;
		\ForEach {branch $b$ in $B$}{
			\ForEach {table $t$ in $b$}{
				\lIf{$t.Card >f.Card$}{
					$p'\leftarrow Join(p', t)$		
				}
				\lElse{
					$p' \leftarrow Join(t, p')$
				}
			}
		}
		\Return{$p'$}
	}

\cut{
\end{algorithm}

\begin{algorithm}
	\DontPrintSemicolon
	\SetAlgoLined
	\LinesNumbered
	
	\caption{Sort branches based on heuristics}
	\label{algorithm:sort_branches}
}
	
	\SetKwProg{FuncSortBranches}{}{:}{}
	\FuncSortBranches{\textbf{SortBranches($G$)}}{
		\KwIn{Join graph $G$}
		\KwOut{Sorted branches $sortedBranches$}
		$groups\leftarrow GroupBranches(G)$\;
		$sortedG\leftarrow SortBySizeDesc(groups)$\;
		$priority\leftarrow []$\;
		\For{$i = 0; i < groups.Count(); i++$}{
			\lIf{$sortedG[i].Size > 1$}{
				$priority[i]\leftarrow sortedG[i].Size$
			}
			\Else{
				\changed{\lIf{$IsNonUniqueKeyJoin(g[0], f)$}{
					$priority[i]\leftarrow 0$
				}}
				\lIf{$g[0].Card <f.Card$}{
					$priority[i]\leftarrow \changed{1}$
				}
				\lElse{
					$priority[i]\leftarrow \changed{|G|+1}$
				}
			}
		}
		$sortedG\leftarrow SortByPriorityDesc(groups, priority)$\;	
		$sortedBranches \leftarrow []$\;
		\ForEach{$group$ in $sortedG$}{
			$branches\leftarrow \changed{SortBySelectivityDesc(group)}$\;
			\lForEach{$b$ in $branches$}{
				$sortedBranches.Add(b)$
			}
		}
		\Return{$sortedBranches$}	
	}
\end{algorithm}

Based on the analysis in Section~\ref{sec:snowflake}, we construct the candidate plans \changed{by two cases}. If $R_0$ is the right most leaf, we join all the branches with the fact table \changed{(line 2)}; otherwise, for each branch, we optimize the branch based on the analysis in \changed{Section~\ref{sec:branch}}, join the remaining branches to complete the plan, and update the best plan if the estimated cost of the new plan is lower (line 3-7).

\cut{
To join the branches to complete an initial plan, we swap the build and probe side if the dimension table has larger cardinality than the fact table (line 12-13). Since putting the dimension table on the build side will eliminate the bitvector filter created from the table and can impact multiple joins in the plan, as a heuristic, we only swap the build and probe side when it is clearly suboptimal.
}

\cut{
To assign priorities to branches, we first group the branches such that the branches with join conditions are grouped together ($SortBranches$). The groups are sorted by the number of tables in the group in descending order (line 18). Then we assign higher priorities to groups with more than one branch or a branch that contains a table with cardinality larger than the fact table (line 20-26). For groups with more than one branch, we prioritize groups with smaller number of tables to eliminate unnecessary tuples earlier in the plan. For branches with the same priority, we order them by their cardinality from small to large. Finally, we order the branches based on their priorities, and the branches from the same group will be put consecutive (line 27-34).
}
	
\changed{\subsection{Queries with multiple fact tables}}

\point{Can apply to arbitrary join graphs}

\changed{
	In addition to snowflakes with a single fact table, complex decision support queries can include multiple fact tables. We further extend our algorithm to arbitrary join graphs by iteratively extracting and optimizing snowflake join graphs.
}

\point{High level intuition: snowflake expansion and assign higher priorities to dimension tables that violate snowflake}

\changed{
	At a high level, our algorithm produces a join order for a join graph by alternating two stages iteratively as shown in Algorithm~\ref{algorithm:general_qo}. In the snowflake extraction stage (line 2), we extract a snowflake subgraph from a join graph by identifying a single fact table and its related dimension tables, potentially with non-PKFK joins. In the snowflake optimization stage (line 3), we use Algorithm~\ref{algorithm:general_snowflake} to produce a join order for the extracted subgraph. The resulting snowflake will be marked as 'optimized' and considered as a new relation in the updated join graph (line 4-5). Our algorithm alternates the two stages until the full join graph is optimized (line 1).

}

\begin{algorithm}
\DontPrintSemicolon
\SetAlgoLined
\LinesNumbered

\caption{\changed{Construct a join order for a decision support query with an arbitrary join graph}}
\label{algorithm:general_qo}

\SetKwProg{FuncOptJoinGraph}{}{:}{}
\FuncOptJoinGraph{\textbf{OptimizeJoinGraph($G$)}}{
	\KwIn{Join graph $G$}
	\KwOut{Query plan $plan$}
	\While{$|G| > 1$}{
		$G' \leftarrow ExtractSnowflake(G)$\;
		$p \leftarrow OptimizeSnowflake(G')$\;
		$G \leftarrow UpdateJoinGraph(G, G')$\;
		$plan \leftarrow UpdateQueryPlan(plan, p)$\;
	}
	\Return{$plan$}
}

\SetKwProg{FuncExtractSnowflake}{}{:}{}
\FuncExtractSnowflake{\textbf{ExtractSnowflake($G$)}}{
	\KwIn{Join graph $G$}
	\KwOut{Snowflake $G'$}
	$n \leftarrow 0$\;
	$G_{sorted} \leftarrow SortByCardinalityAsc(G)$\;
	\ForEach{$g$ in $G_{sorted}$}{
		\If{$g$ is an unoptimized fact table}{
			\If{$n == 0$}{
				$G' \leftarrow ExpandSnowflake(g)$
			}
			$n \leftarrow n + 1$
		}
	}
	\lIf{$n == 1$}{
		$G'\leftarrow G$
	}
	\Return{$G'$}
}

\end{algorithm}

\changed{
Specifically, when extracting a snowflake (line 8-19), a relation is considered as a fact table if it does not join with any other table where the join predicate is an equi-join on its key columns. Among all the unoptimized fact tables in $G$, we find the one with the smallest cardinality and expand from this table recursively to include all related dimension relations (line 4-9). If there is only one fact table in $G$, we simply return the original join graph (line 11).
}

\changed{
\subsection{Cost-based Bitvector Filter}
\label{subsec:cost_based_bitvector}

\point{Bitvectors are not free. Challenges in holistic cost model. Use a local cost model}

In practice, creating and applying bitvector filers has overheads. Consider a hash join with build side $R$ and probe side $S$. Assume the bitvector filter eliminates $\lambda$ percent of the tuples from $S$. The ratio $\lambda$ can be estimated by the optimizer the same way as an anti-semi join operator, and it can include the estimated false positive rate of the bitvector filter.

Assume the cost of a hash join consists of 
building the hash table $g_{b}$, probing the hash table $g_{p}$, and outputing the resulting tuples $g_{o}$. Let the cost of creating and applying a bitvector filter be $h$ and $f$. The cost difference of the hash join with and without using the bitvector filter is
}
\cut{
\vspace{-.5em}
\begin{align*}
\label{eq:cost_diff}
Cost_{\Delta}=g(|R|,|S|) - (g(|R|, \lambda |S|)+f(|R|) + h(|S|))
\end{align*}
\changed{
where $g$ is the cost of the hash join, $f$ is the cost of creating a bitvector, and $h$ is the cost of applying a bitvector filter.

Assume $g(|R|,|S|)$ consists of building the hash table $g_{b}$, probing the hash table $g_{p}$, and outputing the resulting tuples $g_{o}$. Then the cost difference becomes}
\cut{\begin{align*}
Cost_{\Delta}&=g_{b}(|R|) + g_{p}(|S|)+g_{o}(|R\bowtie S|) \\
	&-(g_{b}(|R|) + g_{p}(\lambda|S|)-g_{o}(|R\bowtie S|)+f(|R|)+h(|S|)) \\
	&= g_{p}(|S|)-g_{p}(\lambda|S|)-f(|R|)-h(|S|)
\end{align*}
}
}

\vspace{-1.5em}
\begin{align*}
Cost_{\Delta}=g_{p}(|S|)-g_{p}(\lambda|S|)-f(|R|)-h(|S|)
\end{align*}
\vspace{-1.5em}

\changed{
Assume the cost of probing a tuple is $C_p$, the cost of checking a tuple against a bitvector filter is $C_f$, and creating a bitvector filter is relatively cheap, i.e., $f(|R|)<<h(|S|)$. Then}
\vspace{-1.5em}
\begin{align*}
Cost_{\Delta}=|S|((1-\lambda)C_p-C_f)- f(|R|)\sim |S|((1-\lambda)C_p-C_f)
\end{align*}
\vspace{-1.5em}

\changed{
Using a bitvector filter reduces the cost of a hash join if}
\vspace{-.5em}
\begin{align*}
Cost_{\Delta} < 0 \sim |S|((1-\lambda)C_p - C_f) < 0 \Leftrightarrow \lambda > 1-C_f/C_p
\end{align*}
\vspace{-1.5em}

\changed{
\point{The analysis is independent of the size of the relations}
Let $\lambda_{thresh}=1-C_f/C_p$. Note that $\lambda_{thresh}$ is independent of $R$ and $S$. We can run a micro-benchmark to profile $C_f$ and $C_p$ and compute $\lambda_{thresh}$. When the bitvector filter is pushed down below the root of the probe side, a more detailed analysis is needed to account for the cascading effect of tuple elimination. Empirically, choosing a threshold that is slightly smaller than $1-C_f/C_p$ works well.
}

\changed{
\subsection{Integration}
\label{sec:integration}

\point{At a high level, our algorithm can be used as a transformation rule}
Our algorithm can transform a query plan by optimizing the join order with the underlying join graph. Thus, our algorithm can be used as a new transformation rule in a Volcano / Cascades query optimization framework upon detecting a snowflake join (sub)graph. There are three integration options depending on how the underlying optimizer accounts for the impact of bitvector filters:}
\cut{
\point{Our algorithm can be triggered as a subpattern of snowflake or a general pattern for an arbitrary join graph}
This new rule can be triggered in two ways. First, upon detecting , with our algorithm for optimizing general snowflake join graphs (Algorithm~\ref{algorithm:general_snowflake}), the rule can be triggered if a subplan matches the general snowflake join graph pattern, i.e., \textit{snowflake transformation}. Second, with our algorithm for optimizing arbitrary join graphs (Algorithm~\ref{algorithm:general_qo}), the rule can be triggered by any query plan, i.e., \textit{full transformation}.
}

\vspace{-.2em}
\begin{itemize}[leftmargin=*]
\item
\changed{
\point{If the QO is aware of bitvectors, then any further QO can be performed}
\textit{Full integration:} When applying join order transformation to a (sub)plan, the placement of bitvector filters and their selectivity can change. If the underlying Volcano / Cascades query optimization framework can correctly account for the placement and the selectivity of bitvector filters \emph{during} query optimization, the new transformation rule can be transparently integrated into the query optimizer the same way as any existing transformation rule.
}
\item
\changed{\textit{Alternative-plan integration:}
	If the query optimizer can account for the placement and the selectivity of bitvector filters in a \textit{final plan} \emph{after} query optimization, the new transformation rule can be used to produce an alternative plan. The optimizer can then choose the plan with the cheaper estimated cost from the alternative plan and the plan produced by the original query optimization.}
\item 
\changed{
\point{If the QO is not aware of bitvectors, do not change the reorder}

\textit{Shallow integration:} We mark a (sub)plan after it is transformed by our new transformation rule. The underlying query optimization framework works as usual, except additional join reordering on marked (sub)plans is disabled. \cut{ When full transformation is used, the join order of the query is decided by this transformation without further join reordering.}
}
\end{itemize}

\cut{
\todo{Describe the algorithm}

\subparagraph{Example of the algorithm}

Figure~\ref{fig:general_qo} shows an example of how to construct the query plan with a general join graph with Algorithm~\ref{algorithm:general_qo}.
	
\todo{Describe the example}

\begin{figure}
	\begin{subfigure}[b]{.24\textwidth}
		\centering
		\includegraphics[width=.9\linewidth]{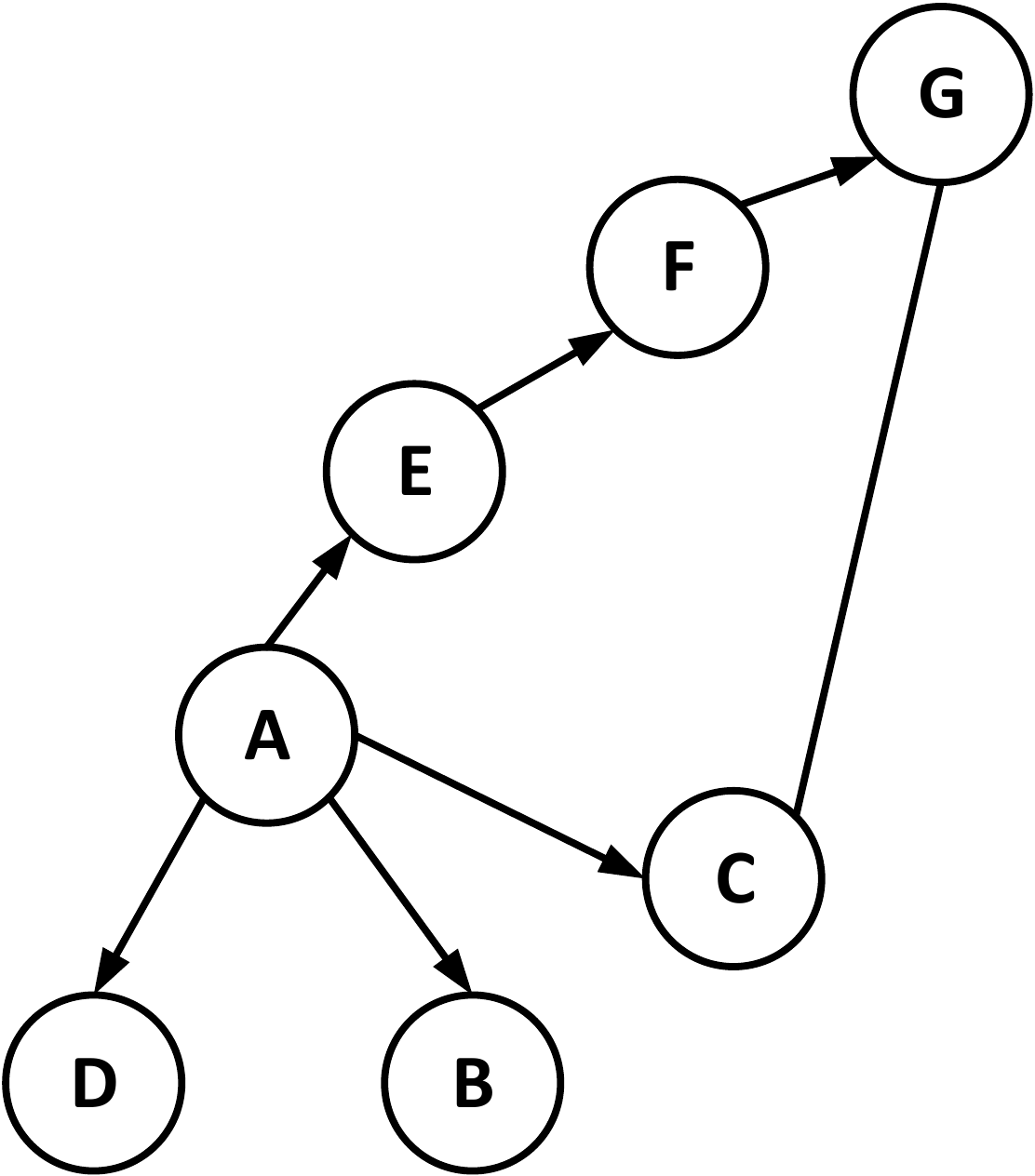}
		\caption{Join graph}
		\label{fig:general_qo:join_graph}
	\end{subfigure}
	\begin{subfigure}[b]{.23\textwidth}
		\centering
		\includegraphics[width=.9\linewidth]{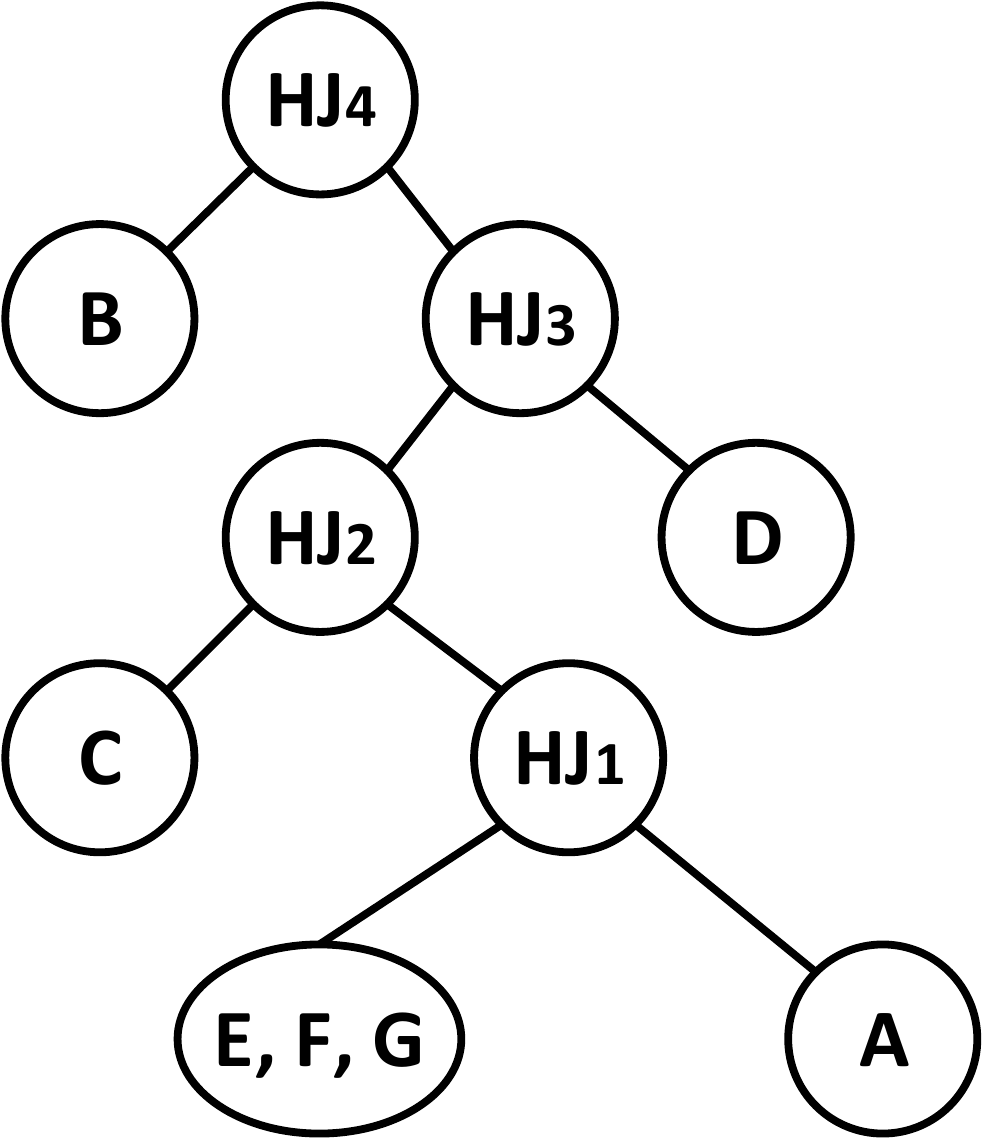}
		\caption{Query plan}
		\label{fig:general_qo:plan}
	\end{subfigure}
	\caption{Example of constructing the plan with bitvector-aware query optimization}
	\label{fig:general_qo}
\end{figure}
}
	
\cut{
\subsection{Limitations}
\label{sec:limitation}

\paragraph{Limitations}
\subparagraph{Cost model 1: column size vary too much}
\subparagraph{Cost model 2: Expensive predicate evaluation}
\subparagraph{Cost model 3: Unable to push down bitvector filter in pages}

\subparagraph{Join graph 1: Non decision support queries}
\subparagraph{Join graph 2: Simple query graph and full enumeration of join orders}
\subparagraph{Join graph 3: Subject to how the snowflakes are expanded when there are multiple fact tables}
}
	\section{Evaluation}
\label{sec:experiment}

\subsection{Implementation}
\label{sec:implementation}
\point{Implement on top of \sqlserver}

We implement Algorithm~\ref{algorithm:general_qo} in \sqlserver as a transformation rule.\cut{, leveraging its existing logic to detect snowflake patterns and estimate selectivity of bitvector filters.}
\point{Describe the snowflake pattern detection}
\sqlserver has a cost-based, \changed{Volcano / Cascades} style query optimizer. Starting from an initial query plan, the optimizer detects various patterns in the plan and fires the corresponding transformation rules. Due to the importance of decision support queries\cut{, as with other commercial DBMSs\todo{citation needed}}, \sqlserver has implemented heuristics to detect snowflake patterns and transform the corresponding subplans\cut{optimize the corresponding subplans with transformation rules accordingly}.

\changed{
We leverage the snowflake detection in \sqlserver and transform the corresponding subplan as described in Algorithm~\ref{algorithm:general_qo}. We implement a shallow integration (Section~\ref{sec:integration}), where join reordering is disabled on the transformed subplan. The subplan is subject to other transformations in \sqlserver. We use the original cardinality estimator and cost modeling in \sqlserver, and the selectivity of a bitvector filter is estimated the same way as the existing semi-join operator. We implement the cost-based bitvector filter as described in Section~\ref{subsec:cost_based_bitvector}, and we will discuss how we profile the elimination threshold $\lambda_{thresh}$ in Section~\ref{sec:profile_bitvector}. The final plan is chosen with the existing cost-based query optimization framework.} 

\point{Disable join order reordering about the snowflake transformation}
\cut{Because the impact of bitvector filters is not accounted properly in the cost model of \sqlserver, we disable transformation rules of join reordering to enforce the join orders produced by our algorithm. All other transformation rules are allowed.
}
\point{Leverage cardinality estimation for bitvector filters}
\cut{We leverage the existing cardinality estimation mechanism in \sqlserver to estimate the selectivity of applying a bitvector filter. Logically, a bitvector filter implements the relational operator semi-join, which has already been supported in \sqlserver. Our technique uses the existing cardinality estimation logic for semi join operator to estimate the selectivity of bitvector filters.}

\point{Use $C_{out}$ as the cost model}
\cut{Since the transformation rule on snowflake patterns is an exploration rule, the physical information of operators and their cost is not accessible due to the limitation of \sqlserver's implementation. Instead, we use $C_{out}$ as the cost function to choose the plan with the lowest cost among the plan candidates as described in Section~\ref{sec:star}, Section~\ref{sec:chain}, and Section~\ref{sec:snowflake}, as the output of our transformation rule. Incorporating the cost model from the underlying DBMS engine is future work.
}
\subsection{Experimental Setup}
\paragraph{Workload}

\begin{table}
\caption{Statistics of workloads, including database size, the number of tables, queries, indexes (B+ trees and columnstores), and joins.}
\label{table:workload}
\begin{tabular}{|m{3cm}|c|c|c|}
\hline
Statistics                & \tpcds & \job & \mssales \\
\hline
DB Size                   & 100GB  & 7GB    & 700GB   \\
Tables              & 25     & 21     & 475     \\
Queries             & 99     & 113    & 100     \\
B+ trees / columnstores    & 0 / 20      & 44 / 20       & 680 / 0   \\
Joins avg / max           & 7.9 / 48   & 7.7 / 16      & 30.3 / 80   \\
\hline
\end{tabular}
\end{table}

We evaluation our technique on three workloads: \tpcds~\cite{tpcds} 100GB with columnstores, \job~\cite{leis2018query} with columnstores, primary key indexes, and foreign key indexes, and a customer workload (\mssales) with B+-tree indexes.
\point{Describe the properties of the workloads}
Table~\ref{table:workload} summarizes the statistics of our workloads. \cut{Among the three workloads, \job and \mssales are complex decision support queries, where most of the queries have snowflake patterns.} In particular, \mssales has the highest number of average joins per query, and \job has the most complex join graphs, including joining multiple fact tables, large dimension tables, and joins between dimension tables.
Our workloads also cover the range of different physical configurations, with B+ trees (\mssales), columnstores (\tpcds), or both (\job).

\paragraph{\changed{Baseline}}
We use the query plans produced by the original \sqlserver as our baseline. 
\changed{Bitvector filters are widely used in the query plans of \sqlserver. As shown in Appendix~\ref{sec:bitvector_effectiveness}, 97\% queries in \job, 98\% queries in \tpcds, and 100\% queries in \mssales have bitvector filters in their original plans. A bitvector filter can be created from a hash join operator, and it is pushed down to the lowest level on the probe side as described in Algorithm~\ref{algorithm:bitvector_pushdown}. The query optimizer in \sqlserver uses heuristics to selectively add bitvector filters to the query plan without fully accounting for the impact of bitvector filters during the query optimization stage. In particular, the heuristics used in its snowflake transformation rules neglect the impact of bitvector filters.
We use a generous timeout for the query optimizer in \sqlserver so that it can explore a large fraction of the relevant plan search.\cut{, where the optimizer sometimes spends up to 25 seconds CPU time in query optimization.}
}
\cut{We compare them with the plans produced by \sqlserver enabling the new transformation rule based on our technique. }

\changed{
\paragraph{Overhead}
Our technique adds very low overhead to query optimization. In fact, since we disable join reordering on the snowflake subplan after it is optimized by our transformation rule, the query optimization time with our transformation rule is one third of that with the original \sqlserver in average. We also measure the memory consumption for query execution. We observe some increase in memory consumption with our technique, since it favors right deep trees. The overall increase in memory consumption is not significant.
}

\paragraph{Environment}
All the experiments are run on a machine with Intel Xeon CPU E5 - 2660 v3 2.6GHz, 192GB memory, a 6.5TB hard disk, and Windows Server 2012 R2. To reduce runtime variance, all the queries are running in isolation at the same level of parallelism. The query CPU time reported is an average over ten warm runs. 

\changed{
\subsection{Overhead of bitvector filters}
\label{sec:profile_bitvector}
}

\cut{
\begin{figure}
	\centering
	\caption{\changed{Micro-benchmark to profile the overhead of bitvector filtering. The bitvector filter is beneficial if it eliminates $>10$\% tuples}}
	\label{fig:exp:bitvector_overhead}
	\includegraphics[width=\linewidth]{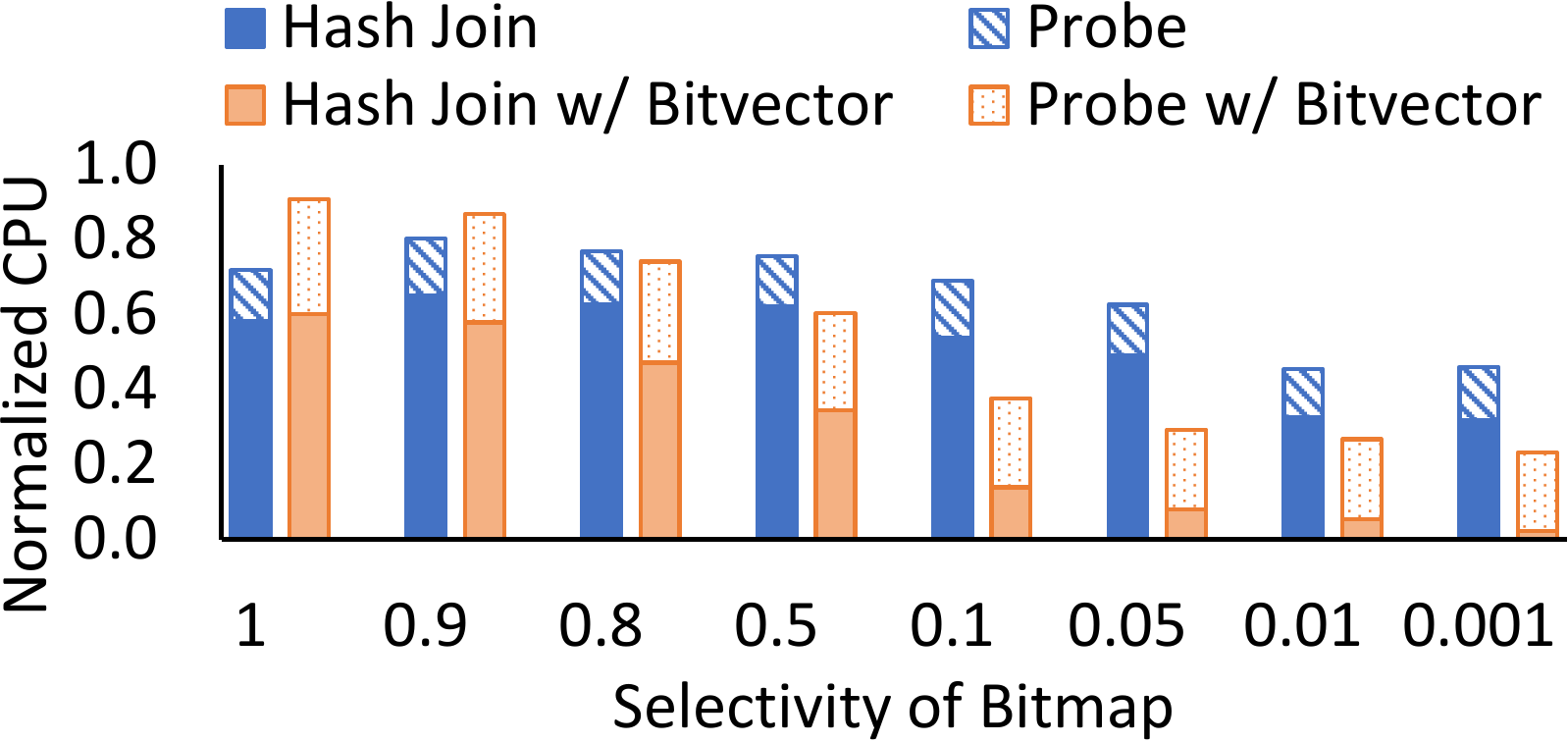}
\end{figure}
}

\begin{figure*}
	\begin{minipage}[b]{.32\textwidth}
		\centering
		\caption{\changed{Profile bitvector filters. A bitvector filter reduces overall cost if it eliminates $>10$\% tuples}}
		\label{fig:exp:bitvector_overhead}
		\includegraphics[width=.95\linewidth]{./exp/bitvector_overhead}
	\end{minipage}
\vspace{.2em}
	\begin{minipage}[b]{.33\textwidth}
		\centering
		\caption{\changed{Total query execution CPU time for a workload, breaking down by query selectivity}}
		\label{fig:exp:workload}
		\centering
		\includegraphics[width=.95\linewidth]{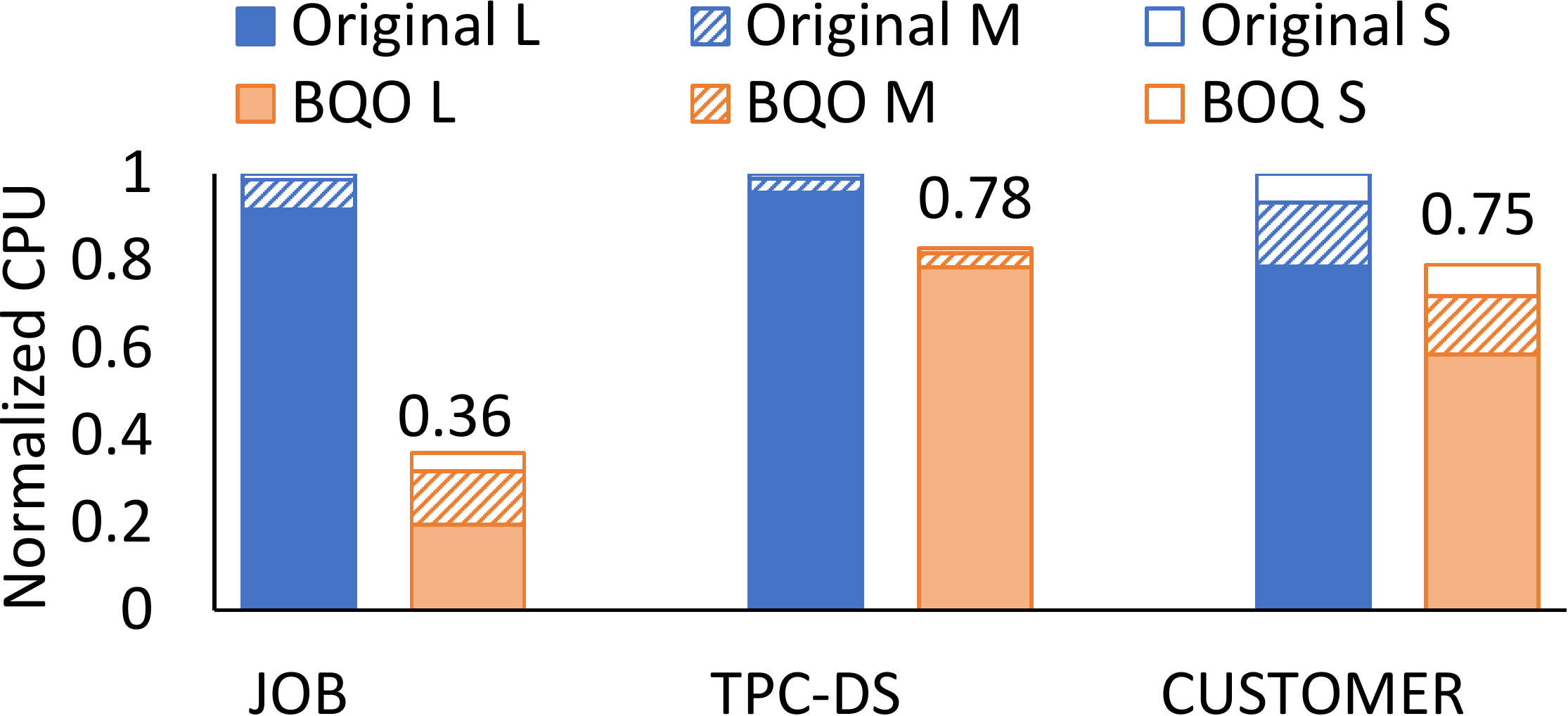}
	\end{minipage}
\vspace{.2em}
	\begin{minipage}[b]{.32\textwidth}
			\centering
		\caption{\changed{Total number of tuples output by operators in a workload, breaking down by operator types}}
		\label{fig:exp:bitvector_tuple_ratio}
		\includegraphics[width=.95\linewidth]{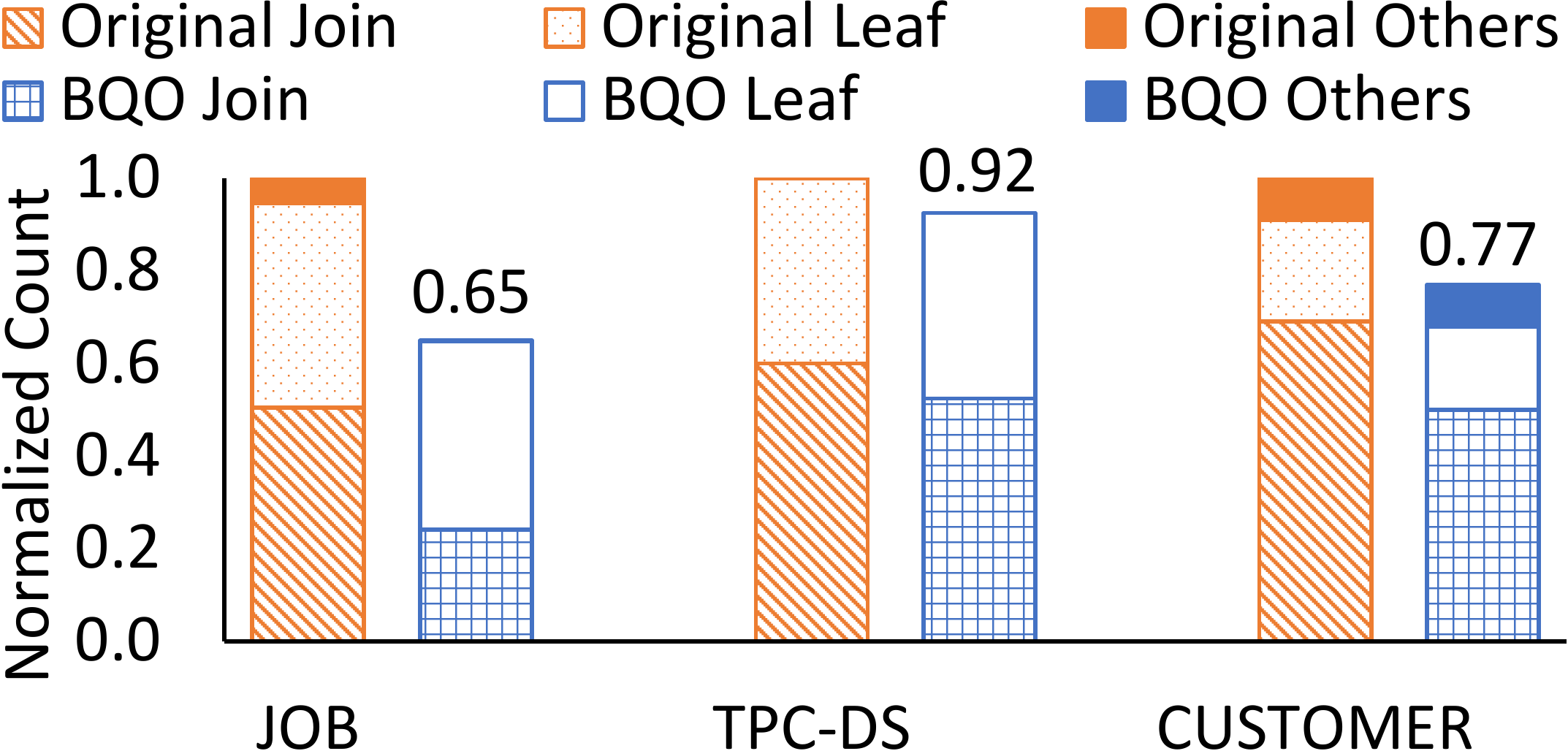}
	\end{minipage}
\end{figure*}

\changed{
\outline{Micro-benchmark on overhead of bitvector filters}

\point{Describe the micro-benchmark}
As discussed in Section~\ref{subsec:cost_based_bitvector}, we can choose a tuple elimination threshold to selectively create bitvector filters. We profile the overhead of bitvector filters with a micro-benchmark by running the following query in \tpcds:}

\begin{verbatim}
SELECT COUNT(*)
FROM store_sales, customer
WHERE ss_customer_sk = c_customer_sk
AND c_customer_sk % 1000 < @P
\end{verbatim}

\changed{
The query plan joins \emph{customer} and \emph{store\_sales} with a hash join. A bitvector filter is created from \emph{customer} on the build side and pushed down to \emph{store\_sales} on the probe side, where tuples are eliminated before the join. We control the selectivity of the bitvector filter with the parameter @P.

\point{Describe what the figure is showing}

Figure~\ref{fig:exp:bitvector_overhead} shows the CPU time of execution of the query varying its selectivity with and without bitvector filtering, normalized by the same constant. We further break down the CPU time by the hash join operator, the probe side, and the build side. Since the CPU time for reading \emph{customer} is very small, we omit it in Figure~\ref{fig:exp:bitvector_overhead} for readability.

\point{Takeaways from the figure}
With selectivity $1$, no tuples are eliminated by the bitvector. With bitvector filtering, the hash join operator is slightly more expensive due to creating the bitvector filter, and the probe side operator has higher execution CPU due to the overhead of checking the tuples from \emph{store\_sales} against the bitvector filter. As the selectivity increases, the bitvector filter eliminates more tuples from the probe side and the execution cost of the hash join operator reduces. The plan with bitvector filtering becomes cheaper than the other plan once the bitvector filter eliminates more than $10\%$ of the tuples. The cost reduction can be even more with queries of multiple joins. Empirically, we find $5\%$ to be a good threshold\cut{ to account for the cascading impact of bitvector filters}, and we set $\lambda_{thresh}$ to $5\%$ in our implementation.

\point{Additional effectiveness of bitvector filters}
In Appendix~\ref{sec:bitvector_effectiveness}, we further evaluate the effectiveness and applicability of bitvector filters as a query processing technique. As shown in Table~\ref{table:bitvector_effectiveness}, \sqlserver uses bitvector filters for $97\%-100\%$ queries in the benchmarks, with $10\%-80\%$ workload-level execution CPU cost reduction. This confirms that bitvector filters is a widely applicable query processing technique, and thus bitvector-aware query optimization can potentially impact a wide range of queries.

}

\subsection{Evaluation on bitvector-aware query optimization}

\cut{
\begin{figure*}
	\caption{Total query execution time, query optimization time, and memory usage for all queries in a workload consumed by the bitvector-aware plans over these of the original plans}
	\label{fig:exp:workload}
	\centering
	\begin{subfigure}{.33\linewidth}
		\centering
		\includegraphics[width=\linewidth]{./exp/overall_cpu_ratio}
		\caption{Normalized execution CPU time}
		\label{fig:exp:workload_cpu}
	\end{subfigure}
	\begin{subfigure}{.33\linewidth}
		\centering
		\includegraphics[width=\linewidth]{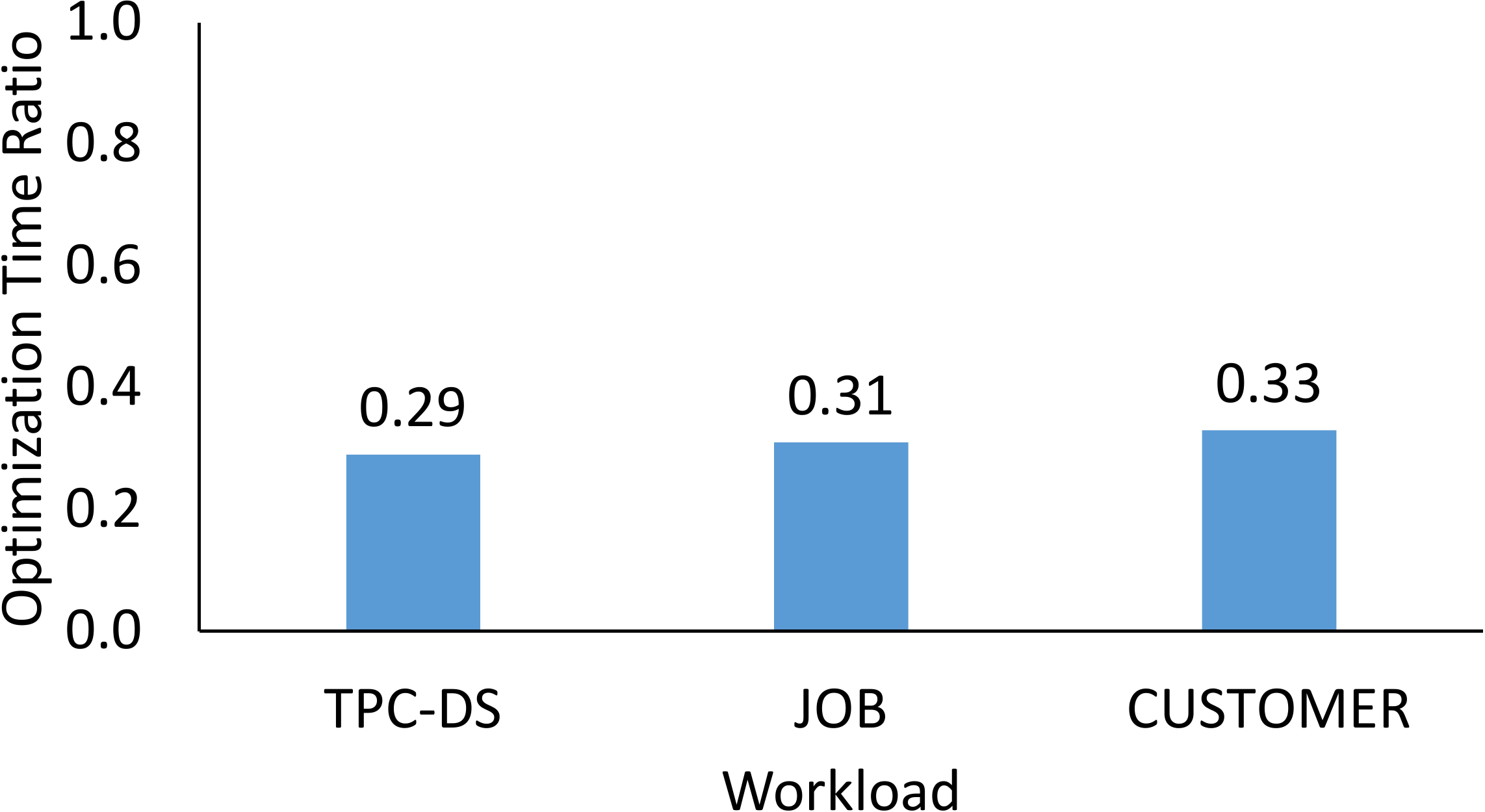}
		\caption{Normalized optimization CPU time}
		\label{fig:exp:workload_optimization}
	\end{subfigure}
	\begin{subfigure}{.33\linewidth}
		\centering
		\includegraphics[width=\linewidth]{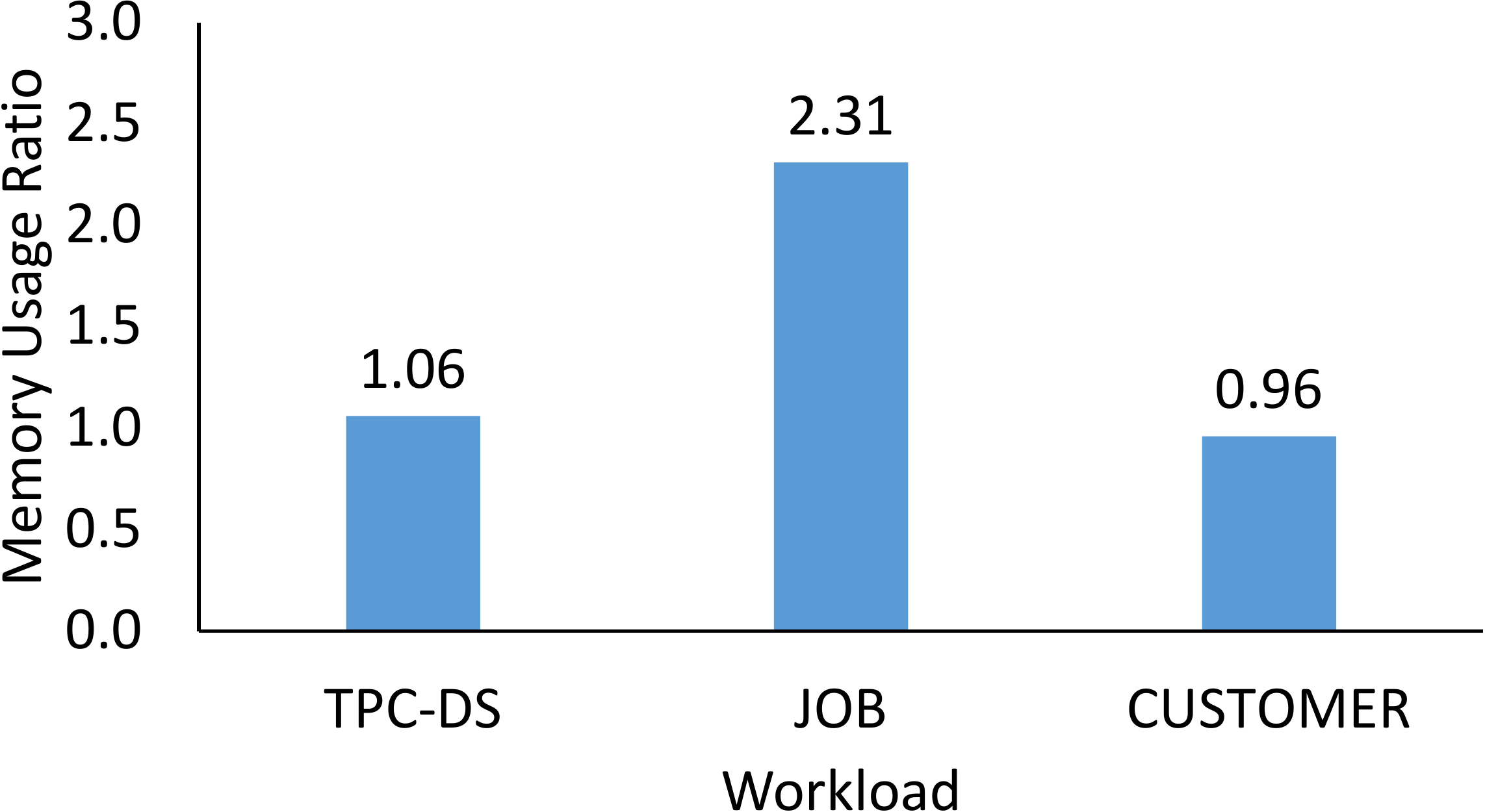}
		\caption{Normalized memory usage}
		\label{fig:exp:workload_memory}
	\end{subfigure}
\end{figure*}
}

\cut{
\begin{figure}
	\centering
	\caption{Total query execution CPU time for all queries in a workload consumed by the plans with our technique over that of the original plans}
	\label{fig:exp:workload}
	\centering
	\includegraphics[width=.9\linewidth]{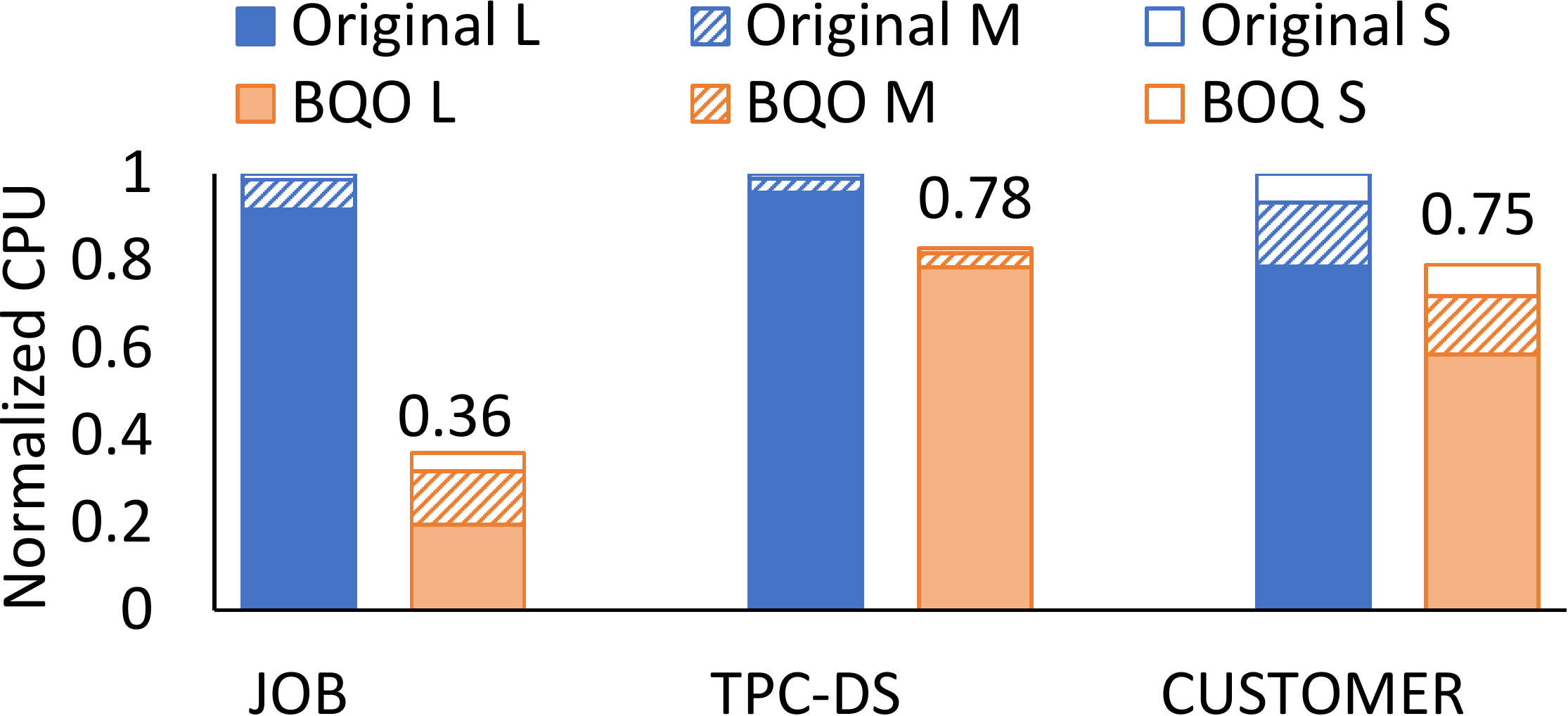}
\end{figure}

\begin{figure}
	\centering
	\caption{\changed{Number of tuples output by operators comparing the original and the bitvectore-aware plans in a workload}}
	\label{fig:exp:bitvector_tuple_ratio}
	\includegraphics[width=.9\linewidth]{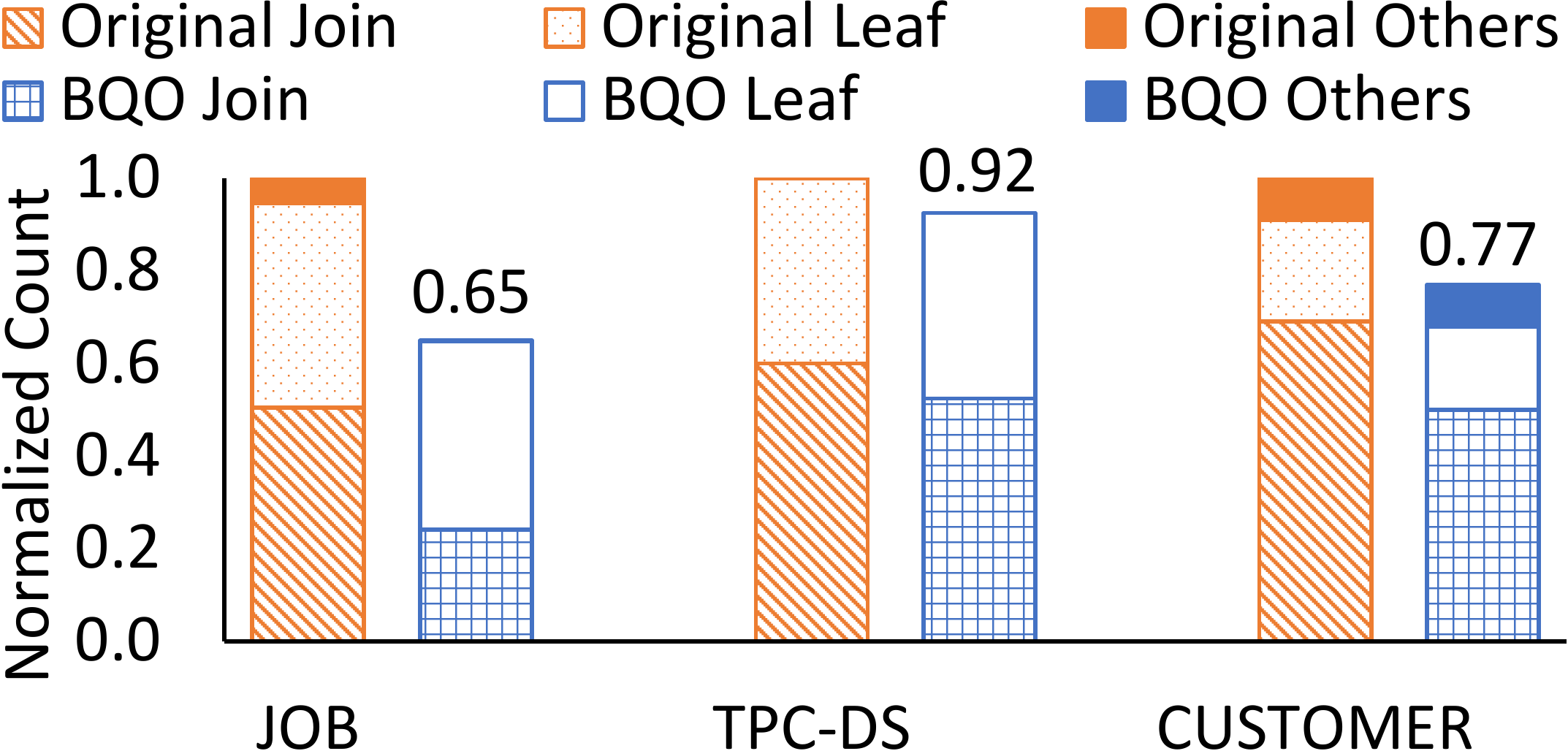}
\end{figure}
}

Figure~\ref{fig:exp:workload} shows the total amount of CPU execution time reduction with our technique. We sum up the total CPU execution time of the plans produced by \sqlserver with our technique and divide it by that of the plans produced by the original \sqlserver. On average, the total workload execution CPU time has been reduced by 37\%. We observe that workloads with more complicated decision support queries benefit more from our technique, with the highest reduction of \maxworkloadreduction in CPU execution time for \job. Since \sqlserver has been heavily tuned to optimize for these benchmarks, the degree of reductions in CPU execution time is very significant.

\outline{break down the CPU cost by query selectivity}
\changed{We break down the CPU execution cost by query types. We divide the queries into three groups based on their selectivity, i.e., high ($S$), moderate ($M$), low ($L$). We approximate the query selectivity by the execution CPU cost of the original query plans, with the cheapest $33.3\%$ queries in group $S$, the $33.3\%$ most expensive queries in group $L$, and the rest in group $M$. We showed that, our technique is especially effective in reducing CPU execution cost for expensive queries or queries with low selectivity, i.e., with execution CPU reduced by $4.8\times$ for expensive queries in \job benchmark. This is because that right deep trees is a preferable plan space for queries with low selectivities (\cite{chen1997applying, galindo2008optimizing}), and our technique produces a better join order for right deep trees.}

\outline{Compare the number of tuples output by each operator}
\point{(a) Describe what the figure shows}
\point{(b) Explain why we show the figure}
\point{(c) Describe the conclusions from this figure: both join and leaf reduces. join reduces more significantly as join order changes}
\changed{Figure~\ref{fig:exp:bitvector_tuple_ratio} shows the total number of tuples output by operators in the query plans produced by the original query optimizer (\emph{Original}) and the bitvector-aware query optimizer (\emph{BQO}), normalized by the total number of tuples output by the original query plans in each workload. We sum up the number of tuples by the type of operators, including leaf operators, join operators, and other operators. \cut{The ratio of tuples output by other operators with \emph{BQO} is too small and thus not visible in the figure.} Figure~\ref{fig:exp:bitvector_tuple_ratio} sheds some insight on the amount of logical work done by operators and thus the quality of query plans. With \emph{BQO}, both the number of tuples processed by join operators as well as leaf operators reduces.\cut{, with more significant reduction for join operators. This confirms that \emph{BQO} improves query plan quality by producing a better join order with smaller intermediate result sizes.} In particular, for \maxjoinreductionworkload benchmark, \emph{BQO} reduces the normalized number of tuples output by join operators from \maxjoinreductionworkloadbefore to \maxjoinreductionworkloadafter, i.e., a \maxjoinreduction reduction.} This again confirms that \emph{BQO} improves query plan quality by producing a better join order.

\point{Individual queries}

\begin{figure*}
	\caption{\changed{Individual query CPU time}}
	\label{fig:exp:query}
	
	\begin{subfigure}{\linewidth}
		\centering
		\includegraphics[width=\linewidth]{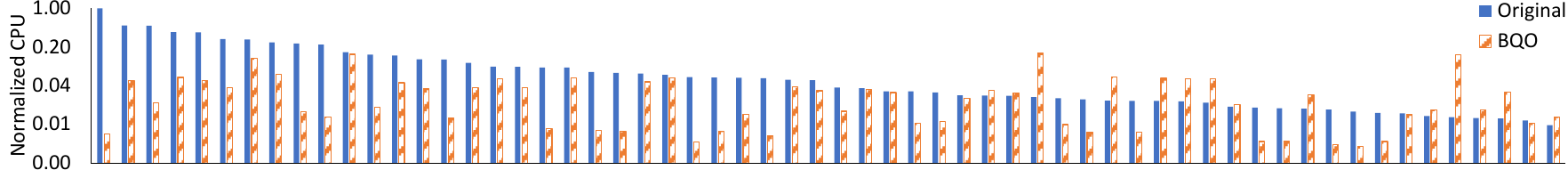}
		\caption{JOB}
		\label{fig:exp:job_pkfk_query}
	\end{subfigure}
	\begin{subfigure}{\linewidth}
		\centering
		\includegraphics[width=\linewidth]{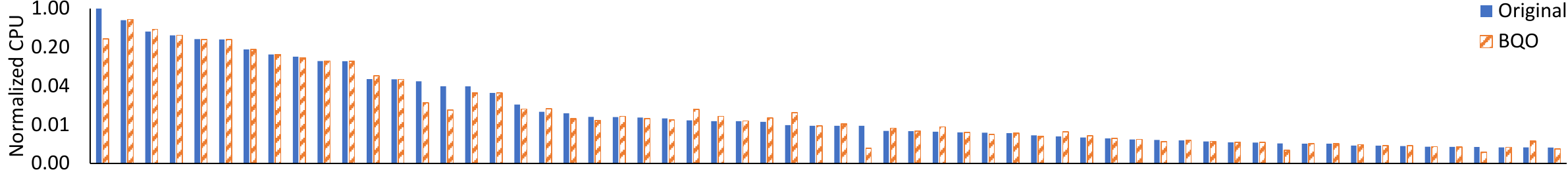}
		\caption{TPC-DS}
		\label{fig:exp:tcpds_query}
	\end{subfigure}
	\begin{subfigure}{\linewidth}
		\centering
		\includegraphics[width=\linewidth]{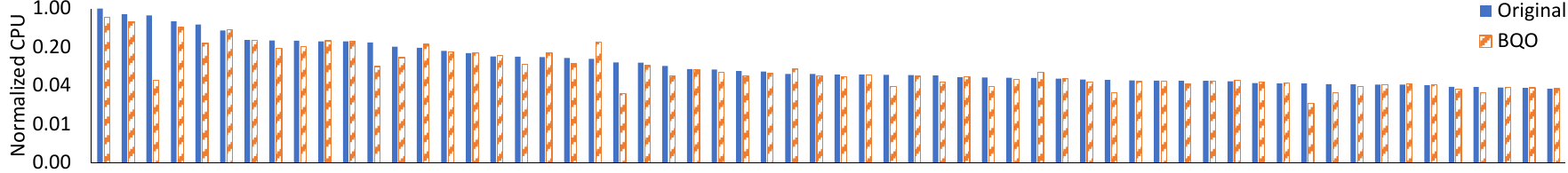}
		\caption{\mssales}
		\label{fig:exp:mssales_query}
	\end{subfigure}
\end{figure*}

Figure~\ref{fig:exp:query} shows the normalized CPU execution time for individual queries with the plans using our technique and these from the original \sqlserver. The queries are sorted by the CPU execution time of their original query plans, and the top 60 most expensive queries are shown for readability. Note that the Y axis uses a logarithmic scale.
We observe a reduction of up to two orders of magnitude in CPU execution time for individual queries. Again, Figure~\ref{fig:exp:query} confirms that our technique is especially effective in reducing the CPU execution time for expensive decision support queries.

\changed{
Our technique can improve plan quality for two reasons. First, if a query optimizer does not fully integrate bitvector filters into query optimization, it can consider the best plan with bitvector filters as 'sub-optimal' as shown in Figure~\ref{fig:bitvector_plan}. Second, due to the importance of decision support queries, many commercial DBMSs have developed dedicated heuristics to identify and optimize snowflake queries~\cite{Weininger:2002:EEJ:564691.564754, galindo2008optimizing, Antova:2014:OQO:2588555.2595640}. If these heuristics do not consider the impact of bitvector filters, they can explore a different plan space which does not even contain the plans considered by our technique.}

Inevitably, there are regressions compared with the original plans. We investigate such regressions and discover three major reasons. First, our cost function $C_{out}$ does not capture the physical information of operators and can be inaccurate. Second, our technique favors right deep trees, which can become suboptimal when the query is highly selective. Finally, our algorithm uses heuristics to extend to complex decision support queries, which can be suboptimal in some cases.

\cut{
\changed{
\subsection{Case study}
Here we show two detailed case studies of how our technique improves over our baseline.
}
}

\cut{

\subsection{Regression analysis and limitations}
Inevitably, a bitvector-aware query plan can regress compared with the original query plan. We have investigated such regressions and classify the major reasons as the following:

\paragraph{Inadequate cost function} The cost function used in the theoretical analysis and algorithms assumes the cost is proportional to the intermediate join cardinalities. While this is a good approximation for the actual execution cost function in many cases, there are operators that violate its assumptions (see Section~\ref{subsection:cost_model}). In addition, using cardinality along does not account for the implementation and execution cost of different physical operators. For example, the cost of accessing the same number of tuples from row indexes and columnstores is very different. As another example, the filter predicates can become expensive and significant for the query CPU time, e.g., n.name LIKE '\%An\%' in \job\ workload, which is completely ignored by the cost function.

Such issues can be addressed by leveraging the more accurate cost function from the underlying DBMS engine. \done{}

\paragraph{The ordering of the groups is wrong}

\paragraph{Inaccurate cardinality estimation}
}
	\section{Related Work}
\label{sec:related_work}
We discuss two lines of related work: plan search and bitvector filters.
\vspace{-.5em}
\paragraph{Plan search}
Many query optimization (QO) frameworks in DBMSs are based on either top-down~\cite{Soliman:2014:OMQ:2588555.2595637, graefe1995cascades, graefe1993volcano} or bottom-up~\cite{astrahan1976system} dynamic programming (DP). \cut{Join ordering has been one of the most important and challenging problems in query optimization. }There has been a large body of prior work on join ordering and plan space complexity analysis with such QO frameworks~\cite{Neumann:2009:QSG:1559845.1559889,moerkotte2006analysis, fender2012effective, moerkotte2008dynamic, ono1990measuring}.

Due to the importance of decision support queries, many commercial DBMSs have developed dedicated heuristics for optimizing complex decision support queries~\cite{Weininger:2002:EEJ:564691.564754, galindo2008optimizing, Antova:2014:OQO:2588555.2595640} based on the plan space of snowflake queries~\cite{Karayannidis:2002:PSQ:1287369.1287432}.

In this work, we adapt the cost function used in analyzing join order enumeration~\cite{neumann2009query, neumann2013taking} for our analysis. We analyze the space of right deep trees without cross products, which has been shown to be a favorable plan space for decision support queries and bitvector filters~\cite{galindo2008optimizing, chen1997applying, zhu2017looking}.
\vspace{-.5em}
\paragraph{Bitvector filter and its variants}
Semi-join is first introduced to reduce communication cost of distributed queries~\cite{bernstein1981using}. Efficient implementation of semi-joins have been heavily studied in the past~\cite{graefe1993query, bratbergsengen1984hashzng, valduriez1984join}. Several prior work has explored different schedules of bitvector filters for various types of query plan trees~\cite{chen1993applying, chen1997applying, chen1992interleaving}. Sideways information passing and magic sets transformation generalize the concept of bitvector filters and combines them with query rewriting~\cite{seshadri1996cost, beeri1991power}. 

Many variants of bitvector filters have also been studied in the past, such as \changed{Bloom} filters~\cite{bloom1970space}, bitvector indexes~\cite{chan1998bitvector}, cuckoo filters~\cite{fan2014cuckoo}, \changed{performance-optimal filters~\cite{lang2019performance} and others~\cite{putze2007cache,almeida2007scalable}.} The focus of this line of research is on the trade-off between space and accuracy, \changed{the efficiency of filter operations, and the extensions of Bloom filter}.

Due to the effectiveness of bitvector filters in reducing query execution cost, several commercial DBMSs have implemented bitvector filter or its variants as query processing techniques for decision support queries~\cite{galindo2008optimizing, Das:2015:QOO:2824032.2824074, Hsiao:1994:PEM:191839.191879, lahiri2015oracle}.

In this work, our analysis is based on the classic bitvector filter algorithm described in~\cite{graefe1993query}. We mainly study the interaction between bitvector filters and query optimization, which is orthogonal to the prior work on bitvector filters as query processing techniques.

Lookahead Information Passing (LIP)~\cite{zhu2017looking} is the closest prior work to our work. LIP studies the star schema where \changed{Bloom} filters created from dimension tables are all applied to the fact table. The focus is on the order of applying \changed{Bloom} filters, and they observe such query plans are robust with different permutations of dimension tables. Compared with LIP, our work systematically analyzes a much broader range of decision support queries and plan search space. Their conclusion on plan robustness can be derived from our analysis.
	\section{Conclusion}
\label{sec:conclusion}
\cut{In this work, we analyze the impact of bitvector filters on query optimization.}\cut{We show that, integrating bitvector filters naively into a top-down or bottom-up dynamic programing based query optimization framework can increase the plan space complexity by an exponential factor w.r.t. the number of relations in the query. We show that, surprisingly, for star and snowflake queries \changed{with primary key foreign key joins} and the plan search space of right deep trees without cross products, we can reduce the plan space complexity from exponential to linear w.r.t. the number of relations in the query with some simplifying assumptions. Motivated by our analysis, we propose an algorithm to optimize the \changed{join order} to decision support queries \changed{with arbitrary join graphs}. \changed{We integrate our algorithm to a commercial DBMS.} Our evaluation shows that, compared with the plans produced by a commercial database \sqlserver, our technique reduces the CPU execution time of benchmarks by \avgworkloadreduction in average, with max workload-level CPU execution time reduction of \maxworkloadreduction, and up to two orders of magnitude CPU execution time reduction for individual queries.}

In this work, we systematically analyze the impact of bitvector filters on query optimization. Based on our analysis, we propose an algorithm to optimize the join order for arbitrary decision support queries. Our evaluation shows that, instead of using bitvector filters only as query processing techniques, there is great potential to improve query plan quality by integrating bitvector filters into query optimization for commercial databases.

This work is the first step to understand the interaction between bitvector filters and query optimization, and it opens new opportunities for query optimization with many open challenges. \changed{Extending the analysis to additional plan space, query patterns, operators beyond hash joins, and more complex cost modeling is challenging. Efficient full integration of bitvector filters for commercial databases with various architectures remains an open problem. Since our analysis shows that bitvector filters result in more robust query plans, which is also observed in~\cite{zhu2017looking}, understanding how bitvector filters impact robust and interleaved query optimization is also an interesting direction.}

	\appendix
\changed{
\section{Additional evaluation}
\label{sec:bitvector_effectiveness}

\outline{Bitmap vs. no bitmap}

\begin{table}[t]
\caption{\changed{Query plans with and without bitvector filters}}
\label{table:bitvector_effectiveness}
\begin{tabular}{|m{1.7cm}|m{.7cm}|m{1.9cm}|m{1.2cm}|m{1.25cm}|}
\hline
Workload & CPU ratio & Ratio of quer-ies w/ bitvector filters & Improved queries & Regressed queries \\
\hline
JOB      & 0.20      & 0.97                       & 0.58             & 0.00              \\
TPC-DS   & 0.53      & 0.98                       & 0.88             & 0.00              \\
CUSTOMER & 0.90      & 1.00                       & 0.42             & 0.00             \\
\hline
\end{tabular}
\end{table}

\point{Experiment setting}
We evaluate the effectiveness of bitvector filters by executing the same query plan with and without bitvector filtering. We use the original \sqlserver to produce a query plan $p$ with bitvector filters. \sqlserver provides an option to ignore bitvector filters during query processing. For comparison, we execute the same plan $p$ with bitvector filters ignored.

\point{Explain the numbers in the table}
Table~\ref{table:bitvector_effectiveness} shows the performance of the plans with and without bitvector filters for the three benchmarks. At a workload level, using bitvector filters reduces the execution CPU cost by $10\% - 80\%$ (\emph{CPU ratio}). In addition, for $97\%-100\%$ of the queries (\emph{Ratio of queries w/ bitvectof filters}), the original query plan uses bitvector filters. At an individual query level, $48\%-88\%$ of the queries has CPU execution cost reduced by more than $20\%$ (\emph{Improved queries}), with no regression on CPU execution cost by more than $20\%$ (\emph{Regressed queries}).

This confirms that bitvector filtering is a widely applicable query processing technique, and thus bitvector-aware query optimization can potentially impact a wide range of queries.
}

\section{Additional proofs}
\label{sec:moreproofs}

\lemmastartrees*

\vspace{-.5em}
\begin{proof}
	Assume $X_0=R_i, X_1=R_j, i\neq 1, j\neq 1$. Then $R_i$ and $R_j$ do not have a join condition based on Definition~\ref{def:star_query}. Thus, $T(X_0, X_1)$ has a cross product, which is a contradiction.
	
	If $X_0=R_0$ or $X_1=R_0$, since $R_0$ joins with $R_1, \cdots, R_n$, then $\mathcal{T}=T(X_0, X_1, \cdots, X_n)$ does not contain any cross product.
	
\end{proof}
\vspace{-.5em}

\lemmastarabsorption*

\vspace{-.5em}
\begin{proof}
	By Property~\ref{property:associativity}, $R_0/(R_1,R_2)=(R_0/R_1)/R_2$.
	Since $R_0\to R_1$, by Lemma~\ref{rule:absorption}, $R_0/R_1=\prod_{R_0} (R_0 \Join R_1)$. 
	Since $R_0\to R_2$, $R_0 \Join R_1\to R_2$.
	By applying Lemma~\ref{rule:absorption} again, we have $(R_0 \Join R_1) / R_2=\prod_{R_0}(R_0 \Join R_1 \Join R_2)$.
	Thus, $R_0/(R_1,R_2)=\prod_{R_0} (R_0 \Join R_1 \Join R_2)$.
	
	By induction, we can prove $R_0/(R_1,R_2,\cdots, R_n)=\prod_{R_0} (R_0 \Join R_1 \Join \cdots \Join R_n)$.	
\end{proof}
\vspace{-.5em}

\lemmastarmincostrightmosttwo*

\vspace{-.5em}
\begin{proof}
	Because $R_1, \cdots, R_n$ only connects to $R_0$, the bitvector filters created from $R_1, \cdots, R_{k-1}, R_{k+1}, \cdots, R_n$ will be pushed down to $R_0$, and the bitvector created from $R_0$ will be pushed down to $R_k$. Thus, $C_{out}(R_0)=|R_0/(R_1, \cdots, R_{k-1}, R_{k+1}, \cdots, R_n)|$. Let $R_0'=R_0/(R_1, \cdots, R_{k-1}, R_{k+1}, \cdots, R_n)$, then $C_{out}(R_k)=|R_k/R_0'|$.
	
	By Lemma~\ref{lemma:star_absorption} and Property~\ref{property:redundancy}, $|S(R_k/R_0', R_0', X_1, X_2, \cdots, \break X_k)|=|S(R_0, R_1, \cdots, R_n)|$. Thus, the total cost of the plan is $C_{out}(T(R_k, R_0, X_1, X_2, \cdots, X_{n-1}))=\sum_{i=1, i\neq k}^{n}|R_i|+C_{out}(R_0)+C_{out}(R_k)+(n-1)\cdot |S(R_0, R_1, \cdots, R-n)|$. Thus, $C_{out}(T(R_k, R_0, X_1, X_2, \cdots, X_{n-1}))$ is the same for every permutation $X_1, X_2, \cdots, X_{n-1}$ of $R_1, R_2, \cdots, R_{k-1}, R_{k+1}, \cdots, R_n$.
	
\end{proof}
\vspace{-.5em}

\lemmasnowflakebitvectorpushdown*

\vspace{-.5em}
\begin{proof}
	Because $\mathcal{T}$ is partially ordered, for every relation $X_k=R_{i,j}, j>1$, there exists one and only one relation $X_p, p<k$ such that $X_k$ connects to $X_p$. Thus, the bitvector filter created from $X_k$ will be pushed down to $X_p$. If $X_k=R_{i,1}$, it only connects to $R_0$. Thus, the bitvector filter created from $X_k$ will be pushed down to $R_0$.
\end{proof}
\vspace{-.5em}

\lemmasnowflakequalcostrightdeeptree*

\vspace{-.5em}
\begin{proof}
	Consider the bitvector filters created in both $\mathcal{T}$ and $\mathcal{T'}$. BY Lemma~\ref{lemma:snowflake:bitvector_pushdown}, the bitvector filters created from $\mathcal{T}$ and $\mathcal{T'}$ from the same relation $R_{i, j}$ will be pushed down to the same relation $R_{i,j-1}$ if $j>1$ or $R_0$ if $j=1$. 
	
	Since, $S(R_{i,n_i})$ is the same in $\mathcal{T}$ and $\mathcal{T'}$. By induction, we can show that $S(R_{i,j})$ is the same in $\mathcal{T}$ and $\mathcal{T'}$. Since, $S(R_0)=R_0/(S(R_{1,1}), S(R_{2,1}), \cdots, S(R_{n,1}))$, $S(R_0)$ is the same in $\mathcal{T}$ and $\mathcal{T'}$.
	
	Now consider the join cardinality in $\mathcal{T}$ and $\mathcal{T'}$. By Lemma~\ref{lemma:star_absorption}, $S(R_{i,j})=S(R_{i,j}, R_{i,j+1}, \cdots, R_{i, n_i})$. Thus, $S(R_0)=R_0/(S(R_{1,1}), S(R_{2,1}), \cdots, S(R_{n,1}))=S(R_0,R_{1,1}, R_{1,2}, \cdots, R{1,n_1}, R{2,1}, \cdots, R_{m,1}, R_{m,2}, \cdots, R_{m,n_m})$. Thus, $S(R_0, X_1, X_2, \cdots, X_u)=S(R_0), 1\leq u\leq n$ and $S(R_0, Y_1, Y_2, \cdots, Y_v)=S(R_0), 1\leq v\leq n$. Thus, $C_{out}(S(R_0, X_1, \cdots, X_u))=C_{out}(S(R_0, Y_1, \cdots, Y_v)), 1\leq u, v\leq n$.
	
	Thus, $C_{out}(\mathcal{T})=C_{out}(\mathcal{T'})$.
	
\end{proof}

\lemmachaintreenocp*

\vspace{-.5em}

\begin{proof}
	If $R_n$ is the right most leaf and there is no cross product in the query plan, then $R_n$ can only join with $R_{n-1}$. Thus, the right most subplan with two relations is $T_2=T(R_n, R_{n-1})$. Similarly, if the right most subplan is $T_k=T(R_n, R_{n-1}, \cdots, R_{n-k+1})$ and there is no cross product, then $T_k$ can only join with $R_{n-k}$. By induction, $T(R_n, R_{n-1}, \cdots, R_0)$ is the only right deep tree without cross products where $R_n$ is the right most leaf.
\end{proof}
\vspace{-.5em}

\lemmachainpushdownone*

\vspace{-.5em}
\begin{proof}
	Since there is no cross product in $\mathcal{T}=T(X_0, X_1,\cdots, X_{k-1}, R_n, X_{k+1}, \cdots, X_n)$, one relation in $\mathcal{A}=\{X_0, X_1, \cdots, X_{k-1}\}$ must connect to $R_n$. Since $R_{n-1}$ is the only relation that connects to $R_n$ in the join graph, $R_{n-1}\in \{X_0, X_1, \cdots, X_{k-1}\}$. By induction, we can show that $ R_{n-2}, R_{n-3},\cdots, R_{n-k} \in \mathcal{A}$. Thus, $\mathcal{A}=\{R_{n-k}, R_{n-k+1}, \cdots, R_{n-1}\}$.
	
	If $X_{k-1}\neq R_{n-1}$, then $X_{k-1}=R_{n-k}$; otherwise, the join graph of $X_0, X_1, \cdots, X_{k-2}$ is not connected and the subplan $T(X_0, X_1, \cdots, X_{k-2})$ has cross products.
	
	Now consider the relations $\{X_{k+1}, X_{k+2}, \cdots, X_n\}$. Because $X_{k+1}$ joins with $\{X_0, X_1,\cdots, X_k\}=\{R_n, R_{n-1}, \cdots, R_{n-k}\}$, $X_{k+1}=R_{n-k-1}$. Similarly, we can show that $X_i=R_{n-i}$ for $k<i\leq n$.
	
	If we swap $R_n$ and $R_{n-k}$, we get a new plan $\mathcal{T'}=T(X_0, X_1, \cdots, X_{k-2}, R_n, R_{n-k}, X_{k+1}, \cdots, X_n)$. Because $\{X_0, X_1, \cdots, X_{k-2}\}=\mathcal{A}\setminus \{R_{n-k}\}=\{R_{n-k+1}, R_{n-k+2}, \cdots, R_{n-1}\}$. Thus, $\mathcal{T'}$ has no cross product.
	
	Now we prove $C_{out}(\mathcal{T'})\leq C_{out}(\mathcal{T})$.
	
	\cut{
		First, consider $X_i$ for $k< i\leq n$. Because there is no change from the root to $X_{k+1}$, the bitvector filters created from $X_{k+1}, \cdots, X_n$ and the bitvector filters pushed down to $X_{k+1}, \cdots, X_n$ are the same. Thus, $C_{out}(X_i)$ is the same for $\mathcal{T}$ and $\mathcal{T'}$.
	}
	
	First, consider $X_i$ for $k< i\leq n$. Since there is no change in bitvector filters, it is easy to see that $C_{out}(X_i)$ is the same for $\mathcal{T}$ and $\mathcal{T'}$.
	
	\cut{
		Next, consider $X_i$ for $0\leq i < k$. Since $\mathcal{B}=\{X_0, X_1, \cdots, X_{k-1}\}=\{R_{n-k+1}, \cdots, R_{n-1}\}$, only $R_{n-k}$ and $R_n$ will create bitvector filters that can be pushed down to subplans of $\mathcal{B}$. Because $R_{n-1}\in \mathcal{B}$, no bitvector filter will be pushed down to $R_n$. Thus, the bitvector filter created from $R_n$ is the same for $\mathcal{T}$ and $\mathcal{T'}$, and the same bitvector filter will be pushed down to $R_{n-1}$ the same way in $\mathcal{T}$ and $\mathcal{T'}$. Similarly, the bitvector filters pushed down to and created from $R_{n-k}$ are the same in $\mathcal{T}$ and $\mathcal{T'}$, and the bitvector filters created from $R_{n-k}$ will be pushed down to $R_{n-k+1}$ the same way in $\mathcal{T}$ and $\mathcal{T'}$.
	}
	
	Next, consider $X_i$ for $0\leq i < k$. Since $\mathcal{B}=\{X_0, X_1, \cdots, X_{k-1}\}=\{R_{n-k+1}, \cdots, R_{n-1}\}$, only $R_{n-k}$ and $R_n$ will create bitvector filters that can be pushed down to subplans of $\mathcal{B}$. Because $R_{n-1}\in \mathcal{B}$, no bitvector filter will be pushed down to $R_n$. Thus, the bitvector filter created from $R_n$ is the same for $\mathcal{T}$ and $\mathcal{T'}$, and the same bitvector filter will be pushed down to $R_{n-1}$ the same way in $\mathcal{T}$ and $\mathcal{T'}$. Similarly, the bitvector filters created from and pushed down to $R_{n-k}$ and $R_{n-k+1}$ are the same in $\mathcal{T}$ and $\mathcal{T'}$.
	
	Thus, we have proved $C_{out}(X_i)$ is the same for $0\leq i\leq n$.
	
	Next, we show that the intermediate join sizes in $\mathcal{T'}$ is equal to or smaller than these in $\mathcal{T}$.
	
	Since $\mathcal{T}$ and $\mathcal{T'}$ share the same subplan $\mathcal{T}_{j}=T(X_0, X_1, \cdots, X_j), 0\leq j\leq k-2$, and we have shown the bitvector filters pushed down to $\mathcal{T}_{j}$ is the same in $\mathcal{T}$ and $\mathcal{T'}$, the intermediate join sizes are the same in $\mathcal{T}_{j}$ for both plans.
	
	Consider the cardinalities of the join $S(T_{k-2}, R_{n-k})$ in $\mathcal{T}$ and $S(T_{k-2}, R_n)$ in $\mathcal{T'}$. Since $R_\{n-1\} \in\mathcal{A}$, $S(T_{k-2}, R_n)$ is a PKFK join. By \absorption,  $|S(T_{k-2}, R_n)|=|T_{k-2}|$. Since $R_{n-k-1}=X_{k+1} \notin \mathcal{A}$, $S(T_{k-2}, R_{n-k})$ is not a PKFK join. By \reduction, $|T_{k-2} \Join R_{n-k}|\geq |T_{k-2}|$. Thus, $|T_{k-2} \Join R_n|=|T_{k-2}|\leq |T_{k-2} \Join R_{n-k}|$.
	
	Now consider the cardinalities for $S(T_{k-2}, R_{n-k}, R_n)$ in $\mathcal{T}$ and $S(T_{k-2}, R_n, R_{n-k})$ in $\mathcal{T'}$. Since the set of bitvector filters from $B$ pushed down to $S(T_{k-2}, R_{n-k}, R_n)$ is the same as those pushed down to $S(T_{k-2}, R_n, R_{n-k})$ and the join relations are the same, $|S(T_{k-2}, R_{n-k}, R_n)|=|S(T_{k-2}, R_n, R_{n-k})|$. Similarly, we can show that $|S(T_{k-2}, R_{n-k}, R_n, X_{k+1}, \cdots, X_i)| =|S(T_{k-2}, R_{n}, R_{n-k}, X_{k+1}, \cdots, X_i)|$ for $k+1\leq i\leq n$.
	
	Thus, $C_{out}(\mathcal{T})=\sum_{i=1}^n C_{out}(X_i)+\sum_{i=0, i\neq k-1, k}^n |S(X_0, \cdots, X_i)|+|S(T_{k-2}, R_{n-k})| +|S(T_{k-2}, R_{n-k}, R_n)| \geq \sum_{i=1}^n C_{out}(X_i)+\sum_{i=0, i\neq k-1, k}^n |S(X_0, \cdots, X_i)|+|S(T_{k-2}, R_n)| +|S(T_{k-2}, R_n, R_{n-k})|=C_{out}(\mathcal{T'})$.
	
\end{proof}
\vspace{-.5em}

\lemmachainpushdowntwo*

\vspace{-.5em}
	\begin{proof}
	Similar to the proof of Lemma~\ref{lemma:chain_pushdown1}, we can show that $X_{k-m-1}=R_{n-k}$ if $X_{k-m-1}\neq R_{n-m-1}$, $\mathcal{A}=\{X_{0}, X_{1}, \cdots, X_{k-m-1}\}=\{R_{n-k}, R_{n-k+1}, \cdots, R_{n-m-1}\}$, and $X_i=R_{n-i}$ for $k<i\leq n$.
	
	Now consider swapping $R_n, R_{n-1}, \cdots, R_{n-m}$ with $R_{n-k}$, the resulting plan is $\mathcal{T'}$. Similar to the proof of Lemma~\ref{lemma:chain_pushdown1}, we can show that $\mathcal{T'}$ has no cross product.
	
	Consider $C_{out}$ for $X_0, X_1, \cdots, X_n$. Similar to the proof of Lemma~\ref{lemma:chain_pushdown1}, we can show that $C_{out}(X_i)$ is the same for $X_0, X_1, \cdots, X_n$ in $\mathcal{T}$ and $\mathcal{T'}$.
	
	Next, consider the intermediate join sizes. Since both $\mathcal{T}$ and $\mathcal{T'}$ share the same subplan $\mathcal{T}_{j}(X_0, X_1, \cdots, X_j), 0\leq j\leq k-m-2$, similar to the proof of  Lemma~\ref{lemma:chain_pushdown1}, we can show that $\mathcal{T}_{j}, 0\leq j\leq k-m-2$ is the same for $\mathcal{T}$ and $\mathcal{T'}$.
	
	Now consider the cardinality of joins $S(\mathcal{T}_{k-m-2}, R_{k-m-1})$ and $S(\mathcal{T}_{k-m-2}, R_{n-k})$, similar to Lemma~\ref{lemma:chain_pushdown1}, we can show $|S(\mathcal{T}_{k-m-2}, R_{k-m-1})| \leq S(\mathcal{T}_{k-m-2}, R_{n-k})$.
	
	Now consider the cardinality of joins $S(\mathcal{T}_{k-m-2}, R_{k-m-1}, R_{k-m})$ and $S(\mathcal{T}_{k-m-2}, R_{n-k}, R_{k-m-1})$. Since $R_{k-m-2}$ is a PKFK join with $S(\mathcal{T}_{k-m-2}, R_{k-m-1})$, $|S(\mathcal{T}_{k-m-2}, R_{k-m-1}, R_{k-m})|=S(\mathcal{T}_{k-m-2}, R_{k-m-1}$. Similarly, since $R_{k-m-1}$ is a PKFK join with $S(\mathcal{T}_{k-m-2}, R_{n-k}, R_{k-m-1})$, we have $|S(\mathcal{T}_{k-m-2}, R_{n-k}, R_{k-m-1})|=|S(\mathcal{T}_{k-m-2}, R_{n-k})|$. Thus, $|S(\mathcal{T}_{k-m-2}, R_{k-m-1}, R_{k-m})|\leq S(\mathcal{T}_{k-m-2}, R_{n-k}, R_{k-m-1})$.
	
	By similar reasoning, we can show that $|S(\mathcal{T}_{k-m-2}, R_{k-m-1}, R_{k-m}, \cdots, R_j)|\leq |S(\mathcal{T}_{k-m-2}, R_{n-k}, R_{k-m-1}, R_{k-m}, \cdots, R_{j-1})|, k-m-1\leq j\leq n$.
	
	Finally, we can show that $|S(\mathcal{T}_{k-m-2}, R_{k-m-1}, R_{k-m}, \cdots, R_n, R_{n-k})|=|S(\mathcal{T}_{k-m-2}, R_{n-k}, R_{k-m-1}, R_{k-m}, \cdots, R_n)|$ and $|S(\mathcal{T}_{k-m-2}, R_{k-m-1}, R_{k-m}, \cdots, R_n, R_{n-k}, R_{n-k-1}, \cdots, R_j)|=|S(\mathcal{T}_{k-m-2}, R_{n-k}, R_{k-m-1}, R_{k-m}, \cdots, R_n, R_{n-k-1}, \cdots, R_j)|, 0\leq j\leq n-k-1$ as in Lemma~\ref{lemma:chain_pushdown1}.
	
	By summing up everything together, we have $C_{out}(\mathcal{T})\geq C_{out}(\mathcal{T'})$.
	
\end{proof}
\vspace{-.5em}

\lemmasnowflakesinglebranch*

\vspace{-.5em}
\begin{proof}
	Assume there exists $X_u=R_{i_1,j_1}$ and $X_v=R_{i_2,j_2}$ such that $0\leq u, v\leq k-1$ and $i_1\neq i_2$. Because $X_k=R_0$, $X_u$ does not connect to $X_v$ by joining with $X_0, X_1, \cdots, X_{k-1}$. Thus, there must be a cross product, \changed{which is a contradiction.}\cut{ This contradicts that $\mathcal{T}$ has no cross product. So $\{X_0, X_1, \cdots, X_{k-1}\}\subseteq \mathcal{R_i}$ for some $1\leq i\leq m$.}

Since $X_0, X_1, \cdots, X_{k-1}$ has a join condition with $R_0$, $R_{i,1}\in \{X_0, X_1, \cdots, X_{k-1}\}$. Because $T(X_0, X_1, \cdots, X_{k-1})$ has no cross product, $\{X_0, X_1, \cdots, X_{k-1}\}=\{R_{i,1}, R_{i,2}, \cdots, R_{i,k}\}$. Thus, $X_{0}, X_{1}, \cdots, X_{k-1}$ is a permutation of $R_{i, 1}, R_{i, 2}, \cdots, R_{i, k}$. 
\end{proof}
\vspace{-.5em}

\lemmasnowflakepartiallyorderedtreetwo*

\vspace{-.5em}
\begin{proof}
	\cut{We first show that the join graph of $\{R_0', X_{k+1}, X_{k+2}, \cdots, X_n\}$ is a snowflake as defined in Definition~\ref{def:snowflake_query}.
	}
	By Lemma~\ref{lemma:snowflake:single_branch}, $X_{0}, X_{1}, \cdots, X_{k-1}$ is a permutation of $R_{i, 1}, R_{i, 2}, \cdots, R_{i, k}$ for some $1\leq i\leq m$. Let's create a new relation $R_0'=Join(X_0, X_1, \cdots, X_{k-1}, R_0)$. For $X_j, k<j\leq n$, if $X_j=R_{i,k+1}$, $R_{i,k}\to X_j$ and thus $R_0' \to X_j$; if $X_j=R_{u, 1}$, $R_0 \to X_j$ and thus $X_j\to R_0'$; if $X_j=R_{u, v}, v>1$, then there exists $R_{u, v-1}\in \{X_{k+1}, X_{k+2}, \cdots, X_n\}$ such that $R_{u,v-1}\to X_j$. Thus, $\{R_0', X_{k+1}, X_{k+2}, \cdots, X_n\}$ is a snowflake query. By Lemma~\ref{lemma:snowflake:right_deep_tree_nocp}, and $X_{k+1}, X_{k+2}, \cdots, X_n$ is a partially ordered right deep tree of the new snowflake query.
\end{proof}
\vspace{-.5em}

\lemmasnowflakecostreductionpushdown*

\vspace{-.5em}
\begin{proof}
		By Lemma~\ref{lemma:snowflake:partially_ordered_tree2}, $\mathcal{T}$ is a partially-ordered subtree. Let $\mathcal{T}_p=T(X_0, X_1, \cdots, X_{k-1}, R_0, R_{i,k+1}, R_{i,k+2}, \cdots, R_{i,n_i}, Y_1, Y_2, \break\cdots, Y_{n-n_i-1})$, where $Y_1, Y_2,\cdots, Y_{n-n_i-1}$ is a permutation of $\mathcal{A}=\{X_{k+1}, X_{k+2}, \cdots, X_n\}\setminus \{R_{i,k+1}, R_{i,k+2},\cdots, R_{i,n_i}\}$, and $Y_1, Y_2, \cdots, Y_{n-n_i-1}$ is partially ordered. By Theorem~\ref{lemma:snowflake:equal_cost_right_deep_tree}, $C_{out}(\mathcal{T}_p)=C_{out}(\mathcal{T})$.
		
		Now consider $\mathcal{T'}= T(X_0, X_1, \cdots,  X_{k-1}, R_{i,k+1}, R_{i,k+2},\break \cdots, R_{i,n_i}, R_0, Y_1, Y_2, Y_{n-n_i-1})$. Let $R_0'=R_0/(Y_1, Y_2, \cdots, \break Y_{n-n_i-1})$. Since $X_0, X_1, \cdots, X_{k-1}$ is a permutation of $\{R_{i,1}, R_{i,2},\cdots, R_{i,k}\}$, joining $\{X_0, X_1, \cdots, X_{k-1}, R_{i,k+1},\break  R_{i,k+2}, \cdots, R_{i,n_i},  R_0'\}$ is a branch of a snowflake. By Lemma~\ref{lemma:chain_pushdown2}, $C_{cout}(T(X_0, X_1, \cdots, X_{k-1}, R_{i,k+1}, R_{i,k+2},\cdots, R_{i,n_i}, R_0'))\leq C_{cout}(T(X_0, X_1, \cdots, X_{k-1}, R_0', R_{i,k+1}, R_{i,k+2}, \cdots, R_{i,n_i}))$.
		
		Consider $\mathcal{T'}$ and $\mathcal{T}_p$. Because $\mathcal{T}_p$ is a partially-ordered subtree, $C_{out}(\mathcal{T}_p)=C_{out}(T(X_0, X_1, \cdots, X_{k-1}, \break R_0', R_{i,k+1}, R_{i,k+2}),\cdots, R_{i,n_i})+\sum_{j=1}^{n-n_i-1}C_{out}(Y_j)+(n-n_i-1)\cdot |S(R_0, R_{i,1}, R_{i,2}, \cdots, R_{i, n_i}, Y_1, Y_2, \cdots, Y_{n-n_i-1})|$. Thus, $C_{out}(\mathcal{T}_p)\geq C_{out}(T(X_1, X_2, \cdots, X_{k-1}, R_{i,k+1}, R_{i,k+2} ,\cdots,\break R_{i,n_i}, R_0'))+(n-n_i-1)\cdot |S(R_0, R_{i,1}, R_{i,2}, \cdots, R_{i, n_i}, Y_1, Y_2, \cdots, \break Y_{n-n_i-1})|=C_{out}(\mathcal{T'})$.
		
		Thus, $C_{out}(\mathcal{T'})\leq C_{out}(\mathcal{T}_p)=C_{out}(\mathcal{T})$.
	\end{proof}
	\vspace{-.5em}

	
	\bibliographystyle{ACM-Reference-Format}
	\bibliography{reference}


\begin{thebibliography}{38}


\ifx \showCODEN    \undefined \def \showCODEN     #1{\unskip}     \fi
\ifx \showDOI      \undefined \def \showDOI       #1{#1}\fi
\ifx \showISBNx    \undefined \def \showISBNx     #1{\unskip}     \fi
\ifx \showISBNxiii \undefined \def \showISBNxiii  #1{\unskip}     \fi
\ifx \showISSN     \undefined \def \showISSN      #1{\unskip}     \fi
\ifx \showLCCN     \undefined \def \showLCCN      #1{\unskip}     \fi
\ifx \shownote     \undefined \def \shownote      #1{#1}          \fi
\ifx \showarticletitle \undefined \def \showarticletitle #1{#1}   \fi
\ifx \showURL      \undefined \def \showURL       {\relax}        \fi
\providecommand\bibfield[2]{#2}
\providecommand\bibinfo[2]{#2}
\providecommand\natexlab[1]{#1}
\providecommand\showeprint[2][]{arXiv:#2}

\bibitem[\protect\citeauthoryear{??}{tpc}{2012}]%
        {tpcds}
 \bibinfo{year}{2012}\natexlab{}.
\newblock \bibinfo{booktitle}{\emph{{TPC-DS}}}.
\newblock
\urldef\tempurl%
\url{http://www.tpc.org/tpcds/}
\showURL{%
\tempurl}


\bibitem[\protect\citeauthoryear{Almeida, Baquero, Pregui{\c{c}}a, and
  Hutchison}{Almeida et~al\mbox{.}}{2007}]%
        {almeida2007scalable}
\bibfield{author}{\bibinfo{person}{Paulo~S{\'e}rgio Almeida},
  \bibinfo{person}{Carlos Baquero}, \bibinfo{person}{Nuno Pregui{\c{c}}a},
  {and} \bibinfo{person}{David Hutchison}.} \bibinfo{year}{2007}\natexlab{}.
\newblock \showarticletitle{Scalable bloom filters}.
\newblock \bibinfo{journal}{\emph{Inform. Process. Lett.}}
  \bibinfo{volume}{101}, \bibinfo{number}{6} (\bibinfo{year}{2007}),
  \bibinfo{pages}{255--261}.
\newblock


\bibitem[\protect\citeauthoryear{Antova, El-Helw, Soliman, Gu, Petropoulos, and
  Waas}{Antova et~al\mbox{.}}{2014}]%
        {Antova:2014:OQO:2588555.2595640}
\bibfield{author}{\bibinfo{person}{Lyublena Antova}, \bibinfo{person}{Amr
  El-Helw}, \bibinfo{person}{Mohamed~A. Soliman}, \bibinfo{person}{Zhongxian
  Gu}, \bibinfo{person}{Michalis Petropoulos}, {and} \bibinfo{person}{Florian
  Waas}.} \bibinfo{year}{2014}\natexlab{}.
\newblock \showarticletitle{Optimizing Queries over Partitioned Tables in MPP
  Systems}. In \bibinfo{booktitle}{\emph{Proceedings of the 2014 ACM SIGMOD
  International Conference on Management of Data}}
  \emph{(\bibinfo{series}{SIGMOD'14})}. \bibinfo{publisher}{Association for
  Computing Machinery}, \bibinfo{address}{New York, NY, USA},
  \bibinfo{pages}{373--384}.
\newblock
\showISBNx{9781450323765}
\urldef\tempurl%
\url{https://doi.org/10.1145/2588555.2595640}
\showDOI{\tempurl}


\bibitem[\protect\citeauthoryear{Astrahan, Blasgen, Chamberlin, Eswaran, Gray,
  Griffiths, King, Lorie, McJones, Mehl, and et~al.}{Astrahan
  et~al\mbox{.}}{1976}]%
        {astrahan1976system}
\bibfield{author}{\bibinfo{person}{M.~M. Astrahan}, \bibinfo{person}{M.~W.
  Blasgen}, \bibinfo{person}{D.~D. Chamberlin}, \bibinfo{person}{K.~P.
  Eswaran}, \bibinfo{person}{J.~N. Gray}, \bibinfo{person}{P.~P. Griffiths},
  \bibinfo{person}{W.~F. King}, \bibinfo{person}{R.~A. Lorie},
  \bibinfo{person}{P.~R. McJones}, \bibinfo{person}{J.~W. Mehl}, {and}
  \bibinfo{person}{et al.}} \bibinfo{year}{1976}\natexlab{}.
\newblock \showarticletitle{System R: Relational Approach to Database
  Management}.
\newblock \bibinfo{journal}{\emph{ACM Trans. Database Syst.}}
  \bibinfo{volume}{1}, \bibinfo{number}{2}, \bibinfo{pages}{97--137}.
\newblock
\showISSN{0362-5915}
\urldef\tempurl%
\url{https://doi.org/10.1145/320455.320457}
\showDOI{\tempurl}


\bibitem[\protect\citeauthoryear{Beeri and Ramakrishnan}{Beeri and
  Ramakrishnan}{1991}]%
        {beeri1991power}
\bibfield{author}{\bibinfo{person}{Catriel Beeri} {and} \bibinfo{person}{Raghu
  Ramakrishnan}.} \bibinfo{year}{1991}\natexlab{}.
\newblock \showarticletitle{On the power of magic}.
\newblock \bibinfo{journal}{\emph{The journal of logic programming}}
  \bibinfo{volume}{10}, \bibinfo{number}{3-4}, \bibinfo{pages}{255--299}.
\newblock


\bibitem[\protect\citeauthoryear{Bernstein and Chiu}{Bernstein and
  Chiu}{1981}]%
        {bernstein1981using}
\bibfield{author}{\bibinfo{person}{Philip~A. Bernstein} {and}
  \bibinfo{person}{Dah-Ming~W. Chiu}.} \bibinfo{year}{1981}\natexlab{}.
\newblock \showarticletitle{Using Semi-Joins to Solve Relational Queries}.
\newblock \bibinfo{journal}{\emph{J. ACM}} \bibinfo{volume}{28},
  \bibinfo{number}{1}, \bibinfo{pages}{25--40}.
\newblock
\showISSN{0004-5411}
\urldef\tempurl%
\url{https://doi.org/10.1145/322234.322238}
\showDOI{\tempurl}


\bibitem[\protect\citeauthoryear{Bloom}{Bloom}{1970}]%
        {bloom1970space}
\bibfield{author}{\bibinfo{person}{Burton~H. Bloom}.}
  \bibinfo{year}{1970}\natexlab{}.
\newblock \showarticletitle{Space/Time Trade-Offs in Hash Coding with Allowable
  Errors}.
\newblock \bibinfo{journal}{\emph{Commun. ACM}} \bibinfo{volume}{13},
  \bibinfo{number}{7}, \bibinfo{pages}{422--426}.
\newblock
\showISSN{0001-0782}
\urldef\tempurl%
\url{https://doi.org/10.1145/362686.362692}
\showDOI{\tempurl}


\bibitem[\protect\citeauthoryear{Bratbergsengen}{Bratbergsengen}{1984}]%
        {bratbergsengen1984hashzng}
\bibfield{author}{\bibinfo{person}{Kjell Bratbergsengen}.}
  \bibinfo{year}{1984}\natexlab{}.
\newblock \showarticletitle{Hashing Methods and Relational Algebra Operations}.
  In \bibinfo{booktitle}{\emph{Tenth International Conference on Very Large
  Data Bases, August 27-31, 1984, Singapore, Proceedings}}.
  \bibinfo{publisher}{Morgan Kaufmann}, \bibinfo{pages}{323--333}.
\newblock


\bibitem[\protect\citeauthoryear{Chan and Ioannidis}{Chan and
  Ioannidis}{1998}]%
        {chan1998bitvector}
\bibfield{author}{\bibinfo{person}{Chee-Yong Chan} {and}
  \bibinfo{person}{Yannis~E. Ioannidis}.} \bibinfo{year}{1998}\natexlab{}.
\newblock \showarticletitle{Bitmap Index Design and Evaluation}. In
  \bibinfo{booktitle}{\emph{Proceedings of the 1998 ACM SIGMOD International
  Conference on Management of Data}} \emph{(\bibinfo{series}{SIGMOD'98})}.
  \bibinfo{publisher}{Association for Computing Machinery},
  \bibinfo{address}{New York, NY, USA}, \bibinfo{pages}{355--366}.
\newblock
\showISBNx{0897919955}
\urldef\tempurl%
\url{https://doi.org/10.1145/276304.276336}
\showDOI{\tempurl}


\bibitem[\protect\citeauthoryear{{Chen} and {Yu}}{{Chen} and {Yu}}{1992}]%
        {chen1992interleaving}
\bibfield{author}{\bibinfo{person}{M.~. {Chen}} {and} \bibinfo{person}{P.~S.
  {Yu}}.} \bibinfo{year}{1992}\natexlab{}.
\newblock \showarticletitle{Interleaving a join sequence with semijoins in
  distributed query processing}.
\newblock \bibinfo{journal}{\emph{IEEE Transactions on Parallel and Distributed
  Systems}} \bibinfo{volume}{3}, \bibinfo{number}{5},
  \bibinfo{pages}{611--621}.
\newblock
\showISSN{2161-9883}
\urldef\tempurl%
\url{https://doi.org/10.1109/71.159044}
\showDOI{\tempurl}


\bibitem[\protect\citeauthoryear{Chen, Hsiao, and Yu}{Chen
  et~al\mbox{.}}{1993}]%
        {chen1993applying}
\bibfield{author}{\bibinfo{person}{Ming{-}Syan Chen}, \bibinfo{person}{Hui{-}I
  Hsiao}, {and} \bibinfo{person}{Philip~S. Yu}.}
  \bibinfo{year}{1993}\natexlab{}.
\newblock \showarticletitle{Applying Hash Filters to Improving the Execution of
  Bushy Trees}. In \bibinfo{booktitle}{\emph{19th International Conference on
  Very Large Data Bases, August 24-27, 1993, Dublin, Ireland, Proceedings}}.
  \bibinfo{publisher}{Morgan Kaufmann}, \bibinfo{pages}{505--516}.
\newblock


\bibitem[\protect\citeauthoryear{Chen, Hsiao, and Yu}{Chen
  et~al\mbox{.}}{1997}]%
        {chen1997applying}
\bibfield{author}{\bibinfo{person}{Ming{-}Syan Chen}, \bibinfo{person}{Hui{-}I
  Hsiao}, {and} \bibinfo{person}{Philip~S. Yu}.}
  \bibinfo{year}{1997}\natexlab{}.
\newblock \showarticletitle{On Applying Hash Filters to Improving the Execution
  of Multi-Join Queries}.
\newblock \bibinfo{journal}{\emph{{VLDB} J.}} \bibinfo{volume}{6},
  \bibinfo{number}{2}, \bibinfo{pages}{121--131}.
\newblock
\urldef\tempurl%
\url{https://doi.org/10.1007/s007780050036}
\showDOI{\tempurl}


\bibitem[\protect\citeauthoryear{Das, Yan, Za{\"{\i}}t, Valluri, Vyas,
  Krishnamachari, Gaharwar, Kamp, and Mukherjee}{Das et~al\mbox{.}}{2015}]%
        {Das:2015:QOO:2824032.2824074}
\bibfield{author}{\bibinfo{person}{Dinesh Das}, \bibinfo{person}{Jiaqi Yan},
  \bibinfo{person}{Mohamed Za{\"{\i}}t}, \bibinfo{person}{Satyanarayana~R.
  Valluri}, \bibinfo{person}{Nirav Vyas}, \bibinfo{person}{Ramarajan
  Krishnamachari}, \bibinfo{person}{Prashant Gaharwar}, \bibinfo{person}{Jesse
  Kamp}, {and} \bibinfo{person}{Niloy Mukherjee}.}
  \bibinfo{year}{2015}\natexlab{}.
\newblock \showarticletitle{Query Optimization in Oracle 12c Database
  In-Memory}.
\newblock \bibinfo{journal}{\emph{{PVLDB}}} \bibinfo{volume}{8},
  \bibinfo{number}{12}, \bibinfo{pages}{1770--1781}.
\newblock
\urldef\tempurl%
\url{https://doi.org/10.14778/2824032.2824074}
\showDOI{\tempurl}


\bibitem[\protect\citeauthoryear{Ding, Chaudhuri, and Narasayya}{Ding
  et~al\mbox{.}}{2020}]%
        {bqo20sigmod}
\bibfield{author}{\bibinfo{person}{Bailu Ding}, \bibinfo{person}{Surajit
  Chaudhuri}, {and} \bibinfo{person}{Vivek Narasayya}.}
  \bibinfo{year}{2020}\natexlab{}.
\newblock \showarticletitle{Bitvector-aware query optimization for decision
  support queries}. In \bibinfo{booktitle}{\emph{Proceedings of the 2020 ACM
  SIGMOD International Conference on Management of Data}}.
\newblock


\bibitem[\protect\citeauthoryear{Fan, Andersen, Kaminsky, and Mitzenmacher}{Fan
  et~al\mbox{.}}{2014}]%
        {fan2014cuckoo}
\bibfield{author}{\bibinfo{person}{Bin Fan}, \bibinfo{person}{Dave~G.
  Andersen}, \bibinfo{person}{Michael Kaminsky}, {and}
  \bibinfo{person}{Michael~D. Mitzenmacher}.} \bibinfo{year}{2014}\natexlab{}.
\newblock \showarticletitle{Cuckoo Filter: Practically Better Than {B}loom}. In
  \bibinfo{booktitle}{\emph{Proceedings of the 10th ACM International on
  Conference on Emerging Networking Experiments and Technologies}}
  \emph{(\bibinfo{series}{CoNEXT'14})}. \bibinfo{publisher}{Association for
  Computing Machinery}, \bibinfo{address}{New York, NY, USA},
  \bibinfo{pages}{75--88}.
\newblock
\showISBNx{9781450332798}
\urldef\tempurl%
\url{https://doi.org/10.1145/2674005.2674994}
\showDOI{\tempurl}


\bibitem[\protect\citeauthoryear{{Fender}, {Moerkotte}, {Neumann}, and
  {Leis}}{{Fender} et~al\mbox{.}}{2012}]%
        {fender2012effective}
\bibfield{author}{\bibinfo{person}{P. {Fender}}, \bibinfo{person}{G.
  {Moerkotte}}, \bibinfo{person}{T. {Neumann}}, {and} \bibinfo{person}{V.
  {Leis}}.} \bibinfo{year}{2012}\natexlab{}.
\newblock \showarticletitle{Effective and Robust Pruning for Top-Down Join
  Enumeration Algorithms}. In \bibinfo{booktitle}{\emph{2012 IEEE 28th
  International Conference on Data Engineering}}. \bibinfo{pages}{414--425}.
\newblock
\showISSN{1063-6382}
\urldef\tempurl%
\url{https://doi.org/10.1109/ICDE.2012.27}
\showDOI{\tempurl}


\bibitem[\protect\citeauthoryear{{Galindo-Legaria}, {Grabs}, {Gukal},
  {Herbert}, {Surna}, {Wang}, {Yu}, {Zabback}, and {Zhang}}{{Galindo-Legaria}
  et~al\mbox{.}}{2008}]%
        {galindo2008optimizing}
\bibfield{author}{\bibinfo{person}{C.~A. {Galindo-Legaria}},
  \bibinfo{person}{T. {Grabs}}, \bibinfo{person}{S. {Gukal}},
  \bibinfo{person}{S. {Herbert}}, \bibinfo{person}{A. {Surna}},
  \bibinfo{person}{S. {Wang}}, \bibinfo{person}{W. {Yu}}, \bibinfo{person}{P.
  {Zabback}}, {and} \bibinfo{person}{S. {Zhang}}.}
  \bibinfo{year}{2008}\natexlab{}.
\newblock \showarticletitle{Optimizing Star Join Queries for Data Warehousing
  in Microsoft SQL Server}. In \bibinfo{booktitle}{\emph{2008 IEEE 24th
  International Conference on Data Engineering}}. \bibinfo{pages}{1190--1199}.
\newblock
\showISSN{2375-026X}
\urldef\tempurl%
\url{https://doi.org/10.1109/ICDE.2008.4497528}
\showDOI{\tempurl}


\bibitem[\protect\citeauthoryear{Graefe}{Graefe}{1993}]%
        {graefe1993query}
\bibfield{author}{\bibinfo{person}{Goetz Graefe}.}
  \bibinfo{year}{1993}\natexlab{}.
\newblock \showarticletitle{Query Evaluation Techniques for Large Databases}.
\newblock \bibinfo{journal}{\emph{ACM Comput. Surv.}} \bibinfo{volume}{25},
  \bibinfo{number}{2}, \bibinfo{pages}{73--169}.
\newblock
\showISSN{0360-0300}
\urldef\tempurl%
\url{https://doi.org/10.1145/152610.152611}
\showDOI{\tempurl}


\bibitem[\protect\citeauthoryear{Graefe}{Graefe}{1995}]%
        {graefe1995cascades}
\bibfield{author}{\bibinfo{person}{Goetz Graefe}.}
  \bibinfo{year}{1995}\natexlab{}.
\newblock \showarticletitle{The {C}ascades framework for query optimization}.
\newblock \bibinfo{journal}{\emph{IEEE Data Eng. Bull.}} \bibinfo{volume}{18},
  \bibinfo{number}{3}, \bibinfo{pages}{19--29}.
\newblock


\bibitem[\protect\citeauthoryear{{Graefe} and {McKenna}}{{Graefe} and
  {McKenna}}{1993}]%
        {graefe1993volcano}
\bibfield{author}{\bibinfo{person}{G. {Graefe}} {and} \bibinfo{person}{W.~J.
  {McKenna}}.} \bibinfo{year}{1993}\natexlab{}.
\newblock \showarticletitle{The Volcano optimizer generator: extensibility and
  efficient search}. In \bibinfo{booktitle}{\emph{Proceedings of IEEE 9th
  International Conference on Data Engineering}}. \bibinfo{pages}{209--218}.
\newblock
\showISSN{null}
\urldef\tempurl%
\url{https://doi.org/10.1109/ICDE.1993.344061}
\showDOI{\tempurl}


\bibitem[\protect\citeauthoryear{Hsiao, Chen, and Yu}{Hsiao
  et~al\mbox{.}}{1994}]%
        {Hsiao:1994:PEM:191839.191879}
\bibfield{author}{\bibinfo{person}{Hui-I Hsiao}, \bibinfo{person}{Ming-Syan
  Chen}, {and} \bibinfo{person}{Philip~S. Yu}.}
  \bibinfo{year}{1994}\natexlab{}.
\newblock \showarticletitle{On Parallel Execution of Multiple Pipelined Hash
  Joins}. In \bibinfo{booktitle}{\emph{Proceedings of the 1994 ACM SIGMOD
  International Conference on Management of Data}}
  \emph{(\bibinfo{series}{SIGMOD'94})}. \bibinfo{publisher}{Association for
  Computing Machinery}, \bibinfo{address}{New York, NY, USA},
  \bibinfo{pages}{185--196}.
\newblock
\showISBNx{0897916395}
\urldef\tempurl%
\url{https://doi.org/10.1145/191839.191879}
\showDOI{\tempurl}


\bibitem[\protect\citeauthoryear{Karayannidis, Tsois, Sellis, Pieringer, Markl,
  Ramsak, Fenk, Elhardt, and Bayer}{Karayannidis et~al\mbox{.}}{2002}]%
        {Karayannidis:2002:PSQ:1287369.1287432}
\bibfield{author}{\bibinfo{person}{Nikos Karayannidis}, \bibinfo{person}{Aris
  Tsois}, \bibinfo{person}{Timos~K. Sellis}, \bibinfo{person}{Roland
  Pieringer}, \bibinfo{person}{Volker Markl}, \bibinfo{person}{Frank Ramsak},
  \bibinfo{person}{Robert Fenk}, \bibinfo{person}{Klaus Elhardt}, {and}
  \bibinfo{person}{Rudolf Bayer}.} \bibinfo{year}{2002}\natexlab{}.
\newblock \showarticletitle{Processing Star Queries on Hierarchically-Clustered
  Fact Tables}. In \bibinfo{booktitle}{\emph{Proceedings of 28th International
  Conference on Very Large Data Bases, {VLDB} 2002, Hong Kong, August 20-23,
  2002}}. \bibinfo{publisher}{Morgan Kaufmann}, \bibinfo{pages}{730--741}.
\newblock
\urldef\tempurl%
\url{https://doi.org/10.1016/B978-155860869-6/50070-6}
\showDOI{\tempurl}


\bibitem[\protect\citeauthoryear{{Lahiri}, {Chavan}, {Colgan}, {Das}, {Ganesh},
  {Gleeson}, {Hase}, {Holloway}, {Kamp}, {Lee}, {Loaiza}, {Macnaughton},
  {Marwah}, {Mukherjee}, {Mullick}, {Muthulingam}, {Raja}, {Roth}, {Soylemez},
  and {Zait}}{{Lahiri} et~al\mbox{.}}{2015}]%
        {lahiri2015oracle}
\bibfield{author}{\bibinfo{person}{T. {Lahiri}}, \bibinfo{person}{S. {Chavan}},
  \bibinfo{person}{M. {Colgan}}, \bibinfo{person}{D. {Das}},
  \bibinfo{person}{A. {Ganesh}}, \bibinfo{person}{M. {Gleeson}},
  \bibinfo{person}{S. {Hase}}, \bibinfo{person}{A. {Holloway}},
  \bibinfo{person}{J. {Kamp}}, \bibinfo{person}{T. {Lee}}, \bibinfo{person}{J.
  {Loaiza}}, \bibinfo{person}{N. {Macnaughton}}, \bibinfo{person}{V. {Marwah}},
  \bibinfo{person}{N. {Mukherjee}}, \bibinfo{person}{A. {Mullick}},
  \bibinfo{person}{S. {Muthulingam}}, \bibinfo{person}{V. {Raja}},
  \bibinfo{person}{M. {Roth}}, \bibinfo{person}{E. {Soylemez}}, {and}
  \bibinfo{person}{M. {Zait}}.} \bibinfo{year}{2015}\natexlab{}.
\newblock \showarticletitle{Oracle Database In-Memory: A dual format in-memory
  database}. In \bibinfo{booktitle}{\emph{2015 IEEE 31st International
  Conference on Data Engineering}}. \bibinfo{pages}{1253--1258}.
\newblock
\showISSN{2375-026X}
\urldef\tempurl%
\url{https://doi.org/10.1109/ICDE.2015.7113373}
\showDOI{\tempurl}


\bibitem[\protect\citeauthoryear{Lang, Neumann, Kemper, and Boncz}{Lang
  et~al\mbox{.}}{2019}]%
        {lang2019performance}
\bibfield{author}{\bibinfo{person}{Harald Lang}, \bibinfo{person}{Thomas
  Neumann}, \bibinfo{person}{Alfons Kemper}, {and} \bibinfo{person}{Peter~A.
  Boncz}.} \bibinfo{year}{2019}\natexlab{}.
\newblock \showarticletitle{Performance-Optimal Filtering: Bloom overtakes
  Cuckoo at High-Throughput}.
\newblock \bibinfo{journal}{\emph{{PVLDB}}} \bibinfo{volume}{12},
  \bibinfo{number}{5}, \bibinfo{pages}{502--515}.
\newblock
\urldef\tempurl%
\url{https://doi.org/10.14778/3303753.3303757}
\showDOI{\tempurl}


\bibitem[\protect\citeauthoryear{Leis, Radke, Gubichev, Mirchev, Boncz, Kemper,
  and Neumann}{Leis et~al\mbox{.}}{2018}]%
        {leis2018query}
\bibfield{author}{\bibinfo{person}{Viktor Leis}, \bibinfo{person}{Bernhard
  Radke}, \bibinfo{person}{Andrey Gubichev}, \bibinfo{person}{Atanas Mirchev},
  \bibinfo{person}{Peter~A. Boncz}, \bibinfo{person}{Alfons Kemper}, {and}
  \bibinfo{person}{Thomas Neumann}.} \bibinfo{year}{2018}\natexlab{}.
\newblock \showarticletitle{Query optimization through the looking glass, and
  what we found running the Join Order Benchmark}.
\newblock \bibinfo{journal}{\emph{{VLDB} J.}} \bibinfo{volume}{27},
  \bibinfo{number}{5}, \bibinfo{pages}{643--668}.
\newblock
\urldef\tempurl%
\url{https://doi.org/10.1007/s00778-017-0480-7}
\showDOI{\tempurl}


\bibitem[\protect\citeauthoryear{Moerkotte and Neumann}{Moerkotte and
  Neumann}{2006}]%
        {moerkotte2006analysis}
\bibfield{author}{\bibinfo{person}{Guido Moerkotte} {and}
  \bibinfo{person}{Thomas Neumann}.} \bibinfo{year}{2006}\natexlab{}.
\newblock \showarticletitle{Analysis of Two Existing and One New Dynamic
  Programming Algorithm for the Generation of Optimal Bushy Join Trees without
  Cross Products}. In \bibinfo{booktitle}{\emph{Proceedings of the 32nd
  International Conference on Very Large Data Bases, Seoul, Korea, September
  12-15, 2006}}. \bibinfo{publisher}{{ACM}}, \bibinfo{pages}{930--941}.
\newblock
\urldef\tempurl%
\url{http://dl.acm.org/citation.cfm?id=1164207}
\showURL{%
\tempurl}


\bibitem[\protect\citeauthoryear{Moerkotte and Neumann}{Moerkotte and
  Neumann}{2008}]%
        {moerkotte2008dynamic}
\bibfield{author}{\bibinfo{person}{Guido Moerkotte} {and}
  \bibinfo{person}{Thomas Neumann}.} \bibinfo{year}{2008}\natexlab{}.
\newblock \showarticletitle{Dynamic Programming Strikes Back}. In
  \bibinfo{booktitle}{\emph{Proceedings of the 2008 ACM SIGMOD International
  Conference on Management of Data}} \emph{(\bibinfo{series}{SIGMOD'08})}.
  \bibinfo{publisher}{Association for Computing Machinery},
  \bibinfo{address}{New York, NY, USA}, \bibinfo{pages}{539--552}.
\newblock
\showISBNx{9781605581026}
\urldef\tempurl%
\url{https://doi.org/10.1145/1376616.1376672}
\showDOI{\tempurl}


\bibitem[\protect\citeauthoryear{Neumann}{Neumann}{2009a}]%
        {neumann2009query}
\bibfield{author}{\bibinfo{person}{Thomas Neumann}.}
  \bibinfo{year}{2009}\natexlab{a}.
\newblock \showarticletitle{Query Simplification: Graceful Degradation for
  Join-Order Optimization}. In \bibinfo{booktitle}{\emph{Proceedings of the
  2009 ACM SIGMOD International Conference on Management of Data}}
  \emph{(\bibinfo{series}{SIGMOD'09})}. \bibinfo{publisher}{Association for
  Computing Machinery}, \bibinfo{address}{New York, NY, USA},
  \bibinfo{pages}{403--414}.
\newblock
\showISBNx{9781605585512}
\urldef\tempurl%
\url{https://doi.org/10.1145/1559845.1559889}
\showDOI{\tempurl}


\bibitem[\protect\citeauthoryear{Neumann}{Neumann}{2009b}]%
        {Neumann:2009:QSG:1559845.1559889}
\bibfield{author}{\bibinfo{person}{Thomas Neumann}.}
  \bibinfo{year}{2009}\natexlab{b}.
\newblock \showarticletitle{Query Simplification: Graceful Degradation for
  Join-Order Optimization}. In \bibinfo{booktitle}{\emph{Proceedings of the
  2009 ACM SIGMOD International Conference on Management of Data}}
  \emph{(\bibinfo{series}{SIGMOD'09})}. \bibinfo{publisher}{Association for
  Computing Machinery}, \bibinfo{address}{New York, NY, USA},
  \bibinfo{pages}{403--414}.
\newblock
\showISBNx{9781605585512}
\urldef\tempurl%
\url{https://doi.org/10.1145/1559845.1559889}
\showDOI{\tempurl}


\bibitem[\protect\citeauthoryear{Neumann and Galindo{-}Legaria}{Neumann and
  Galindo{-}Legaria}{2013}]%
        {neumann2013taking}
\bibfield{author}{\bibinfo{person}{Thomas Neumann} {and}
  \bibinfo{person}{C{\'{e}}sar~A. Galindo{-}Legaria}.}
  \bibinfo{year}{2013}\natexlab{}.
\newblock \showarticletitle{Taking the Edge off Cardinality Estimation Errors
  using Incremental Execution}. In \bibinfo{booktitle}{\emph{Datenbanksysteme
  f{\"{u}}r Business, Technologie und Web (BTW), 15. Fachtagung des
  GI-Fachbereichs "Datenbanken und Informationssysteme" (DBIS), 11.-15.3.2013
  in Magdeburg, Germany. Proceedings}} \emph{(\bibinfo{series}{{LNI}})},
  Vol.~\bibinfo{volume}{{P-214}}. \bibinfo{publisher}{{GI}},
  \bibinfo{pages}{73--92}.
\newblock
\urldef\tempurl%
\url{https://dl.gi.de/20.500.12116/17356}
\showURL{%
\tempurl}


\bibitem[\protect\citeauthoryear{Ono and Lohman}{Ono and Lohman}{1990}]%
        {ono1990measuring}
\bibfield{author}{\bibinfo{person}{Kiyoshi Ono} {and} \bibinfo{person}{Guy~M.
  Lohman}.} \bibinfo{year}{1990}\natexlab{}.
\newblock \showarticletitle{Measuring the Complexity of Join Enumeration in
  Query Optimization}. In \bibinfo{booktitle}{\emph{16th International
  Conference on Very Large Data Bases, August 13-16, 1990, Brisbane,
  Queensland, Australia, Proceedings}}. \bibinfo{publisher}{Morgan Kaufmann},
  \bibinfo{pages}{314--325}.
\newblock


\bibitem[\protect\citeauthoryear{Putze, Sanders, and Singler}{Putze
  et~al\mbox{.}}{2007}]%
        {putze2007cache}
\bibfield{author}{\bibinfo{person}{Felix Putze}, \bibinfo{person}{Peter
  Sanders}, {and} \bibinfo{person}{Johannes Singler}.}
  \bibinfo{year}{2007}\natexlab{}.
\newblock \showarticletitle{Cache-, hash-and space-efficient bloom filters}. In
  \bibinfo{booktitle}{\emph{International Workshop on Experimental and
  Efficient Algorithms}}. Springer, \bibinfo{pages}{108--121}.
\newblock


\bibitem[\protect\citeauthoryear{Seshadri, Hellerstein, Pirahesh, Leung,
  Ramakrishnan, Srivastava, Stuckey, and Sudarshan}{Seshadri
  et~al\mbox{.}}{1996}]%
        {seshadri1996cost}
\bibfield{author}{\bibinfo{person}{Praveen Seshadri},
  \bibinfo{person}{Joseph~M. Hellerstein}, \bibinfo{person}{Hamid Pirahesh},
  \bibinfo{person}{T.~Y.~Cliff Leung}, \bibinfo{person}{Raghu Ramakrishnan},
  \bibinfo{person}{Divesh Srivastava}, \bibinfo{person}{Peter~J. Stuckey},
  {and} \bibinfo{person}{S. Sudarshan}.} \bibinfo{year}{1996}\natexlab{}.
\newblock \showarticletitle{Cost-Based Optimization for Magic: Algebra and
  Implementation}. In \bibinfo{booktitle}{\emph{Proceedings of the 1996 ACM
  SIGMOD International Conference on Management of Data}}
  \emph{(\bibinfo{series}{SIGMOD'96})}. \bibinfo{publisher}{Association for
  Computing Machinery}, \bibinfo{address}{New York, NY, USA},
  \bibinfo{pages}{435--446}.
\newblock
\showISBNx{0897917944}
\urldef\tempurl%
\url{https://doi.org/10.1145/233269.233360}
\showDOI{\tempurl}


\bibitem[\protect\citeauthoryear{Simmen, Shekita, and Malkemus}{Simmen
  et~al\mbox{.}}{1996}]%
        {Simmen:1996:FTO:235968.233320}
\bibfield{author}{\bibinfo{person}{David Simmen}, \bibinfo{person}{Eugene
  Shekita}, {and} \bibinfo{person}{Timothy Malkemus}.}
  \bibinfo{year}{1996}\natexlab{}.
\newblock \showarticletitle{Fundamental Techniques for Order Optimization}. In
  \bibinfo{booktitle}{\emph{Proceedings of the 1996 ACM SIGMOD International
  Conference on Management of Data}} \emph{(\bibinfo{series}{SIGMOD'96})}.
  \bibinfo{publisher}{Association for Computing Machinery},
  \bibinfo{address}{New York, NY, USA}, \bibinfo{pages}{57--67}.
\newblock
\showISBNx{0897917944}
\urldef\tempurl%
\url{https://doi.org/10.1145/233269.233320}
\showDOI{\tempurl}


\bibitem[\protect\citeauthoryear{Soliman, Antova, Raghavan, El-Helw, Gu, Shen,
  Caragea, Garcia-Alvarado, Rahman, Petropoulos, and et~al.}{Soliman
  et~al\mbox{.}}{2014}]%
        {Soliman:2014:OMQ:2588555.2595637}
\bibfield{author}{\bibinfo{person}{Mohamed~A. Soliman},
  \bibinfo{person}{Lyublena Antova}, \bibinfo{person}{Venkatesh Raghavan},
  \bibinfo{person}{Amr El-Helw}, \bibinfo{person}{Zhongxian Gu},
  \bibinfo{person}{Entong Shen}, \bibinfo{person}{George~C. Caragea},
  \bibinfo{person}{Carlos Garcia-Alvarado}, \bibinfo{person}{Foyzur Rahman},
  \bibinfo{person}{Michalis Petropoulos}, {and} \bibinfo{person}{et al.}}
  \bibinfo{year}{2014}\natexlab{}.
\newblock \showarticletitle{Orca: A Modular Query Optimizer Architecture for
  Big Data}. In \bibinfo{booktitle}{\emph{Proceedings of the 2014 ACM SIGMOD
  International Conference on Management of Data}}
  \emph{(\bibinfo{series}{SIGMOD'14})}. \bibinfo{publisher}{Association for
  Computing Machinery}, \bibinfo{address}{New York, NY, USA},
  \bibinfo{pages}{337--348}.
\newblock
\showISBNx{9781450323765}
\urldef\tempurl%
\url{https://doi.org/10.1145/2588555.2595637}
\showDOI{\tempurl}


\bibitem[\protect\citeauthoryear{Valduriez and Gardarin}{Valduriez and
  Gardarin}{1984}]%
        {valduriez1984join}
\bibfield{author}{\bibinfo{person}{Patrick Valduriez} {and}
  \bibinfo{person}{Georges Gardarin}.} \bibinfo{year}{1984}\natexlab{}.
\newblock \showarticletitle{Join and Semijoin Algorithms for a Multiprocessor
  Database Machine}.
\newblock \bibinfo{journal}{\emph{ACM Trans. Database Syst.}}
  \bibinfo{volume}{9}, \bibinfo{number}{1}, \bibinfo{pages}{133--161}.
\newblock
\showISSN{0362-5915}
\urldef\tempurl%
\url{https://doi.org/10.1145/348.318590}
\showDOI{\tempurl}


\bibitem[\protect\citeauthoryear{Weininger}{Weininger}{2002}]%
        {Weininger:2002:EEJ:564691.564754}
\bibfield{author}{\bibinfo{person}{Andreas Weininger}.}
  \bibinfo{year}{2002}\natexlab{}.
\newblock \showarticletitle{Efficient Execution of Joins in a Star Schema}. In
  \bibinfo{booktitle}{\emph{Proceedings of the 2002 ACM SIGMOD International
  Conference on Management of Data}} \emph{(\bibinfo{series}{SIGMOD'02})}.
  \bibinfo{publisher}{Association for Computing Machinery},
  \bibinfo{address}{New York, NY, USA}, \bibinfo{pages}{542--545}.
\newblock
\showISBNx{1581134975}
\urldef\tempurl%
\url{https://doi.org/10.1145/564691.564754}
\showDOI{\tempurl}


\bibitem[\protect\citeauthoryear{Zhu, Potti, Saurabh, and Patel}{Zhu
  et~al\mbox{.}}{2017}]%
        {zhu2017looking}
\bibfield{author}{\bibinfo{person}{Jianqiao Zhu}, \bibinfo{person}{Navneet
  Potti}, \bibinfo{person}{Saket Saurabh}, {and} \bibinfo{person}{Jignesh~M.
  Patel}.} \bibinfo{year}{2017}\natexlab{}.
\newblock \showarticletitle{Looking Ahead Makes Query Plans Robust}.
\newblock \bibinfo{journal}{\emph{{PVLDB}}} \bibinfo{volume}{10},
  \bibinfo{number}{8}, \bibinfo{pages}{889--900}.
\newblock
\urldef\tempurl%
\url{https://doi.org/10.14778/3090163.3090167}
\showDOI{\tempurl}


\end{thebibliography}
	
\end{document}